\documentclass[12pt,english]{article}
\pdfoutput=1

\usepackage[authordate,
backend=biber,
doi=only,
isbn=false,
sorting=nyt,
maxcitenames=3,
minbibnames=7,
maxbibnames=7,
uniquename=false,
sortcites=true]{biblatex-chicago}

\bibliography{bib/references.bib}

\AtEveryBibitem{\clearlist{note}\clearlist{language}\clearlist{issn}} 
\AtEveryBibitem{%
	\ifentrytype{online}{%
		\clearfield{urlyear}
		\clearfield{urlmonth}
		\clearfield{urlday}
		\clearfield{note}
		\clearlist{language}
	}{%
		\clearfield{eprint}%
		\clearfield{urlyear}
		\clearfield{urlmonth}
		\clearfield{urlday}
		\clearfield{note}
		\clearlist{language}
	}
}%

\usepackage{mathpazo}
\usepackage{newtxtext}
\usepackage{microtype}
\everypar{\looseness=-1}
\usepackage{amsmath}
\usepackage{amssymb}
\usepackage[utf8]{inputenc}
\usepackage[T1]{fontenc}
\usepackage{babel}
\usepackage{setspace}
\usepackage{graphicx}
\usepackage{longtable,booktabs,threeparttablex}
\usepackage{array}
\usepackage{multirow}
\usepackage[usenames,dvipsnames,svgnames,table]{xcolor}
\usepackage[export]{adjustbox}[2011/08/13]
\usepackage{enumitem}
\usepackage[text={16cm,24cm}]{geometry}
\usepackage{ragged2e}
\usepackage{csquotes}

\usepackage[hang, flushmargin, bottom, symbol]{footmisc}
\usepackage[colorlinks=true, linkcolor=blue, citecolor=blue, plainpages=false, pdfpagelabels=true, urlcolor=blue]{hyperref}
\usepackage{float}

\usepackage{mathtools}
\usepackage{accents} 
\newcommand{\dbtilde}[1]{\accentset{\approx}{#1}}
\usepackage{xspace}
\usepackage{calculator}
\newcommand{\rowgroup}[1]{\hspace{-0.5em}#1}
\usepackage{makecell}
\usepackage{tikz}
\usetikzlibrary{positioning}

\usepackage{subcaption}
\usepackage{caption}
\makeatletter
\date{September 8, 2021}

\usepackage{footnotebackref}
\newcommand{\CISESIZE}{\scriptsize}
\newcommand{\CISESIZEMT}{\scriptstyle}

\newcommand{\CI}[3]{
{\vspace{+0.5mm}\CISESIZE(\hspace{0.5mm}#3\hspace{0.5mm})}
}
\newcommand{\CIT}[3]{
\hspace{0.1mm}
{\CISESIZE(\hspace{0.5mm}#3\hspace{0.5mm})}
}
\newcommand{\CII}[3]{
\vspace{+0.5mm}
{\CISESIZE $\left[\thinspace #1 \text{,} \thinspace #2 \thinspace \right]_{\thinspace\text{95\%}}^{\thinspace\text{c.i.}}$}
}

\newcommand{\TTSS}[2]{
${\CISESIZEMT \overset{\text{data}}{#1}} \thinspace\thinspace \bigl\rvert $ \hspace{-2mm} #2
}

\newcommand{\psep}{;}
\newcommand{\nextt}{'}
\newcommand{\opti}{*}

\newcommand{\condi}{\mid}

\newcommand{\suchthat}{\mid}
\newcommand{\nextsup}{^{\nextt}}

\newcommand{\ITG}{\int}
\newcommand{\ITGD}{\mathrm{d}}
\newcommand{\SUMA}{\sum}
\newcommand{\PMULT}{\prod}

\newcommand{\HATT}{\widehat}
\newcommand{\LOGG}{\log}
\newcommand{\EXPP}{\exp}
\newcommand{\INDI}{\mathbb{I}}

\newcommand{\LESEQ}{\leqslant}
\newcommand{\GRTEQ}{\geqslant}

\newcommand{\INTEGER}{\mathbb{N}}
\newcommand{\REAL}{\mathbb{R}}

\newcommand{\PDF}{f}
\newcommand{\CDF}{F}
\newcommand{\PROB}{P}
\newcommand{\E}{\mathrm{E}}
\newcommand{\MEAN}{mean}

\newcommand{\J}{j}
\newcommand{\JJ}{J}
\newcommand{\OtS}{1,...,7} \newcommand{\JinOtS}{\J\in\left\{\OtS\right\}}
\newcommand{\F}{F}
\newcommand{\I}{I}

\newcommand{\FFBB}{\text{FB}}
\newcommand{\FFSS}{\text{FS}}
\newcommand{\IIBB}{\text{IB}}
\newcommand{\IILL}{\text{IS}}
\newcommand{\FFBBIIBB}{\text{FB+IB}}
\newcommand{\FFBBIILL}{\text{FB+IS}}

\newcommand{\MINTXT}{\text{min}}
\newcommand{\MAXTXT}{\text{max}}
\DeclareMathOperator*{\argmax}{arg\,max}
\DeclareMathOperator*{\argmin}{arg\,min}

\newcommand{\discalpha}{q}

\newcommand{\totk}{N_\K}
\newcommand{\totb}{N_\B}
\newcommand{\disck}{n_\K}
\newcommand{\discb}{n_\B}

\newcommand{\totcoh}{N_\COH}
\newcommand{\disccoh}{{n_\COH}}

\newcommand{\totA}{Q}
\newcommand{\discA}{\discalpha}
\newcommand{\eachSolveR}{G}
\newcommand{\eachSolveReach}{g}
\newcommand{\totR}{M}
\newcommand{\discR}{m}

\newcommand{\EstiTime}{\tau}
\newcommand{\EstiRegion}{r}
\newcommand{\ETR}{_{\EstiRegion\tau}}

\newcommand{\totAWSdraws}{\mathcal{N}}
\newcommand{\eachAWSdraws}{\mathit{n}}
\newcommand{\totAWSobjs}{\mathcal{M}}
\newcommand{\eachAWSobjs}{\mathit{m}}

\newcommand{\totAWSparam}{\mathcal{Q}}
\newcommand{\eachAWSparam}{\mathit{q}}

\newcommand{\eachAWSdrawsONE}{\eachAWSparam_1}
\newcommand{\eachAWSdrawsTWO}{\eachAWSparam_2}
\newcommand{\eachAWSdrawsTHREE}{\eachAWSparam_3}

\newcommand{\totAWSpolyDegree}{\mathcal{D}}

\newcommand{\totAWSpoly}{\mathcal{P}}

\newcommand{\modelObjMatrix}{\mathbb{\mathcal{O}}}
\newcommand{\modelObjMatrixHat}{\hat{\mathbb{\mathcal{O}}}}

\newcommand{\regCoefMatrix}{\mathbb{\mathcal{R}}}
\newcommand{\regCoefMatrixHat}{\hat{\mathbb{\mathcal{R}}}}
\newcommand{\dataMatrix}{\mathbb{\mathcal{X}}}

\newcommand{\residualEle}{\mathit{e}}

\newcommand{\ONEOBJ}{
\modelObjMatrix_{\eachAWSdraws}^{\eachAWSobjs}
\left(
  \left\{
    \ESTIeach_{\eachAWSdraws\eachAWSparam}
  \right\}_{\eachAWSparam = 1}^{\totAWSparam}
  \condi
  \dataMatrix_\eachAWSobjs
\right)
}

\newcommand{\zoom}{\left(\iota\right)}
\newcommand{\zoomopt}[1]{\left(\iota=#1\right)}
\newcommand{\eqitnobracket}{\tau}
\newcommand{\eqit}{\left(\eqitnobracket\right)}
\newcommand{\eqitopt}[1]{\left(\eqitnobracket+#1\right)}
\newcommand{\eqitoptequal}[1]{\left(\eqitnobracket=#1\right)}

\newcommand{\HH}{i}
\newcommand{\ET}{t}

\newcommand{\HT}{_{\HH\ET}}

\newcommand{\estiobs}{N_\EstiRegion}
\newcommand{\estiobsi}{i}
\newcommand{\estitime}{T_\EstiTime}
\newcommand{\estitimei}{t}
\newcommand{\EIT}{_{\estiobsi\estitimei}}

\newcommand{\subhhj}{\HH\J}

\newcommand{\ESTIeach}{\theta}
\newcommand{\ESTIALL}{\Theta}
\newcommand{\ESTIALLETR}{\ESTIALL_{\ETR}}

\newcommand{\ESTIAPPROX}[1]{\Theta^{\opti}_{#1}}

\newcommand{\PBETA}{\beta}
\newcommand{\PCRRA}{\rho}
\newcommand{\YMIN}{\Y^{\MINTXT}}
\newcommand{\PPHISIGMA}{\sigma_{\widehat{\phi}}}

\newcommand{\PPHISIGMASub}[1]{\sigma_{\widehat{\phi#1}}}

\newcommand{\PALPHA}{\alpha}
\newcommand{\PDELTA}{\delta}

\newcommand{\PASIGMA}{\sigma_{\A}}
\newcommand{\PASIGMASub}[1]{\sigma_{\A#1}}
\newcommand{\PAMU}{\mu_{\A}}
\newcommand{\PAMUSub}[1]{\mu_{\A{#1}}}

\newcommand{\EESIGMA}{\sigma_{\SHK}}
\newcommand{\EESIGMASub}[1]{\sigma_{\SHK{#1}}}

\newcommand{\FXC}{\Psi}
\newcommand{\FXCj}{\psi_\J}
\newcommand{\FXCjfb}{\psi_{\FFBB}}
\newcommand{\FXCjfs}{\psi_{\FFSS}}
\newcommand{\FXCjib}{\psi_{\IIBB}}
\newcommand{\FXCjil}{\psi_{\IILL}}

\newcommand{\FXCjfbM}{\(\psi_{\FFBB}\)}
\newcommand{\FXCjfsM}{\(\psi_{\FFSS}\)}
\newcommand{\FXCjibM}{\(\psi_{\IIBB}\)}
\newcommand{\FXCjilM}{\(\psi_{\IILL}\)}

\newcommand{\PKPPA}{\gamma}

\newcommand{\MINBORRj}{\B^{\text{loan-smallest}}_\J}

\newcommand{\MINjfb}{\B_{\MAXTXT}^{\FFBB}}
\newcommand{\MINjfs}{\B_{\MINTXT}^{\FFSS}}
\newcommand{\MINjib}{\B_{\MAXTXT}^{\IIBB}}
\newcommand{\MINjil}{\B_{\MINTXT}^{\IILL}}

\newcommand{\PZETA}{\zeta}
\newcommand{\BMsrErr}{\PDF_{\PZETA^{\B}}}
\newcommand{\KMsrErr}{\PDF_{\PZETA^{\K}}}
\newcommand{\CCMsrErr}{\PDF_{\PZETA^{\CC}}}

\newcommand{\MEASURE}{\pmb{\mu}}

\newcommand{\PR}{r}
\newcommand{\PRj}{\PR_\J}

\newcommand{\R}{r}
\newcommand{\RI}{\R^\I}
\newcommand{\Rj}{\R_\J}

\newcommand{\Rjfb}{\R^\FFBB}
\newcommand{\Rjfs}{\R^\FFSS}

\newcommand{\A}{\mathcal{A}}
\newcommand{\Ahat}{\HATT{\mathcal{A}}}

\newcommand{\STATEset}{S}
\newcommand{\STATEsetEle}{s}
\newcommand{\Y}{y}
\newcommand{\YMSR}{\widehat{\Y}}
\newcommand{\K}{k}
\newcommand{\KSET}{\mathcal{K}}
\newcommand{\B}{b}
\newcommand{\BSET}{\mathcal{B}}
\newcommand{\COH}{\mathcal{W}}
\newcommand{\COHMSR}{\widehat{\mathcal{W}}}
\newcommand{\COHi}{\mathcal{W}_\HH}
\newcommand{\SHK}{\epsilon}
\newcommand{\SHKnext}{\SHK\nextsup}
\newcommand{\SHKU}{\Phi}
\newcommand{\SHKUj}{\phi_\J}
\newcommand{\SHKUp}{\SHKU\nextsup}
\newcommand{\SHKUjopt}[1]{\phi_{#1}}
\newcommand{\SHKUdist}{\pi_{\SHKU}}

\newcommand{\STATES}{\A, \K, \B, \SHK, \SHKU}
\newcommand{\STATESnext}{\A, \Kp, \Bp, \SHKnext, \SHKUp}
\newcommand{\STATESj}{\A, \K, \B, \SHK}
\newcommand{\STATESjuj}{\A, \K, \B, \SHK, \SHKUj}
\newcommand{\STATESjujopt}[1]{\A, \K, \B, \SHK, \SHKUjopt{#1}}

\newcommand{\OO}{o} \newcommand{\CC}{c}

\newcommand{\CCMSR}{\widehat{\CC}}
\newcommand{\Kp}{\K\nextsup}
\newcommand{\KMSR}{\widehat{\K}}
\newcommand{\KpMSR}{\widehat{\K}\nextsup}

\newcommand{\KpSET}[1]{\KSET_{#1}\nextsup}

\newcommand{\Bp}{\B\nextsup}
\newcommand{\BMSR}{\widehat{\B}}
\newcommand{\BpMSR}{\widehat{\B}\nextsup}

\newcommand{\COHp}{\COH\nextsup}
\newcommand{\PP}{P}

\newcommand{\BIp}{\B^{\nextt\I}}

\newcommand{\BIpSET}[1]{\BSET_{#1}^{\nextt\I}} \newcommand{\BFp}{\B^{\nextt\F}}

\newcommand{\BFpSET}[1]{\BSET_{#1}^{\nextt\F}}

\newcommand{\PPFB}{\PROB^{\FFBB}}
\newcommand{\PPFS}{\PROB^{\FFSS}}
\newcommand{\PPIB}{\PROB^{\IIBB}}
\newcommand{\PPIL}{\PROB^{\IILL}}
\newcommand{\PPFIB}{\PROB^{\FFBBIIBB}}
\newcommand{\PPFBIS}{\PROB^{\FFBBIILL}}

\newcommand{\CAPCAP}{\expandafter\MakeUppercase}
\newcommand{\CZH}{wealth\xspace}
\newcommand{\CHOHH}{cash-on-hand\xspace}
\newcommand{\ICM}{income\xspace}
\newcommand{\OTP}{output\xspace}
\newcommand{\RKI}{risky investment\xspace}
\newcommand{\PYS}{physical capital\xspace}
\newcommand{\FNT}{financial asset\xspace}
\newcommand{\PDT}{productivity type\xspace}
\newcommand{\SKE}{productivity shock\xspace}
\newcommand{\SKEs}{productivity shocks\xspace}
\newcommand{\USK}{credit category utility shocks\xspace}

\newcommand{\SSSSRKI}{savings and risky investment\xspace}

\newcommand{\NLC}{formal\xspace}
\newcommand{\NLCLY}{formally\xspace}
\newcommand{\LLC}{informal\xspace}
\newcommand{\LLCLY}{informally\xspace}
\newcommand{\FBB}{formal borrowing\xspace}
\newcommand{\IBB}{informal borrowing\xspace}
\newcommand{\FJB}{joint borrowing\xspace}
\newcommand{\YMINZERO}{minimum \OTP\xspace}

\newcommand{\FSS}{formal savings\xspace}
\newcommand{\ISS}{informal savings\xspace}
\newcommand{\FJS}{joint formal borrowing and informal savings\xspace}

\newcommand{\SSSS}{savings\xspace}
\newcommand{\SSSnoS}{saving\xspace}
\newcommand{\BBBB}{borrowing\xspace}

\newcommand{\SSshort}{save\xspace}
\newcommand{\BBshort}{borr\xspace}

\newcommand{\KFSS}{Formal Saving\xspace}

\newcommand{\CLC}{collateral constraint\xspace}
\newcommand{\CLCs}{collateral constraints\xspace}

\newcommand{\IORATEFOR}{Asset Outflows\xspace}

\newcommand{\IORATEFORDESC}{excess \NLC \(\B\) supply \% of \OTP}

\newcommand{\EFRATEFOR}{Financing Share\xspace}

\newcommand{\EFRATEDESC}{all borrowing \% of \OTP}

\newcommand{\ITR}{interest rate\xspace}
\newcommand{\ITRs}{interest rates\xspace}
\newcommand{\FCF}{fixed costs\xspace}
\newcommand{\FCFs}{fixed costs\xspace}

\newcommand{\fcoh}{\mathit{\COH}}
\newcommand{\fb}{\mathit{\B\nextsup_{\J}}}
\newcommand{\fbi}{\mathit{\B^{\nextt\I}_\J}}
\newcommand{\fbf}{\mathit{\B^{\nextt\F}_\J}}
\newcommand{\fk}{\mathit{\K\nextsup_\J}}
\newcommand{\fc}{\mathit{\CC_\J}}
\newcommand{\fo}{\mathit{\OO}}
\newcommand{\fp}{\PP_\J}

\newcommand{\utt}{u}

\newcommand{\uallj}{\mathcal{U}_\J}

\newcommand{\FCOH}{\fcoh \left( \STATESj \right)}

\newcommand{\FBcohs}{\fb \left( \A, \fcoh \right)}

\newcommand{\FKcohs}{\fk \left( \A, \fcoh \right)}

\newcommand{\FPcohs}{\fp \left( \A, \fcoh \right)}

\newcommand{\FCcohsESTI}{\fc \left( \A, \fcoh \psep \ESTIALLETR \right)}
\newcommand{\FBcohsESTI}{\fb \left( \A, \fcoh \psep \ESTIALLETR \right)}
\newcommand{\FKcohsESTI}{\fk \left( \A, \fcoh \psep \ESTIALLETR \right)}
\newcommand{\FPcohsESTI}{\fp \left( \A, \fcoh \psep \ESTIALLETR \right)}

\newcommand{\newred}{black}
\newcommand{\Vakbephi}{\textcolor{blue}{v}}
\newcommand{\Vakbejuj}{\textcolor{\newred}{v_{j}}}
\newcommand{\VakbejujOpt}[1]{\textcolor{\newred}{v_{#1}}}
\newcommand{\Vakbej}{\textcolor{\newred}{\hat{v}_{\J}}}
\newcommand{\Vakbeitj}{\textcolor{\newred}{\hat{v}_{\HH\ET\J}}}
\newcommand{\Vakbeitl}{\textcolor{\newred}{\hat{v}_{\HH\ET l}}}

\newcommand{\EVakbe}{\textcolor{blue}{\upsilon}}
\newcommand{\EVacohk}{\textcolor{blue}{E\widetilde{\upsilon}}_{\A}}
\newcommand{\EVakb}{\textcolor{blue}{E\dbtilde{\upsilon}}_{\A}}

\newcommand{\RAWONE}{\left(\STATES \right)}
\newcommand{\RAWTWO}{\left(\STATESjuj \right)}
\newcommand{\RAWTHREE}{\left(\STATESj \right)}

\newcommand{\INTPONES}{\left(\FCOH, \K\right)}
\newcommand{\INTPTWOS}{\left(\B, \K \right)}

\newcommand{\VakbephiStates}{\Vakbephi\left(\STATES\right)}
\newcommand{\VakbejujOptStates}[1]{\VakbejujOpt{#1}\left(\STATESjujopt{#1}\right)}
\newcommand{\EVJEVp}[1]{
 \E_{\SHKnext,\SHKUp}
 \left(\Vakbephi
 \left(\A, \Kp, #1, \SHKnext, \SHKUp \right)
 \right)
}

\newcommand{\EVJKset}{\Kp \GRTEQ 0}
\newcommand{\EVJIBset}{\BIp \LESEQ \MINjib}
\newcommand{\EVJISset}{\BIp \GRTEQ \MINjil}
\newcommand{\EVJFBset}{-\PKPPA\Kp \LESEQ \BFp \LESEQ \MINjfb}
\newcommand{\EVJFSset}{\BFp \GRTEQ \MINjfs}

\newcommand{\UtCBK}[4]{
 \utt\left(
 \FCOH - \Kp
 - \frac{#1}{1+\textcolor{\newred}{#2}}
 - \textcolor{blue}{#3}
 \right) + \SHKUjopt{#4}
}
\newcommand{\UtCBKjoint}[7]{
 \utt\left(
 \FCOH - \Kp
 - \frac{#1}{1+\textcolor{\newred}{#2}}
 - \frac{#3}{1+\textcolor{\newred}{#4}}
 - \textcolor{blue}{#5}
 - \textcolor{blue}{#6}
 \right) + \SHKUjopt{#7}
}

\newcommand{\mmsbeg}{ \begin{array}{c} }
\newcommand{\mmsend}{ \end{array} }

\newcommand{\Bphhjmin}{\B_{\subhhj}^{\nextt\MINTXT}}
\newcommand{\Bphhjmax}{\B_{\subhhj}^{\nextt\MAXTXT}}
\newcommand{\Kphhjbmin}{\K_{\subhhj \left(\Bp\right)}^{\nextt\MINTXT}}
\newcommand{\Kphhjbmax}{\K_{\subhhj \left(\Bp\right)}^{\nextt\MAXTXT}}

\newcommand{\Bphhjopti}{\B_{\subhhj}^{\nextt\opti}}

\newcommand{\disckzoom}{\disck^{\zoom}}

\newcommand{\discbzoom}{\discb^{\zoom}}

\newcommand{\subdisckzoom}{(\disck^{\zoom})}
\newcommand{\subdiscbzoom}{(\discb^{\zoom})}

\newcommand{\disckopti}{n^{\opti\zoom}_k}
\newcommand{\discbopti}{n^{\opti\zoom}_b}

\newcommand{\discboptiopt}[1]{n^{\opti\zoomopt{#1}}_b}

\newcommand{\Kijnbnkzoom}{\Kp_{\subhhj\subdiscbzoom\subdisckzoom}}
\newcommand{\Bijnbzoom}{\Bp_{\subhhj\subdiscbzoom}}
\newcommand{\Bijnbzoomopti}[1]{\Bp_{\subhhj\discboptiopt{#1}}}

\newcommand{\Bphhjminzoom}{\B_{\subhhj}^{\nextt\MINTXT,\zoom}}
\newcommand{\Bphhjmaxzoom}{\B_{\subhhj}^{\nextt\MAXTXT,\zoom}}
\newcommand{\Bphhjminzoomopt}[1]{\B_{\subhhj}^{\nextt\MINTXT,\zoomopt{#1}}}
\newcommand{\Bphhjmaxzoomopt}[1]{\B_{\subhhj}^{\nextt\MAXTXT,\zoomopt{#1}}}

\newcommand{\Kphhjnbminzoom}{\K_{\subhhj\subdiscbzoom}^{\nextt\MINTXT,\zoom}}
\newcommand{\Kphhjnbmaxzoom}{\K_{\subhhj\subdiscbzoom}^{\nextt\MAXTXT,\zoom}}

\newcommand{\subdisccoh}{\disccoh}
\newcommand{\Wnw}{\COH_{\subdisccoh}}

\newcommand{\Wnwp}{\COH_{\subdisccoh}^{\nextt}}

\newcommand{\subdisccohcdf}{\HATT{\disccoh}}
\newcommand{\Wnwcdf}{\COH_{\subdisccohcdf}^{\nextt}}
\newcommand{\Wnwcdfpp}{\COH_{(\subdisccohcdf+1)}^{\nextt}}
\newcommand{\Wnwcdfk}{\COH_{(\subdisccohcdf+k)}}

\newcommand{\Wmin}{\COH^{\MINTXT}}
\newcommand{\Wmax}{\COH^{\MAXTXT}}

\newcommand{\Wphhjmin}{\COH^{\nextt\MINTXT} \left( \A, \COH, \J \right) }

\newcommand{\Adisc}{\A_\discA}
\newcommand{\QUADz}{z_\discalpha}
\newcommand{\QUADWGTz}{\omega_\discalpha}

\newcommand{\discReqit}{\discR^{\eqit}}
\newcommand{\discReqitopt}[1]{\discR^{\eqitopt{#1}}}

\newcommand{\RIstariter}{\R^{\I}_{\discR^{\eqit\opti}}}
\newcommand{\RIstariterpone}{\R^{\I}_{\discR^{\eqit\opti} + 1}}

\newcommand{\RnReqit}{\RI_{\discReqit}}
\newcommand{\RnReqitopt}[1]{\RI_{\discReqitopt{#1}}}

\newcommand{\Requi}{\R^{\I,equi}}
\newcommand{\Requiapp}{\HATT{\R}^{\I,equi}}

\newcommand{\COHd}{\COH_{\disccoh}}

\newcommand{\PROBA}{\left( \frac{1}{\sqrt{\pi}} \cdot \QUADWGTz \right)  }
\newcommand{\AQSUMA}{\SUMA_{\discA=1}^{\totA}}

\newcommand{\FPcohsn}{\fp \left( \Adisc, \COHd \right)}

\newcolumntype{L}[1]{>{\raggedright\let\newline\\\arraybackslash\hspace{0pt}}m{#1}}
\newcolumntype{C}[1]{>{\centering\let\newline\\\arraybackslash\hspace{0pt}}m{#1}}
\newcolumntype{R}[1]{>{\raggedleft\let\newline\\\arraybackslash\hspace{0pt}}m{#1}}

\newcommand{\FITPRBKY}[1]{
\begin{figure}[H]
\makebox[\textwidth][c]{
\fcolorbox{white}{white}{
\centering
\begin{minipage}{1.1\textwidth}
\begin{center}
\newcommand{\dCYKF}{Continuous Choices (in thousands of baht, 2005 northeast rural price):}
\newcommand{\dCC}{\MEAN\xspace \CC}
\newcommand{\dYY}{\MEAN\xspace \Y}
\newcommand{\dKK}{\MEAN\xspace \K}
\newcommand{\dBFRAC}{\MEAN\xspace \(\frac{\text{All Borr.}}{\Y} \) }
\newcommand{\dIBFRAC}{\MEAN\xspace \(\frac{\text{\CAPCAP\LLC Borr.}}{ \text{Borrow}} \)}
\newcommand{\dPROBB}{Discrete Choices (0.01 = 1 percent):}
\newcommand{\dsPIB}{\(\PPIB\)}
\newcommand{\dsPIS}{\(\PPIL\)}
\newcommand{\dsPFB}{\(\PPFB\)}
\newcommand{\dsPFS}{\(\PPFS\)}
\newcommand{\dsPFIB}{\(\PPFIB\)}
\newcommand{\dsPFBIS}{\(\PPFBIS\)}
\newcommand{\COEFCIGAP}{0ex}
\begin{table}[H]\centering
\def\sym#1{\ifmmode^{#1}\else\(^{#1}\)\fi}
\caption{\label{tab:modelFit} Model Fit}
\begin{tabular}{l*{4}{c}}
\toprule
&\multicolumn{2}{c}{Northeast}&\multicolumn{2}{c}{Central}\\
& 1999-2001& 2002-2009& 1999-2001& 2002-2009\\
\midrule
\multicolumn{5}{l}{\footnotesize\textit{\dPROBB}}\\
\midrule
\dsPIB 	&\TTSS{\PIBnNz}{\PIBnN} &\TTSS{\PIBnZz}{\PIBnZ} &\TTSS{\PIBcNz}{\PIBcN} &\TTSS{\PIBcZz}{\PIBcZ} \\[-0.8ex]
                         &\CII{\PIBnNa}{\PIBnNe}{\PIBnNs}
                         &\CII{\PIBnZa}{\PIBnZe}{\PIBnZs}
                         &\CII{\PIBcNa}{\PIBcNe}{\PIBcNs}
                         &\CII{\PIBcZa}{\PIBcZe}{\PIBcZs}\\
\dsPIS 	&\TTSS{\PISnNz}{\PISnN} &\TTSS{\PISnZz}{\PISnZ} &\TTSS{\PIScNz}{\PIScN} &\TTSS{\PIScZz}{\PIScZ} \\[-0.8ex]
                        &\CII{\PISnNa}{\PISnNe}{\PISnNs}
                        &\CII{\PISnZa}{\PISnZe}{\PISnZs}
                        &\CII{\PIScNa}{\PIScNe}{\PIScNs}
                        &\CII{\PIScZa}{\PIScZe}{\PIScZs}\\
\dsPFB 	&\TTSS{\PFBnNz}{\PFBnN} &\TTSS{\PFBnZz}{\PFBnZ} &\TTSS{\PFBcNz}{\PFBcN} &\TTSS{\PFBcZz}{\PFBcZ} \\[-0.8ex]
                        &\CII{\PFBnNa}{\PFBnNe}{\PFBnNs}
                        &\CII{\PFBnZa}{\PFBnZe}{\PFBnZs}
                        &\CII{\PFBcNa}{\PFBcNe}{\PFBcNs}
                        &\CII{\PFBcZa}{\PFBcZe}{\PFBcZs}\\
\dsPFS	&\TTSS{\PFSnNz}{\PFSnN} &\TTSS{\PFSnZz}{\PFSnZ} &\TTSS{\PFScNz}{\PFScN} &\TTSS{\PFScZz}{\PFScZ} \\[-0.8ex]
                        &\CII{\PFSnNa}{\PFSnNe}{\PFSnNs}
                        &\CII{\PFSnZa}{\PFSnZe}{\PFSnZs}
                        &\CII{\PFScNa}{\PFScNe}{\PFScNs}
                        &\CII{\PFScZa}{\PFScZe}{\PFScZs}\\
\dsPFIB	&\TTSS{\PBBnNz}{\PBBnN} &\TTSS{\PBBnZz}{\PBBnZ} &\TTSS{\PBBcNz}{\PBBcN} &\TTSS{\PBBcZz}{\PBBcZ} \\[-0.8ex]
                        &\CII{\PBBnNa}{\PBBnNe}{\PBBnNs}
                        &\CII{\PBBnZa}{\PBBnZe}{\PBBnZs}
                        &\CII{\PBBcNa}{\PBBcNe}{\PBBcNs}
                        &\CII{\PBBcZa}{0.124}{\PBBcZs}\\
\dsPFBIS	&\TTSS{\PBSnNz}{\PBSnN} &\TTSS{\PBSnZz}{\PBSnZ} &\TTSS{\PBScNz}{\PBScN} &\TTSS{\PBScZz}{\PBScZ} \\[-0.8ex]
                        &\CII{\PBSnNa}{\PBSnNe}{\PBSnNs}
                        &\CII{\PBSnZa}{\PBSnZe}{\PBSnZs}
                        &\CII{\PBScNa}{\PBScNe}{\PBScNs}
                        &\CII{\PBScZa}{\PBScZe}{\PBScZs}\\
                        \midrule
\multicolumn{5}{l}{\footnotesize\textit{\dCYKF}}\\
\midrule
\dKK &\TTSS{\KKnNz}{\KKnN} &\TTSS{\KKnZz}{\KKnZ} &\TTSS{\KKcNz}{\KKcN} &\TTSS{\KKcZz}{\KKcZ} \\[-0.8ex]
                         &\CII{\KKnNa}{\KKnNe}{\KKnNs}
                         &\CII{\KKnZa}{\KKnZe}{\KKnZs}
                         &\CII{\KKcNa}{\KKcNe}{\KKcNs}
                         &\CII{\KKcZa}{\KKcZe}{\KKcZs}\\
\dYY  &\TTSS{\YYnNz}{\YYnN} &\TTSS{\YYnZz}{\YYnZ} &\TTSS{\YYcNz}{\YYcN} &\TTSS{\YYcZz}{\YYcZ} \\[-0.8ex]
                          &\CII{\YYnNa}{\YYnNe}{\YYnNs}
                          &\CII{\YYnZa}{\YYnZe}{\YYnZs}
                          &\CII{\YYcNa}{\YYcNe}{\YYcNs}
                          &\CII{\YYcZa}{\YYcZe}{\YYcZs}\\
\dCC 	&\TTSS{\CCnNz}{\CCnN} &\TTSS{\CCnZz}{\CCnZ} &\TTSS{\CCcNz}{\CCcN} &\TTSS{\CCcZz}{\CCcZ} \\[-0.8ex]
                           &\CII{\CCnNa}{\CCnNe}{\CCnNs}
                           &\CII{\CCnZa}{\CCnZe}{\CCnZs}
                           &\CII{\CCcNa}{\CCcNe}{\CCcNs}
                           &\CII{\CCcZa}{\CCcZe}{\CCcZs}\\
\dBFRAC &\TTSS{\IBFBnNz}{\IBFBnN} &\TTSS{\IBFBnZz}{\IBFBnZ} &\TTSS{\IBFBcNz}{\IBFBcN} &\TTSS{\IBFBcZz}{\IBFBcZ} \\[-0.8ex]
                          &\CII{\IBFBnNa}{\IBFBnNe}{\IBFBnNs}
                          &\CII{\IBFBnZa}{\IBFBnZe}{\IBFBnZs}
                          &\CII{\IBFBcNa}{\IBFBcNe}{\IBFBcNs}
                          &\CII{\IBFBcZa}{\IBFBcZe}{\IBFBcZs}\\
\dIBFRAC &\TTSS{\IBBFnNz}{\IBBFnN} &\TTSS{\IBBFnZz}{\IBBFnZ} &\TTSS{\IBBFcNz}{\IBBFcN} &\TTSS{\IBBFcZz}{\IBBFcZ} \\[-0.8ex]
                          &\CII{\IBBFnNa}{\IBBFnNe}{\IBBFnNs}
                          &\CII{\IBBFnZa}{\IBBFnZe}{\IBBFnZs}
                          &\CII{\IBBFcNa}{\IBBFcNe}{\IBBFcNs}
                          &\CII{\IBBFcZa}{\IBBFcZe}{\IBBFcZs}\\
\bottomrule
\end{tabular}
\end{table}
 \end{center}
\vspace*{-10mm}
\small \emph{Notes:} See Section \ref{subsec:estimates} for detail. Cells show data average, model prediction average, and confidence interval based on estimates from Tables \ref{tab:costEsti} and \ref{tab:otherEstimates}.
\end{minipage}
}}
\end{figure}
}

\newcommand{\FIGESTIMATES}[1]{
\begin{table}[H]\centering
\def\sym#1{\ifmmode^{#1}\else\(^{#1}\)\fi}
\caption{\label{tab:costEsti} Credit Access Estimates}
\begin{tabular}{l*{4}{c}}
\toprule
&\multicolumn{2}{c}{Northeast}&\multicolumn{2}{c}{Central}\\
& 1999-2001& 2002-2009& 1999-2001& 2002-2009\\
\midrule
&\multicolumn{4}{c}{\footnotesize\textit{Fixed Costs Levels}}\\
\midrule
Formal Borrow Fixed Cost  	&\FBnN &\FBnZ &\FBcN &\FBcZ \\
                           &\CI{\FBnNa}{\FBnNe}{\FBnNs}
                           &\CI{\FBnZa}{\FBnZe}{\FBnZs}
                           &\CI{\FBcNa}{\FBcNe}{\FBcNs}
                           &\CI{\FBcZa}{\FBcZe}{\FBcZs}\\
Formal Save Fixed Cost     &\FSnN &\FSnZ &\FScN &\FScZ \\
                           &\CI{\FSnNa}{\FSnNe}{\FSnNs}
                           &\CI{\FSnZa}{\FSnZe}{\FSnZs}
                           &\CI{\FScNa}{\FScNe}{\FScNs}
                           &\CI{\FScZa}{\FScZe}{\FScZs}\\
Informal Borrow Fixed Cost &\IBnN &\IBnZ &\IBcN &\IBcZ \\
                           &\CI{\IBnNa}{\IBnNe}{\IBnNs}
                           &\CI{\IBnZa}{\IBnZe}{\IBnZs}
                           &\CI{\IBcNa}{\IBcNe}{\IBcNs}
                           &\CI{\IBcZa}{\IBcZe}{\IBcZs}\\
Informal Save Fixed Cost   &\ILnN &\ILnZ &\ILcN &\ILcZ \\
                           &\CI{\ILnNa}{\ILnNe}{\ILnNs}
                           &\CI{\ILnZa}{\ILnZe}{\ILnZs}
                           &\CI{\ILcNa}{\ILcNe}{\ILcNs}
                           &\CI{\ILcZa}{\ILcZe}{\ILcZs}\\
\midrule
Collateral Constraint      &\KAPPAnN    &\KAPPAnZ    &\KAPPAcN    &\KAPPAcZ \\
                            &\CI{\KAPPAnNa}{\KAPPAnNe}{\KAPPAnNs}
                            &\CI{\KAPPAnNa}{\KAPPAnNe}{\KAPPAnNs}
                            &\CI{\KAPPAcZa}{\KAPPAcZe}{\KAPPAcZs}
                            &\CI{\KAPPAcZa}{\KAPPAcZe}{\KAPPAcZs}\\
\bottomrule
\end{tabular}
\end{table}

 \small \emph{Notes:} See Section \ref{subsec:estimates} for detail.
\newcommand{\dELAS}{Elasticity \(\PALPHA\)}
\newcommand{\dDEPR}{Depreciation \(\PDELTA\)}

\newcommand{\dBETA}{Discount \(\PBETA\)}
\newcommand{\dSHKS}{\(\EESIGMASub{}\)}
\newcommand{\dAAVG}{\(\PAMUSub{}\)}
\newcommand{\dASD}{\(\PASIGMASub{}\)}

\newcommand{\dRHO}{ CRRA \(\PCRRA\)}
\newcommand{\dLGIT}{ Multinomial \(\PPHISIGMASub{}\)}

\begin{table}[H]\centering
\def\sym#1{\ifmmode^{#1}\else\(^{#1}\)\fi}
\caption{\label{tab:otherEstimates} Preference and Production Parameters (See Section \ref{subsec:estimates})}
\begin{tabular}{l*{2}{l}}
\toprule
& Northeast& Central\\
\midrule
\multicolumn{3}{l}{\vspace{-1mm}\footnotesize\textit{Preference:}}\\
\dBETA &\BETAn \CIT{\BETAna}{\BETAne}{\BETAns} &\BETAc \CIT{\BETAca}{\BETAce}{\BETAcs}  \\
\dRHO &\RHOn \CIT{\RHOna}{\RHOne}{\RHOns} &\RHOc \CIT{\RHOca}{\RHOce}{\RHOcs}  \\
\dLGIT &\LOGITSDSCALEn \CIT{\LOGITSDSCALEna}{\LOGITSDSCALEne}{\LOGITSDSCALEns} &\LOGITSDSCALEc \CIT{\LOGITSDSCALEca}{\LOGITSDSCALEce}{\LOGITSDSCALEcs}  \\
\midrule
\multicolumn{3}{l}{\vspace{-1mm}\footnotesize\textit{Production Function:}}\\
\dELAS &\ALPHAKn \CIT{\ALPHAKna}{\ALPHAKne}{\ALPHAKns} &\ALPHAKc \CIT{\ALPHAKca}{\ALPHAKce}{\ALPHAKcs}  \\
\dDEPR &\KDEPRECIATIONn \CIT{\KDEPRECIATIONna}{\KDEPRECIATIONne}{\KDEPRECIATIONns} &\KDEPRECIATIONc \CIT{\KDEPRECIATIONca}{\KDEPRECIATIONce}{\KDEPRECIATIONcs}  \\
\dAAVG (99-01) & \AMUnN \CIT{\AMUnNa}{\AMUnNe}{\AMUnNs}
            & \AMUcN \CIT{\AMUcNa}{\AMUcNe}{\AMUcNs}\\
\dAAVG (02-99) & \AMUnZ \CIT{\AMUnZa}{\AMUnZe}{\AMUnZs}
            & \AMUcZ \CIT{\AMUcZa}{\AMUcZe}{\AMUcZs}\\
\dASD &\ASDn \CIT{\ASDna}{\ASDne}{\ASDns} &\ASDc \CIT{\ASDca}{\ASDce}{\ASDcs}  \\
\dSHKS &\STDEPSn \CIT{\STDEPSna}{\STDEPSne}{\STDEPSns} &\STDEPSc \CIT{\STDEPSca}{\STDEPSce}{\STDEPScs}  \\
\midrule
\multicolumn{3}{l}{\vspace{-1mm}\footnotesize\textit{Measurement Error:}}\\
\( \sigma_{\PZETA^{\B}}\)\footnote{\(\PZETA^{\B}\sim \mathcal{N}\left(\frac{-\sigma^2_{\PZETA^{\B}}}{2}, \sigma^2_{\PZETA^{\B}} \right) \)} & 0.25 \CIT{0}{0}{0.019} & 0.15 \CIT{0}{0}{0.012}  \\
\( \sigma_{\PZETA^{\K}}\) & 0.18 \CIT{0}{0}{0.014} & 0.15 \CIT{0}{0}{0.012}  \\
\( \sigma_{\PZETA^{\Y}}\) & 0.38 \CIT{0}{0}{0.03} &  0.18 \CIT{0}{0}{0.014}  \\
\( \sigma_{\PZETA^{\CC}}\) & 0.16 \CIT{0}{0}{0.012} & 0.68 \CIT{0}{0}{0.53}  \\
\bottomrule
\end{tabular}
\end{table}

 }

\newcommand{\FIGESTIMATESFY}[1]{
\begin{figure}[H]
\vspace*{-3mm}
\makebox[\textwidth][c]{
\fcolorbox{white}{white}{
\centering
\begin{minipage}{1.0\textwidth}
\newcommand{\dFBFC}{\FXCjfbM \hspace{1mm} {\footnotesize (\CAPCAP\NLC \CAPCAP\BBBB)} }
\newcommand{\dFSFC}{\FXCjfsM \hspace{1mm} {\footnotesize (\CAPCAP\NLC \CAPCAP\SSSS)} }
\newcommand{\dIBFC}{\FXCjibM \hspace{1mm} {\footnotesize (\CAPCAP\LLC \CAPCAP\BBBB)} }
\newcommand{\dISFC}{\FXCjilM \hspace{1mm} {\footnotesize (\CAPCAP\LLC \CAPCAP\SSSS)} }
\newcommand{\dCOLL}{\(\PKPPA\)\hspace{1mm} {\footnotesize (\CAPCAP Collateral Constraint)}}

\begin{table}[H]\centering
\def\sym#1{\ifmmode^{#1}\else\(^{#1}\)\fi}
\caption{\label{tab:costEsti} Credit Access Parameters (See Section \ref{subsec:estimates})}
\begin{tabular}{l*{4}{c}}
\toprule
&\multicolumn{2}{c}{Northeast}&\multicolumn{2}{c}{Central}\\
& 1999-2001& 2002-2009& 1999-2001& 2002-2009\\
\midrule
&\multicolumn{4}{c}{\footnotesize\textit{Fixed Costs as a Fraction of 99-01 Average Income}}\\
\dFBFC &\FBnNY &\FBnZY &\FBcNY &\FBcZY \\
          &\CI{\FBnNaY}{\FBnNeY}{\FBnNsY}
          &\CI{\FBnZaY}{\FBnZeY}{\FBnZsY}
          &\CI{\FBcNaY}{\FBcNeY}{\FBcNsY}
          &\CI{\FBcZaY}{\FBcZeY}{\FBcZsY}\\
\dFSFC &\FSnNY &\FSnZY &\FScNY &\FScZY \\
          &\CI{\FSnNaY}{\FSnNeY}{\FSnNsY}
          &\CI{\FSnZaY}{\FSnZeY}{\FSnZsY}
          &\CI{\FScNaY}{\FScNeY}{\FScNsY}
          &\CI{\FScZaY}{\FScZeY}{\FScZsY}\\
\dIBFC &\multicolumn{2}{c}{\IBnNY} &\multicolumn{2}{c}{\IBcNY}\\
          &\multicolumn{2}{c}{\CI{\IBnNaY}{\IBnNeY}{\IBnNsY}}
          &\multicolumn{2}{c}{\CI{\IBcNaY}{\IBcNeY}{\IBcNsY}}\\
\dISFC &\ILnNY &\ILnZY &\ILcNY &\ILcZY \\
          &\CI{\ILnNaY}{\ILnNeY}{\ILnNsY}
          &\CI{\ILnZaY}{\ILnZeY}{\ILnZsY}
          &\CI{\ILcNaY}{\ILcNeY}{\ILcNsY}
          &\CI{\ILcZaY}{\ILcZeY}{\ILcZsY}\\
\midrule
\dCOLL &\KAPPAnN &\KAPPAnZ &\KAPPAcN &\KAPPAcZ \\
          &\CI{\KAPPAnNa}{\KAPPAnNe}{\KAPPAnNs}
          &\CI{\KAPPAnZa}{\KAPPAnZe}{\KAPPAnZs}
          &\CI{\KAPPAcNa}{\KAPPAcNe}{\KAPPAcNs}
          &\CI{\KAPPAcZa}{\KAPPAcZe}{\KAPPAcZs}\\
\bottomrule
\end{tabular}
\end{table}

 \vspace{-4mm}
\newcommand{\dELAS}{Elasticity \(\PALPHA\)}
\newcommand{\dDEPR}{Depreciation \(\PDELTA\)}

\newcommand{\dBETA}{Discount \(\PBETA\)}
\newcommand{\dSHKS}{\(\EESIGMASub{}\)}
\newcommand{\dAAVG}{\(\PAMUSub{}\)}
\newcommand{\dASD}{\(\PASIGMASub{}\)}

\newcommand{\dRHO}{ CRRA \(\PCRRA\)}
\newcommand{\dLGIT}{ Multinomial \(\PPHISIGMASub{}\)}

\begin{table}[H]\centering
\def\sym#1{\ifmmode^{#1}\else\(^{#1}\)\fi}
\caption{\label{tab:otherEstimates} Preference and Production Parameters (See Section \ref{subsec:estimates})}
\begin{tabular}{l*{2}{l}}
\toprule
& Northeast& Central\\
\midrule
\multicolumn{3}{l}{\vspace{-1mm}\footnotesize\textit{Preference:}}\\
\dBETA &\BETAn \CIT{\BETAna}{\BETAne}{\BETAns} &\BETAc \CIT{\BETAca}{\BETAce}{\BETAcs}  \\
\dRHO &\RHOn \CIT{\RHOna}{\RHOne}{\RHOns} &\RHOc \CIT{\RHOca}{\RHOce}{\RHOcs}  \\
\dLGIT &\LOGITSDSCALEn \CIT{\LOGITSDSCALEna}{\LOGITSDSCALEne}{\LOGITSDSCALEns} &\LOGITSDSCALEc \CIT{\LOGITSDSCALEca}{\LOGITSDSCALEce}{\LOGITSDSCALEcs}  \\
\midrule
\multicolumn{3}{l}{\vspace{-1mm}\footnotesize\textit{Production Function:}}\\
\dELAS &\ALPHAKn \CIT{\ALPHAKna}{\ALPHAKne}{\ALPHAKns} &\ALPHAKc \CIT{\ALPHAKca}{\ALPHAKce}{\ALPHAKcs}  \\
\dDEPR &\KDEPRECIATIONn \CIT{\KDEPRECIATIONna}{\KDEPRECIATIONne}{\KDEPRECIATIONns} &\KDEPRECIATIONc \CIT{\KDEPRECIATIONca}{\KDEPRECIATIONce}{\KDEPRECIATIONcs}  \\
\dAAVG (99-01) & \AMUnN \CIT{\AMUnNa}{\AMUnNe}{\AMUnNs}
            & \AMUcN \CIT{\AMUcNa}{\AMUcNe}{\AMUcNs}\\
\dAAVG (02-99) & \AMUnZ \CIT{\AMUnZa}{\AMUnZe}{\AMUnZs}
            & \AMUcZ \CIT{\AMUcZa}{\AMUcZe}{\AMUcZs}\\
\dASD &\ASDn \CIT{\ASDna}{\ASDne}{\ASDns} &\ASDc \CIT{\ASDca}{\ASDce}{\ASDcs}  \\
\dSHKS &\STDEPSn \CIT{\STDEPSna}{\STDEPSne}{\STDEPSns} &\STDEPSc \CIT{\STDEPSca}{\STDEPSce}{\STDEPScs}  \\
\midrule
\multicolumn{3}{l}{\vspace{-1mm}\footnotesize\textit{Measurement Error:}}\\
\( \sigma_{\PZETA^{\B}}\)\footnote{\(\PZETA^{\B}\sim \mathcal{N}\left(\frac{-\sigma^2_{\PZETA^{\B}}}{2}, \sigma^2_{\PZETA^{\B}} \right) \)} & 0.25 \CIT{0}{0}{0.019} & 0.15 \CIT{0}{0}{0.012}  \\
\( \sigma_{\PZETA^{\K}}\) & 0.18 \CIT{0}{0}{0.014} & 0.15 \CIT{0}{0}{0.012}  \\
\( \sigma_{\PZETA^{\Y}}\) & 0.38 \CIT{0}{0}{0.03} &  0.18 \CIT{0}{0}{0.014}  \\
\( \sigma_{\PZETA^{\CC}}\) & 0.16 \CIT{0}{0}{0.012} & 0.68 \CIT{0}{0}{0.53}  \\
\bottomrule
\end{tabular}
\end{table}

 \end{minipage}
}}
\end{figure}
}

\newcommand{\FIGTYPEONE}[5]{
\begin{figure}[!htbp]
\makebox[\textwidth][c]{
\fcolorbox{white}{white}{
\centering
\begin{minipage}{1.0\textwidth}
\caption{#2}
\vspace{-0.25cm}
\begin{center}
\centerline{\includegraphics[scale=#4]{#1/#3.eps}}
\end{center}
\vspace{-1.0cm}
\small #5
\end{minipage}
}}
\end{figure}
}

\newcommand{\FIGTYPETWO}[5]{
\begin{figure}[!htbp]
\makebox[\textwidth][c]{
\fcolorbox{white}{white}{
\centering
\begin{minipage}{1.0\textwidth}
\caption{#1}
\begin{center}
\minipage[t]{0.60\textwidth}
\vspace{5pt}
\vspace{0.20cm}
  \small \emph{Notes:} #5
\endminipage
\hspace{0.01\textwidth}
\minipage[t]{0.38\textwidth}
\vspace{0pt}
  \tcbset{enhanced,colframe=blue!50!black,colback=white}
\begin{tcolorbox}[
  left=5pt,right=1pt,top=1pt,bottom=1pt,
  colback=background,
  colframe=linecolor,
  ]\footnotesize
    \textbf{Row Stats Legend:}\\
    \textit{(Numbers = Percentage Points)}
    \vspace{-0.25cm}
    \tcbline
    \vspace{-0.25cm}
    \parbox{1.0\linewidth}{
    \textit{Row} \textbf{1}: \expandafter\MakeUppercase\LLC \(\textcolor{blue}{\RI}\) \\
    \textit{Row} \textbf{2}: \EFRATEFOR \\
                            \null\hspace{9.5mm} {\scriptsize (\EFRATEDESC)}\\
    \textit{Row} \textbf{3}: \IORATEFOR \\
                            \null\hspace{9.5mm} {\scriptsize (\IORATEFORDESC)}\\
    \textit{Row} \textbf{4}: \( \PPIB \) / \( \PPIL \)\\
                            \null\hspace{9.5mm} {\scriptsize (\LLC \BBshort/\SSshort prob)}
}
\end{tcolorbox}
 \endminipage
\end{center}
\begin{center}
\caption*{Panel \textbf{A}: \textit{Low} Cost \KFSS (\(\FXCjfs = 0, \Rjfs = 1.05\))}
\includegraphics[scale=#4]{graphs/#2}
\vspace{0cm}
\caption*{Panel \textbf{B}: \textit{High} Cost \KFSS (\(\FXCjfs = 1, \Rjfs = 1.025\))}
\includegraphics[scale=#4]{graphs/#3}
\end{center}

\end{minipage}
}}
\end{figure}
}

\newcommand{\FIGTYPETHREE}[6]{
\begin{figure}[!htbp]
    \begin{center}
      \caption{#1}
      \minipage[t]{#3\textwidth}
      \vspace{0pt}
        \includegraphics[width=\textwidth]{graphs/#2}
      \endminipage
      \hspace{0.01\textwidth}
      \minipage[t]{#5\textwidth}
      \vspace{5pt}
      \tcbset{enhanced,colframe=blue!50!black,colback=white}
\begin{tcolorbox}[
  boxsep=5pt,left=1pt,right=1pt,top=1pt,bottom=1pt,
  colback=background,
  colframe=linecolor,
  ]\footnotesize
    \textbf{Row Stats Legend:}\\
    \textit{(Numbers = Percentage Points)}
    \vspace{-0.25cm}
    \tcbline
    \vspace{-0.25cm}
    \parbox{0.89\linewidth}{
    \textit{Row} \textbf{1}: \expandafter\MakeUppercase\LLC \(\textcolor{blue}{\RI}\)\\
    \textit{Row} \textbf{2}: \EFRATEFOR \\
                              \null\hspace{9.5mm} {\scriptsize (\EFRATEDESC)}\\
    \textit{Row} \textbf{3}: \( \PPIB \) / \( \PPIL \)\\
                              \null\hspace{9.5mm} {\scriptsize (\LLC \BBshort/\SSshort prob)}
}
\end{tcolorbox}

       \endminipage
    \end{center}
    \small \emph{Notes:} #6
\end{figure}
}

\newcommand{\FIGLMjfcr}{fig:choicesjfcr}
\newcommand{\FIGRMjfcr}{Figure \ref{\FIGLMjfcr}\xspace}
\newcommand{\FIGTMjfcr}{\label{\FIGLMjfcr}Borrowing Choice Set}
\newcommand{\FIGPMjfcr}{choices_J_fc_r}
\newcommand{\FIGDMjfcr}{\emph{Notes:} The figure shows the effects of wealth, \FCFs, \CLCs, and \ITRs on the \BBBB and \RKI choice set. Within a borrowing discrete choice $j$, households face a three dimensional choice set over $c$, risky investments $k'$, and borrowing $b'$. Given that households cannot have negative consumption today or tomorrow, the domain of the choice set is an individual and choice $j$ specific triangular pyramid. The figure here plots the base of the pyramid where the x-axis is the borrowing choice $b'$ and the y-axis is the risky investment choice $k'$. The implicit but not shown z-axis is consumption $c$, which is chosen to clear budgets. The percentage asset solution algorithm used in this paper relies on taking advantage of the geometry of this figure.}

\newcommand{\FIGLSckb}{fig:choicesckb}
\newcommand{\FIGRSckb}{Figure \ref{\FIGLSckb}\xspace}
\newcommand{\FIGTSckb}{\label{\FIGLSckb}Weighted Continuous Policy Functions}
\newcommand{\FIGPSckb}{choices_CKB}
\newcommand{\FIGDSckb}{\emph{Panels show:} 1. consumption share of \CZH; 2. safe \SSSS (the sum of \ISS, \FSS, and \FJS) share of \SSSSRKI; 3. \RKI as a fraction of \CZH; 4. the share of \RKI financed by \BBBB (the sum of \IBB, \FBB, and \FJB).}

\newcommand{\FIGLSgebenchmarkTone}{fig:gebechmarkone}

\newcommand{\FIGTSgebenchmarkTone}{\label{\FIGLSgebenchmarkTone}Equilibrium with only \expandafter\MakeUppercase\LLC Options}

\newcommand{\FIGLSgebenchmarkTtwosg}{fig:gebechmarktwosg}

\newcommand{\FIGTSgebenchmarkTtwosg}{\label{\FIGLSgebenchmarkTtwosg}Equilibrium with \expandafter\MakeUppercase\LLC and \expandafter\MakeUppercase\NLC Options}

 \newcommand{\AMUcZ}{0} \newcommand{\AMUcZa}{0}  \newcommand{\AMUcZe}{0}   \newcommand{\AMUcZs}{0} \newcommand{\AMUcN}{0} \newcommand{\AMUcNa}{0}  \newcommand{\AMUcNe}{0}   \newcommand{\AMUcNs}{0} \newcommand{\AMUnZ}{0} \newcommand{\AMUnZa}{0}  \newcommand{\AMUnZe}{0}   \newcommand{\AMUnZs}{0} \newcommand{\AMUnN}{0} \newcommand{\AMUnNa}{0}  \newcommand{\AMUnNe}{0}   \newcommand{\AMUnNs}{0} \newcommand{\ASDc}{0} \newcommand{\ASDca}{0}  \newcommand{\ASDce}{0}   \newcommand{\ASDcs}{0} \newcommand{\ASDn}{0} \newcommand{\ASDna}{0}  \newcommand{\ASDne}{0}   \newcommand{\ASDns}{0} \newcommand{\KDEPRECIATIONc}{0} \newcommand{\KDEPRECIATIONca}{0}  \newcommand{\KDEPRECIATIONce}{0}   \newcommand{\KDEPRECIATIONcs}{0} \newcommand{\KDEPRECIATIONn}{0} \newcommand{\KDEPRECIATIONna}{0}  \newcommand{\KDEPRECIATIONne}{0}   \newcommand{\KDEPRECIATIONns}{0} \newcommand{\ALPHAKc}{0} \newcommand{\ALPHAKca}{0}  \newcommand{\ALPHAKce}{0}   \newcommand{\ALPHAKcs}{0} \newcommand{\ALPHAKn}{0} \newcommand{\ALPHAKna}{0}  \newcommand{\ALPHAKne}{0}   \newcommand{\ALPHAKns}{0} \newcommand{\BETAc}{0} \newcommand{\BETAca}{0}  \newcommand{\BETAce}{0}   \newcommand{\BETAcs}{0} \newcommand{\BETAn}{0} \newcommand{\BETAna}{0}  \newcommand{\BETAne}{0}   \newcommand{\BETAns}{0} \newcommand{\KAPPAcZ}{0} \newcommand{\KAPPAcZa}{0}  \newcommand{\KAPPAcZe}{0}   \newcommand{\KAPPAcZs}{0} \newcommand{\KAPPAcN}{0} \newcommand{\KAPPAcNa}{0}  \newcommand{\KAPPAcNe}{0}   \newcommand{\KAPPAcNs}{0} \newcommand{\KAPPAnZ}{0} \newcommand{\KAPPAnZa}{0}  \newcommand{\KAPPAnZe}{0}   \newcommand{\KAPPAnZs}{0} \newcommand{\KAPPAnN}{0} \newcommand{\KAPPAnNa}{0}  \newcommand{\KAPPAnNe}{0}   \newcommand{\KAPPAnNs}{0} \newcommand{\LOGITSDSCALEc}{0} \newcommand{\LOGITSDSCALEca}{0}  \newcommand{\LOGITSDSCALEce}{0}   \newcommand{\LOGITSDSCALEcs}{0} \newcommand{\LOGITSDSCALEn}{0} \newcommand{\LOGITSDSCALEna}{0}  \newcommand{\LOGITSDSCALEne}{0}   \newcommand{\LOGITSDSCALEns}{0} \newcommand{\RHOc}{0} \newcommand{\RHOca}{0}  \newcommand{\RHOce}{0}   \newcommand{\RHOcs}{0} \newcommand{\RHOn}{0} \newcommand{\RHOna}{0}  \newcommand{\RHOne}{0}   \newcommand{\RHOns}{0}       \newcommand{\STDEPSEcs}{0}       \newcommand{\STDEPSEns}{0} \newcommand{\STDEPSc}{0} \newcommand{\STDEPSca}{0}  \newcommand{\STDEPSce}{0}   \newcommand{\STDEPScs}{0} \newcommand{\STDEPSn}{0} \newcommand{\STDEPSna}{0}  \newcommand{\STDEPSne}{0}   \newcommand{\STDEPSns}{0}

\newcommand{\FBcZY}{0} \newcommand{\FBcZaY}{0}  \newcommand{\FBcZeY}{0}   \newcommand{\FBcZsY}{0} \newcommand{\FBcNY}{0} \newcommand{\FBcNaY}{0}  \newcommand{\FBcNeY}{0}   \newcommand{\FBcNsY}{0} \newcommand{\FBnZY}{0} \newcommand{\FBnZaY}{0}  \newcommand{\FBnZeY}{0}   \newcommand{\FBnZsY}{0} \newcommand{\FBnNY}{0} \newcommand{\FBnNaY}{0}  \newcommand{\FBnNeY}{0}   \newcommand{\FBnNsY}{0} \newcommand{\FScZY}{0} \newcommand{\FScZaY}{0}  \newcommand{\FScZeY}{0}   \newcommand{\FScZsY}{0} \newcommand{\FScNY}{0} \newcommand{\FScNaY}{0}  \newcommand{\FScNeY}{0}   \newcommand{\FScNsY}{0} \newcommand{\FSnZY}{0} \newcommand{\FSnZaY}{0}  \newcommand{\FSnZeY}{0}   \newcommand{\FSnZsY}{0} \newcommand{\FSnNY}{0} \newcommand{\FSnNaY}{0}  \newcommand{\FSnNeY}{0}   \newcommand{\FSnNsY}{0} \newcommand{\IBcZY}{0}      \newcommand{\IBcZsY}{0} \newcommand{\IBcNY}{0} \newcommand{\IBcNaY}{0}  \newcommand{\IBcNeY}{0}   \newcommand{\IBcNsY}{0} \newcommand{\IBnZY}{0}      \newcommand{\IBnZsY}{0} \newcommand{\IBnNY}{0} \newcommand{\IBnNaY}{0}  \newcommand{\IBnNeY}{0}   \newcommand{\IBnNsY}{0} \newcommand{\ILcZY}{0} \newcommand{\ILcZaY}{0}  \newcommand{\ILcZeY}{0}   \newcommand{\ILcZsY}{0} \newcommand{\ILcNY}{0} \newcommand{\ILcNaY}{0}  \newcommand{\ILcNeY}{0}   \newcommand{\ILcNsY}{0} \newcommand{\ILnZY}{0} \newcommand{\ILnZaY}{0}  \newcommand{\ILnZeY}{0}   \newcommand{\ILnZsY}{0} \newcommand{\ILnNY}{0} \newcommand{\ILnNaY}{0}  \newcommand{\ILnNeY}{0}   \newcommand{\ILnNsY}{0}

\newcommand{\FBcZ}{0} \newcommand{\FBcZa}{0}  \newcommand{\FBcZe}{0}   \newcommand{\FBcZs}{0} \newcommand{\FBcN}{0} \newcommand{\FBcNa}{0}  \newcommand{\FBcNe}{0}   \newcommand{\FBcNs}{0} \newcommand{\FBnZ}{0} \newcommand{\FBnZa}{0}  \newcommand{\FBnZe}{0}   \newcommand{\FBnZs}{0} \newcommand{\FBnN}{0} \newcommand{\FBnNa}{0}  \newcommand{\FBnNe}{0}   \newcommand{\FBnNs}{0} \newcommand{\FScZ}{0} \newcommand{\FScZa}{0}  \newcommand{\FScZe}{0}   \newcommand{\FScZs}{0} \newcommand{\FScN}{0} \newcommand{\FScNa}{0}  \newcommand{\FScNe}{0}   \newcommand{\FScNs}{0} \newcommand{\FSnZ}{0} \newcommand{\FSnZa}{0}  \newcommand{\FSnZe}{0}   \newcommand{\FSnZs}{0} \newcommand{\FSnN}{0} \newcommand{\FSnNa}{0}  \newcommand{\FSnNe}{0}   \newcommand{\FSnNs}{0} \newcommand{\IBcZ}{0} \newcommand{\IBcZa}{0}  \newcommand{\IBcZe}{0}   \newcommand{\IBcZs}{0} \newcommand{\IBcN}{0} \newcommand{\IBcNa}{0}  \newcommand{\IBcNe}{0}   \newcommand{\IBcNs}{0} \newcommand{\IBnZ}{0} \newcommand{\IBnZa}{0}  \newcommand{\IBnZe}{0}   \newcommand{\IBnZs}{0} \newcommand{\IBnN}{0} \newcommand{\IBnNa}{0}  \newcommand{\IBnNe}{0}   \newcommand{\IBnNs}{0} \newcommand{\ILcZ}{0} \newcommand{\ILcZa}{0}  \newcommand{\ILcZe}{0}   \newcommand{\ILcZs}{0} \newcommand{\ILcN}{0} \newcommand{\ILcNa}{0}  \newcommand{\ILcNe}{0}   \newcommand{\ILcNs}{0} \newcommand{\ILnZ}{0} \newcommand{\ILnZa}{0}  \newcommand{\ILnZe}{0}   \newcommand{\ILnZs}{0} \newcommand{\ILnN}{0} \newcommand{\ILnNa}{0}  \newcommand{\ILnNe}{0}   \newcommand{\ILnNs}{0}  \newcommand{\CCcZ}{0} \newcommand{\CCcZs}{0} \newcommand{\CCcZa}{0}    \newcommand{\CCcZe}{0} \newcommand{\CCcZz}{0} \newcommand{\CCcN}{0} \newcommand{\CCcNs}{0} \newcommand{\CCcNa}{0}    \newcommand{\CCcNe}{0} \newcommand{\CCcNz}{0} \newcommand{\CCnZ}{0} \newcommand{\CCnZs}{0} \newcommand{\CCnZa}{0}    \newcommand{\CCnZe}{0} \newcommand{\CCnZz}{0} \newcommand{\CCnN}{0} \newcommand{\CCnNs}{0} \newcommand{\CCnNa}{0}    \newcommand{\CCnNe}{0} \newcommand{\CCnNz}{0} \newcommand{\KKcZ}{0} \newcommand{\KKcZs}{0} \newcommand{\KKcZa}{0}    \newcommand{\KKcZe}{0} \newcommand{\KKcZz}{0} \newcommand{\KKcN}{0} \newcommand{\KKcNs}{0} \newcommand{\KKcNa}{0}    \newcommand{\KKcNe}{0} \newcommand{\KKcNz}{0} \newcommand{\KKnZ}{0} \newcommand{\KKnZs}{0} \newcommand{\KKnZa}{0}    \newcommand{\KKnZe}{0} \newcommand{\KKnZz}{0} \newcommand{\KKnN}{0} \newcommand{\KKnNs}{0} \newcommand{\KKnNa}{0}    \newcommand{\KKnNe}{0} \newcommand{\KKnNz}{0} \newcommand{\YYcZ}{0} \newcommand{\YYcZs}{0} \newcommand{\YYcZa}{0}    \newcommand{\YYcZe}{0} \newcommand{\YYcZz}{0} \newcommand{\YYcN}{0} \newcommand{\YYcNs}{0} \newcommand{\YYcNa}{0}    \newcommand{\YYcNe}{0} \newcommand{\YYcNz}{0} \newcommand{\YYnZ}{0} \newcommand{\YYnZs}{0} \newcommand{\YYnZa}{0}    \newcommand{\YYnZe}{0} \newcommand{\YYnZz}{0} \newcommand{\YYnN}{0} \newcommand{\YYnNs}{0} \newcommand{\YYnNa}{0}    \newcommand{\YYnNe}{0} \newcommand{\YYnNz}{0}

\newcommand{\PBBcZ}{0} \newcommand{\PBBcZs}{0} \newcommand{\PBBcZa}{0} \newcommand{\PBBcZb}{0} \newcommand{\PBBcZc}{0} \newcommand{\PBBcZd}{0} \newcommand{\PBBcZe}{0} \newcommand{\PBBcZz}{0} \newcommand{\PBBcN}{0} \newcommand{\PBBcNs}{0} \newcommand{\PBBcNa}{0} \newcommand{\PBBcNb}{0} \newcommand{\PBBcNc}{0} \newcommand{\PBBcNd}{0} \newcommand{\PBBcNe}{0} \newcommand{\PBBcNz}{0} \newcommand{\PBBnZ}{0} \newcommand{\PBBnZs}{0} \newcommand{\PBBnZa}{0} \newcommand{\PBBnZb}{0} \newcommand{\PBBnZc}{0} \newcommand{\PBBnZd}{0} \newcommand{\PBBnZe}{0} \newcommand{\PBBnZz}{0} \newcommand{\PBBnN}{0} \newcommand{\PBBnNs}{0} \newcommand{\PBBnNa}{0} \newcommand{\PBBnNb}{0} \newcommand{\PBBnNc}{0} \newcommand{\PBBnNd}{0} \newcommand{\PBBnNe}{0} \newcommand{\PBBnNz}{0} \newcommand{\PBScZ}{0} \newcommand{\PBScZs}{0} \newcommand{\PBScZa}{0} \newcommand{\PBScZb}{0} \newcommand{\PBScZc}{0} \newcommand{\PBScZd}{0} \newcommand{\PBScZe}{0} \newcommand{\PBScZz}{0} \newcommand{\PBScN}{0} \newcommand{\PBScNs}{0} \newcommand{\PBScNa}{0} \newcommand{\PBScNb}{0} \newcommand{\PBScNc}{0} \newcommand{\PBScNd}{0} \newcommand{\PBScNe}{0} \newcommand{\PBScNz}{0} \newcommand{\PBSnZ}{0} \newcommand{\PBSnZs}{0} \newcommand{\PBSnZa}{0} \newcommand{\PBSnZb}{0} \newcommand{\PBSnZc}{0} \newcommand{\PBSnZd}{0} \newcommand{\PBSnZe}{0} \newcommand{\PBSnZz}{0} \newcommand{\PBSnN}{0} \newcommand{\PBSnNs}{0} \newcommand{\PBSnNa}{0} \newcommand{\PBSnNb}{0} \newcommand{\PBSnNc}{0} \newcommand{\PBSnNd}{0} \newcommand{\PBSnNe}{0} \newcommand{\PBSnNz}{0} \newcommand{\PFBcZ}{0} \newcommand{\PFBcZs}{0} \newcommand{\PFBcZa}{0} \newcommand{\PFBcZb}{0} \newcommand{\PFBcZc}{0} \newcommand{\PFBcZd}{0} \newcommand{\PFBcZe}{0} \newcommand{\PFBcZz}{0} \newcommand{\PFBcN}{0} \newcommand{\PFBcNs}{0} \newcommand{\PFBcNa}{0} \newcommand{\PFBcNb}{0} \newcommand{\PFBcNc}{0} \newcommand{\PFBcNd}{0} \newcommand{\PFBcNe}{0} \newcommand{\PFBcNz}{0} \newcommand{\PFBnZ}{0} \newcommand{\PFBnZs}{0} \newcommand{\PFBnZa}{0} \newcommand{\PFBnZb}{0} \newcommand{\PFBnZc}{0} \newcommand{\PFBnZd}{0} \newcommand{\PFBnZe}{0} \newcommand{\PFBnZz}{0} \newcommand{\PFBnN}{0} \newcommand{\PFBnNs}{0} \newcommand{\PFBnNa}{0} \newcommand{\PFBnNb}{0} \newcommand{\PFBnNc}{0} \newcommand{\PFBnNd}{0} \newcommand{\PFBnNe}{0} \newcommand{\PFBnNz}{0} \newcommand{\PFScZ}{0} \newcommand{\PFScZs}{0} \newcommand{\PFScZa}{0} \newcommand{\PFScZb}{0} \newcommand{\PFScZc}{0} \newcommand{\PFScZd}{0} \newcommand{\PFScZe}{0} \newcommand{\PFScZz}{0} \newcommand{\PFScN}{0} \newcommand{\PFScNs}{0} \newcommand{\PFScNa}{0} \newcommand{\PFScNb}{0} \newcommand{\PFScNc}{0} \newcommand{\PFScNd}{0} \newcommand{\PFScNe}{0} \newcommand{\PFScNz}{0} \newcommand{\PFSnZ}{0} \newcommand{\PFSnZs}{0} \newcommand{\PFSnZa}{0} \newcommand{\PFSnZb}{0} \newcommand{\PFSnZc}{0} \newcommand{\PFSnZd}{0} \newcommand{\PFSnZe}{0} \newcommand{\PFSnZz}{0} \newcommand{\PFSnN}{0} \newcommand{\PFSnNs}{0} \newcommand{\PFSnNa}{0} \newcommand{\PFSnNb}{0} \newcommand{\PFSnNc}{0} \newcommand{\PFSnNd}{0} \newcommand{\PFSnNe}{0} \newcommand{\PFSnNz}{0} \newcommand{\PIBcZ}{0} \newcommand{\PIBcZs}{0} \newcommand{\PIBcZa}{0} \newcommand{\PIBcZb}{0} \newcommand{\PIBcZc}{0} \newcommand{\PIBcZd}{0} \newcommand{\PIBcZe}{0} \newcommand{\PIBcZz}{0} \newcommand{\PIBcN}{0} \newcommand{\PIBcNs}{0} \newcommand{\PIBcNa}{0} \newcommand{\PIBcNb}{0} \newcommand{\PIBcNc}{0} \newcommand{\PIBcNd}{0} \newcommand{\PIBcNe}{0} \newcommand{\PIBcNz}{0} \newcommand{\PIBnZ}{0} \newcommand{\PIBnZs}{0} \newcommand{\PIBnZa}{0} \newcommand{\PIBnZb}{0} \newcommand{\PIBnZc}{0} \newcommand{\PIBnZd}{0} \newcommand{\PIBnZe}{0} \newcommand{\PIBnZz}{0} \newcommand{\PIBnN}{0} \newcommand{\PIBnNs}{0} \newcommand{\PIBnNa}{0} \newcommand{\PIBnNb}{0} \newcommand{\PIBnNc}{0} \newcommand{\PIBnNd}{0} \newcommand{\PIBnNe}{0} \newcommand{\PIBnNz}{0} \newcommand{\PIScZ}{0} \newcommand{\PIScZs}{0} \newcommand{\PIScZa}{0} \newcommand{\PIScZb}{0} \newcommand{\PIScZc}{0} \newcommand{\PIScZd}{0} \newcommand{\PIScZe}{0} \newcommand{\PIScZz}{0} \newcommand{\PIScN}{0} \newcommand{\PIScNs}{0} \newcommand{\PIScNa}{0} \newcommand{\PIScNb}{0} \newcommand{\PIScNc}{0} \newcommand{\PIScNd}{0} \newcommand{\PIScNe}{0} \newcommand{\PIScNz}{0} \newcommand{\PISnZ}{0} \newcommand{\PISnZs}{0} \newcommand{\PISnZa}{0} \newcommand{\PISnZb}{0} \newcommand{\PISnZc}{0} \newcommand{\PISnZd}{0} \newcommand{\PISnZe}{0} \newcommand{\PISnZz}{0} \newcommand{\PISnN}{0} \newcommand{\PISnNs}{0} \newcommand{\PISnNa}{0} \newcommand{\PISnNb}{0} \newcommand{\PISnNc}{0} \newcommand{\PISnNd}{0} \newcommand{\PISnNe}{0} \newcommand{\PISnNz}{0}

\newcommand{\IBBFcZ}{0} \newcommand{\IBBFcZs}{0} \newcommand{\IBBFcZa}{0}    \newcommand{\IBBFcZe}{0} \newcommand{\IBBFcZz}{0} \newcommand{\IBBFcN}{0} \newcommand{\IBBFcNs}{0} \newcommand{\IBBFcNa}{0}    \newcommand{\IBBFcNe}{0} \newcommand{\IBBFcNz}{0} \newcommand{\IBFBcZ}{0} \newcommand{\IBFBcZs}{0} \newcommand{\IBFBcZa}{0}    \newcommand{\IBFBcZe}{0} \newcommand{\IBFBcZz}{0} \newcommand{\IBBFnZ}{0} \newcommand{\IBBFnZs}{0} \newcommand{\IBBFnZa}{0}    \newcommand{\IBBFnZe}{0} \newcommand{\IBBFnZz}{0} \newcommand{\IBBFnN}{0} \newcommand{\IBBFnNs}{0} \newcommand{\IBBFnNa}{0}    \newcommand{\IBBFnNe}{0} \newcommand{\IBBFnNz}{0} \newcommand{\IBFBcN}{0} \newcommand{\IBFBcNs}{0} \newcommand{\IBFBcNa}{0}    \newcommand{\IBFBcNe}{0} \newcommand{\IBFBcNz}{0} \newcommand{\IBFBnZ}{0} \newcommand{\IBFBnZs}{0} \newcommand{\IBFBnZa}{0}    \newcommand{\IBFBnZe}{0} \newcommand{\IBFBnZz}{0} \newcommand{\IBFBnN}{0} \newcommand{\IBFBnNs}{0} \newcommand{\IBFBnNa}{0}    \newcommand{\IBFBnNe}{0} \newcommand{\IBFBnNz}{0}

\newcommand{\FBRNbefore}{14.8}
\newcommand{\FBRNafter}{6.1}

\newcommand{\IRNbefore}{27.9}
\newcommand{\IRNafter}{13.9}
\newcommand{\IRNafterFC}{17}
\newcommand{\IRNafterCL}{20.5}
\newcommand{\IRNafterRR}{15.4}

\SUBTRACT{\IRNafterRR}{\IRNafter}{\IRNafterRRdiff}

\SUBTRACT{\IRNbefore}{\IRNafter}{\IRNdiff}
\SUBTRACT{\IRNafterCL}{\IRNafter}{\IRNdiffCL}
\DIVIDE{\IRNdiffCL}{\IRNdiff}{\IRNdiffCLShare}
\MULTIPLY{\IRNdiffCLShare}{100}{\IRNdiffCLShare}
\ROUND[0]{\IRNdiffCLShare}{\IRNdiffCLShareRnd}

\DIVIDE{\IRNdiffCL}{\IRNbefore}{\IRNafterFCPropInc}
\MULTIPLY{\IRNafterFCPropInc}{100}{\IRNafterFCPropInc}
\ROUND[0]{\IRNafterFCPropInc}{\IRNafterFCPropIncRnd}

\newcommand{\FPNbefore}{21.1}
\newcommand{\FPNafter}{44}
\newcommand{\FPNafterFC}{32}
\newcommand{\FPNafterCL}{41}
\SUBTRACT{\FPNafter}{\FPNafterCL}{\FPNafterCLdiff}

\SUBTRACT{\FPNafter}{\FPNbefore}{\FPNdiff}
\SUBTRACT{\FPNafter}{\FPNafterFC}{\FPNdiffFC}
\DIVIDE{\FPNdiffFC}{\FPNdiff}{\FPNafterFCShare}
\MULTIPLY{\FPNafterFCShare}{100}{\FPNafterFCShare}
\ROUND[0]{\FPNafterFCShare}{\FPNafterFCShareRnd}

\DIVIDE{\FPNdiffFC}{\FPNbefore}{\FPNafterFCPropInc}
\MULTIPLY{\FPNafterFCPropInc}{100}{\FPNafterFCPropInc}
\ROUND[0]{\FPNafterFCPropInc}{\FPNafterFCPropIncRnd}

\newcommand{\IPNbefore}{32}
\newcommand{\IPNafter}{11.1}
\newcommand{\IPNafterFC}{23}
\newcommand{\IPNafterCL}{13.1}
\SUBTRACT{\IPNafterCL}{\IPNafter}{\IPNafterCLdiff}

\SUBTRACT{\IPNafter}{\IPNbefore}{\IPNdiff}
\SUBTRACT{\IPNafter}{\IPNafterFC}{\IPNdiffFC}
\DIVIDE{\IPNdiffFC}{\IPNdiff}{\IPNafterFCShare}
\MULTIPLY{\IPNafterFCShare}{100}{\IPNafterFCShare}
\ROUND[0]{\IPNafterFCShare}{\IPNafterFCShareRnd}

\DIVIDE{\IPNdiffFC}{\IPNbefore}{\IPNafterFCPropInc}
\MULTIPLY{\IPNafterFCPropInc}{100}{\IPNafterFCPropInc}
\ROUND[0]{\IPNafterFCPropInc}{\IPNafterFCPropIncRnd}

\newcommand{\FIPNbefore}{52.6}
\newcommand{\FIPNafter}{84.3}
\newcommand{\FIPNafterFC}{65.3}
\newcommand{\FIPNafterCL}{83.3}
\SUBTRACT{\FIPNafter}{\FIPNafterCL}{\FIPNafterCLdiff}

\SUBTRACT{\FIPNafter}{\FIPNbefore}{\FIPNdiff}
\SUBTRACT{\FIPNafter}{\FIPNafterFC}{\FIPNdiffFC}
\DIVIDE{\FIPNdiff}{\FIPNbefore}{\FIPpercInc}
\MULTIPLY{\FIPpercInc}{100}{\FIPpercIncRnd}
\ROUND[0]{\FIPpercIncRnd}{\FIPpercIncRnd}
\ROUND[0]{\FIPNdiff}{\FIPNdiffRnd}

\DIVIDE{\FIPNdiffFC}{\FIPNdiff}{\FIPNafterFCShare}
\MULTIPLY{\FIPNafterFCShare}{100}{\FIPNafterFCShare}
\ROUND[0]{\FIPNafterFCShare}{\FIPNafterFCShareRnd}

\DIVIDE{\FIPNdiffFC}{\FIPNbefore}{\FIPNafterFCPropInc}
\MULTIPLY{\FIPNafterFCPropInc}{100}{\FIPNafterFCPropInc}
\ROUND[0]{\FIPNafterFCPropInc}{\FIPNafterFCPropIncRnd}

\makeatother

\usepackage{xcolor}
\definecolor{ceruleanblue}{rgb}{0.16, 0.32, 0.75}
\definecolor{brightmaroon}{rgb}{0.76, 0.13, 0.28}
\definecolor{bronze}{rgb}{0.8, 0.5, 0.2}
\definecolor{dartmouthgreen}{rgb}{0.05, 0.5, 0.06}

\newcommand{\AlertNote}[1]{}

 \newcommand{\DataCEoneIBprob}{9.2}

\newcommand{\DataCEoneISprob}{3}

\newcommand{\DataCEoneFBprob}{21.5}

\newcommand{\DataCEoneFSprob}{30}

\newcommand{\DataCEoneFBIBprob}{5}

\newcommand{\DataCEoneFBISprob}{1.7}

\newcommand{\DataCEoneNONEprob}{29.7}

\newcommand{\DataCEtwoIBprob}{6.1}

\newcommand{\DataCEtwoISprob}{3.8}

\newcommand{\DataCEtwoFBprob}{24.2}

\newcommand{\DataCEtwoFSprob}{34.1}

\newcommand{\DataCEtwoFBIBprob}{14.3}

\newcommand{\DataCEtwoFBISprob}{3.8}

\newcommand{\DataCEtwoNONEprob}{13.8}

\newcommand{\DataNEoneIBprob}{25}

\newcommand{\DataNEoneISprob}{7}

\newcommand{\DataNEoneFBprob}{14.9}

\newcommand{\DataNEoneFSprob}{6.2}

\newcommand{\DataNEoneFBIBprob}{24.1}

\newcommand{\DataNEoneFBISprob}{7.4}

\newcommand{\DataNEoneNONEprob}{15.5}

\newcommand{\DataNEtwoIBprob}{6.2}

\newcommand{\DataNEtwoISprob}{4.9}

\newcommand{\DataNEtwoFBprob}{29}

\newcommand{\DataNEtwoFSprob}{14}

\newcommand{\DataNEtwoFBIBprob}{32.5}

\newcommand{\DataNEtwoFBISprob}{7.8}

\newcommand{\DataNEtwoNONEprob}{5.6}

\renewcommand{\FBcZY}{0.033} \renewcommand{\FBcNY}{0.044} \renewcommand{\FBnZY}{0.046} \renewcommand{\FBnNY}{0.090} \renewcommand{\FScZY}{0.001} \renewcommand{\FScNY}{0.011} \renewcommand{\FSnZY}{0.071} \renewcommand{\FSnNY}{0.083} \renewcommand{\IBcZY}{0.124} \renewcommand{\IBcNY}{0.124} \renewcommand{\IBnZY}{0.053} \renewcommand{\IBnNY}{0.053} \renewcommand{\ILcZY}{0.156} \renewcommand{\ILcNY}{0.151} \renewcommand{\ILnZY}{0.186} \renewcommand{\ILnNY}{0.134}  \renewcommand{\FBcZsY}{0.0053} \renewcommand{\FBcNsY}{0.044} \renewcommand{\FBnZsY}{0} \renewcommand{\FBnNsY}{0.008} \renewcommand{\FScZsY}{0.00025} \renewcommand{\FScNsY}{0.037} \renewcommand{\FSnZsY}{0.0054} \renewcommand{\FSnNsY}{0.0045} \renewcommand{\IBcZsY}{0.098} \renewcommand{\IBcNsY}{0.098} \renewcommand{\IBnZsY}{0.0044} \renewcommand{\IBnNsY}{0.0044} \renewcommand{\ILcZsY}{0.042} \renewcommand{\ILcNsY}{0.056} \renewcommand{\ILnZsY}{0.011} \renewcommand{\ILnNsY}{0.034}  \renewcommand{\FBnZsY}{0.02}  \renewcommand{\FBcZs}{0.0079} \renewcommand{\FBcNs}{0.015} \renewcommand{\FBnZs}{0} \renewcommand{\FBnNs}{0.095} \renewcommand{\FScZs}{0.0002} \renewcommand{\FScNs}{0.0049} \renewcommand{\FSnZs}{0.029} \renewcommand{\FSnNs}{0.00012} \renewcommand{\IBcZs}{0.021} \renewcommand{\IBcNs}{0.021} \renewcommand{\IBnZs}{0.039} \renewcommand{\IBnNs}{0.039} \renewcommand{\ILcZs}{0.021} \renewcommand{\ILcNs}{0.052} \renewcommand{\ILnZs}{7.2e-05} \renewcommand{\ILnNs}{0.29} \renewcommand{\AMUcZs}{0.21} \renewcommand{\AMUcNs}{0.22} \renewcommand{\AMUnZs}{7.2e-05} \renewcommand{\AMUnNs}{0.045} \renewcommand{\ASDcs}{0.02} \renewcommand{\ASDns}{0} \renewcommand{\KDEPRECIATIONcs}{0.054} \renewcommand{\KDEPRECIATIONns}{0.0048} \renewcommand{\ALPHAKcs}{0.058} \renewcommand{\ALPHAKns}{0.01} \renewcommand{\BETAcs}{0} \renewcommand{\BETAns}{0} \renewcommand{\KAPPAcZs}{0.07} \renewcommand{\KAPPAcNs}{0.14} \renewcommand{\KAPPAnZs}{0.016} \renewcommand{\KAPPAnNs}{0.062} \renewcommand{\LOGITSDSCALEcs}{0.39} \renewcommand{\LOGITSDSCALEns}{0.014} \renewcommand{\RHOcs}{0.11} \renewcommand{\RHOns}{0.0052} \renewcommand{\STDEPSEcs}{0.45} \renewcommand{\STDEPSEns}{0.52} \renewcommand{\STDEPScs}{0.24} \renewcommand{\STDEPSns}{0.052}  \renewcommand{\FBcZs}{0.019} \renewcommand{\FBcNs}{0.03} \renewcommand{\FBnZs}{4.3e-18} \renewcommand{\FBnNs}{0.35} \renewcommand{\FScZs}{8.1e-05} \renewcommand{\FScNs}{0.0083} \renewcommand{\FSnZs}{0.061} \renewcommand{\FSnNs}{0.036} \renewcommand{\IBcZs}{0.064} \renewcommand{\IBcNs}{0.064} \renewcommand{\IBnZs}{0.14} \renewcommand{\IBnNs}{0.14} \renewcommand{\ILcZs}{0.058} \renewcommand{\ILcNs}{0.073} \renewcommand{\ILnZs}{0.01} \renewcommand{\ILnNs}{0.51} \renewcommand{\AMUnZs}{0.12} \renewcommand{\AMUnNs}{0.17}  \renewcommand{\FBcZ}{0.050} \renewcommand{\FBcN}{0.087} \renewcommand{\FBnZ}{0.403} \renewcommand{\FBnN}{1.350} \renewcommand{\FScZ}{0.000} \renewcommand{\FScN}{0.024} \renewcommand{\FSnZ}{1.303} \renewcommand{\FSnN}{1.305} \renewcommand{\IBcZ}{0.363} \renewcommand{\IBcN}{0.363} \renewcommand{\IBnZ}{0.491} \renewcommand{\IBnN}{0.491} \renewcommand{\ILcZ}{0.518} \renewcommand{\ILcN}{0.498} \renewcommand{\ILnZ}{3.041} \renewcommand{\ILnN}{2.319} \renewcommand{\AMUcZ}{0.58} \renewcommand{\AMUcN}{0.34} \renewcommand{\AMUnZ}{0.28} \renewcommand{\AMUnN}{0.18} \renewcommand{\ALPHAKn}{0.32}  \renewcommand{\FBcZ}{0.008} \renewcommand{\FBcZs}{0.0072} \renewcommand{\FBcN}{0.037} \renewcommand{\FBcNs}{0.015} \renewcommand{\FBnZ}{0.000} \renewcommand{\FBnZs}{0} \renewcommand{\FBnN}{0.589} \renewcommand{\FBnNs}{0.073} \renewcommand{\FScZ}{0.000} \renewcommand{\FScZs}{0.00024} \renewcommand{\FScN}{0.019} \renewcommand{\FScNs}{0.0042} \renewcommand{\FSnZ}{1.486} \renewcommand{\FSnZs}{0.023} \renewcommand{\FSnN}{1.500} \renewcommand{\FSnNs}{0.00011} \renewcommand{\IBcZ}{0.145} \renewcommand{\IBcZs}{0.014} \renewcommand{\IBcN}{0.145} \renewcommand{\IBcNs}{0.014} \renewcommand{\IBnZ}{0.435} \renewcommand{\IBnZs}{0.039} \renewcommand{\IBnN}{0.435} \renewcommand{\IBnNs}{0.039} \renewcommand{\ILcZ}{0.573} \renewcommand{\ILcZs}{0.017} \renewcommand{\ILcN}{0.508} \renewcommand{\ILcNs}{0.051} \renewcommand{\ILnZ}{3.500} \renewcommand{\ILnZs}{0} \renewcommand{\ILnN}{2.948} \renewcommand{\ILnNs}{0.27} \renewcommand{\KAPPAcZ}{0.13} \renewcommand{\KAPPAcZs}{0.052} \renewcommand{\KAPPAcN}{0.3} \renewcommand{\KAPPAcNs}{0.15} \renewcommand{\KAPPAnZ}{0.56} \renewcommand{\KAPPAnZs}{0.015} \renewcommand{\KAPPAnN}{0.13} \renewcommand{\KAPPAnNs}{0.047}  \renewcommand{\FBcZ}{0.139} \renewcommand{\FBcZs}{0.15} \renewcommand{\FBcN}{0.183} \renewcommand{\FBcNs}{0.16} \renewcommand{\FBnZ}{1.287} \renewcommand{\FBnZs}{1.2} \renewcommand{\FBnN}{1.482} \renewcommand{\FBnNs}{0.91} \renewcommand{\FScZ}{0.012} \renewcommand{\FScZs}{0.045} \renewcommand{\FScN}{0.024} \renewcommand{\FScNs}{0.037} \renewcommand{\FSnZ}{1.126} \renewcommand{\FSnZs}{0.48} \renewcommand{\FSnN}{1.232} \renewcommand{\FSnNs}{0.48} \renewcommand{\IBcZ}{0.393} \renewcommand{\IBcZs}{0.16} \renewcommand{\IBcN}{0.393} \renewcommand{\IBcNs}{0.16} \renewcommand{\IBnZ}{0.990} \renewcommand{\IBnZs}{1} \renewcommand{\IBnN}{0.990} \renewcommand{\IBnNs}{1} \renewcommand{\ILcZ}{0.490} \renewcommand{\ILcZs}{0.11} \renewcommand{\ILcN}{0.469} \renewcommand{\ILcNs}{0.12} \renewcommand{\ILnZ}{2.489} \renewcommand{\ILnZs}{1.2} \renewcommand{\ILnN}{2.127} \renewcommand{\ILnNs}{0.94} \renewcommand{\ASDn}{0.12} \renewcommand{\ASDns}{0.1} \renewcommand{\ALPHAKc}{0.38} \renewcommand{\ALPHAKcs}{0.18} \renewcommand{\BETAc}{0.92} \renewcommand{\BETAcs}{0.04} \renewcommand{\BETAn}{0.91} \renewcommand{\BETAns}{0.038} \renewcommand{\KAPPAcZ}{0.41} \renewcommand{\KAPPAcZs}{0.23} \renewcommand{\KAPPAcN}{0.36} \renewcommand{\KAPPAcNs}{0.21} \renewcommand{\LOGITSDSCALEc}{1.7} \renewcommand{\LOGITSDSCALEcs}{0.33} \renewcommand{\LOGITSDSCALEn}{1.5} \renewcommand{\LOGITSDSCALEns}{0.45} \renewcommand{\RHOc}{1.3} \renewcommand{\RHOcs}{0.15} \renewcommand{\RHOn}{1.2} \renewcommand{\RHOns}{0.17}  \renewcommand{\FBcZ}{0.025} \renewcommand{\FBcN}{0.064} \renewcommand{\FBnZ}{-0.000} \renewcommand{\FBnN}{0.935} \renewcommand{\FScZ}{0.000} \renewcommand{\FScN}{0.022} \renewcommand{\FSnZ}{1.473} \renewcommand{\FSnN}{1.497} \renewcommand{\IBcZ}{0.212} \renewcommand{\IBcN}{0.212} \renewcommand{\IBnZ}{0.432} \renewcommand{\IBnN}{0.432} \renewcommand{\ILcZ}{0.527} \renewcommand{\ILcN}{0.481} \renewcommand{\ILnZ}{3.499} \renewcommand{\ILnN}{2.644} \renewcommand{\ASDc}{0.13} \renewcommand{\KDEPRECIATIONc}{0.071} \renewcommand{\KDEPRECIATIONn}{0.056} \renewcommand{\BETAn}{0.88} \renewcommand{\STDEPSc}{0.55} \renewcommand{\STDEPSn}{0.37}  \renewcommand{\FBcZ}{0.017} \renewcommand{\FBcN}{0.050} \renewcommand{\FBnZ}{-0.000} \renewcommand{\FBnN}{0.789} \renewcommand{\FScZ}{0.000} \renewcommand{\FScN}{0.019} \renewcommand{\FSnZ}{1.480} \renewcommand{\FSnN}{1.500} \renewcommand{\IBcZ}{0.183} \renewcommand{\IBcN}{0.183} \renewcommand{\IBnZ}{0.439} \renewcommand{\IBnN}{0.439} \renewcommand{\ILcZ}{0.549} \renewcommand{\ILcN}{0.492} \renewcommand{\ILnZ}{3.500} \renewcommand{\ILnN}{2.764} \renewcommand{\KAPPAnZ}{0.56} \renewcommand{\KAPPAnN}{0.28}

\newcommand{\PAPERKEYWORDS}{\textbf{Keywords}: Informal credit, microfinance, structural estimation}
\newcommand{\PAPERJEL}{\textbf{JEL}: D15, D25, G21, O16, O17}

\newcommand{\PAPERTITLE}{An Empirical Equilibrium Model of Formal and Informal Credit Markets in Developing Countries}

\newcommand{\AUTHORWANG}{Fan Wang}
\newcommand{\AUTHORWANGURL}{https://orcid.org/0000-0003-2640-5420}
\newcommand{\AUTHORWANGINFO}{\href{\AUTHORWANGURL}{\AUTHORWANG}: Department of Economics, University of Houston, Houston, Texas, USA (email: fwang26@uh.edu)}

\newcommand{\ACKNOWLEDGMENTS}{
I am grateful to Flavio Cunha and Jere Behrman for their support and guidance throughout the course of this project, and I would like to thank Petra Todd, Dirk Krueger, and Robert Townsend for their invaluable support and extensive feedback. I would also like to thank Andrew Shephard, Jeremy Greenwood, Holger Sieg, Kenneth Wolpin, Esteban Puentes and Eun-Young Shim for useful discussions and comments.}

\newcommand{\PAPERABSTRACT}{
I develop and estimate a dynamic equilibrium model of risky entrepreneurs' borrowing and savings decisions incorporating both formal and local-informal credit markets. Households have access to an exogenous formal credit market and to an informal credit market in which the interest rate is endogenously determined by the local demand and supply of credit. I estimate the model via Simulated Maximum Likelihood using Thai village data during an episode of formal credit market expansion. My estimates suggest that a 49 percent reduction in fixed costs increased the proportion of households borrowing formally by \FIPNafterFCPropIncRnd\xspace percent, and that a doubling of the collateralized borrowing limits lowered informal interest rates by \IRNafterFCPropIncRnd\xspace percent. I find that more productive households benefited from the policies that expanded borrowing access, but less productive households lost in terms of welfare due to diminished savings opportunities. Gains are overall smaller than would be predicted by models that do not consider the informal credit market.\\
\PAPERJEL}

\newcommand{\PAPERDOIURL}{https://doi.org/10.1016/j.red.2021.09.001}
\newcommand{\PAPERINFO}{
This paper has been accepted for publication at the Review of Economic Dynamics: \url{\PAPERDOIURL}.
}

\begin{document}

\title{\PAPERTITLE
\thanks{
\PAPERINFO
}}

\author{
\AUTHORWANG\thanks{
\AUTHORWANGINFO.
\ACKNOWLEDGMENTS
}}

\date{September 17, 2021}

\maketitle

\begin{abstract}
\singlespacing
\PAPERABSTRACT\\
\PAPERKEYWORDS
\end{abstract}
\thispagestyle{empty}
\clearpage

\pagenumbering{arabic}
\setcounter{page}{1}
\renewcommand*{\thefootnote}{\arabic{footnote}}

\section{Introduction}

In recent decades, development banks and micro-finance institutions in developing countries have funded the expansion of formal financial institutions into rural areas. Recent randomized evaluations show that the effects of expanding formal borrowing and savings opportunities are generally positive but limited \autocite{banerjee_six_2015}. One factor that might impact the effectiveness of formal credit market expansions is the availability of local informal alternatives.

In this paper, I structurally evaluate the effects of microfinance interventions in the presence of informal credit markets. Microfinance interventions introduce formal options to households' financial choice set. But microfinance might also impact existing informal choices through equilibrium effects as demand and supply in those informal credit markets shift in response to interventions. The joint effects of microfinance could lead to different welfare consequences for heterogeneous households depending on their borrowing and savings needs. By combining the formal and informal options, my analysis spans prior structural analyses of microfinance. \textcite{kaboski_structural_2011} conduct a structural evaluation of microfinance at the micro level without distinguishing between formal and informal choices, while \textcite{buera_macroeconomics_2012} analyze the macro-equilibrium effects of microfinance without considering the response of local informal credit markets.

To accomplish this, I develop and estimate a dynamic equilibrium risky entrepreneur model that incorporates formal as well as informal borrowing and savings choices.
In the model, households are infinitely lived, risk averse, and have varying productivity and \CZH. Households choose risky capital investments and \SSSS, and they can finance their risky investments through \BBBB.
The formal credit market is characterized by exogenously determined differential borrowing and savings interest rates, and formal credit is subject to a potential collateral constraint. In contrast, in the informal market considered here, full enforcement is assumed, and market clears given a locally determined borrowing and lending interest rate. The model has several key features of formal-informal credit market interactions: First, formal and informal credit market options differ in interest rates, collateral requirements, and access/transaction costs; second, formal and informal options could be substitutes or complements as households sort into seven credit market participation categories;\footnote{Households could borrow formally, save formally, borrow informally, lend (save) informally, borrow formally and informally at the same time, borrow formally and lend informally, or not borrow or save. Households pay per-period \FCF to access any of the different credit market options. If households decide to borrow from the formal credit market, they face a \CLC that limits \BBBB up to a fraction of their \PYS choices.} third, informal credit markets are localized, and exogenous changes in formal credit market conditions could shift equilibrium local informal interest rates.

Given these features, microfinance availability can have redistributive consequences through equilibrium effects on locally determined informal interest rates. The existing structural microfinance evaluation literature generally models improvements in microfinance as a shift in the borrowing collateral constraints \autocite{BKSannualReview}. In the context of this model, subsidizing the centrally set formal borrowing and savings interest rates, reducing access
transaction costs, and relaxing borrowing collateral constraints can separately and jointly improve microfinance availability. The effects of relaxing the collateral constraints could be magnified or dampened by variations in formal borrowing and savings fixed costs and interest rates.

I estimate the model using panel data from the Townsend Thai Monthly Survey, which contains extensive information about the sizes and rates of formal and informal credit market transactions for about 650 households from 16 villages between 1999 and 2009. During this period, in 2001, the Thai government, under Prime Minister Thaksin Shinawatra, implemented a wave of large-scale policies, including the Million Baht Village Fund program, aimed at broadening rural households' access to the formal credit market.\footnote{\textcite{kaboski_structural_2011} studied the Million Baht Fund using a partial equilibrium model without distinguishing between formal and informal choices and modeled the Thaksin policy as shifting the borrowing constraints. They used data from 64 villages of the Townsend Thai Annual Survey. The annual survey has more villages across a wider geographic area but has limited information on the local-informal credit market compared to the monthly survey.} Specifically, the policies reduced interest rates by offering a low fixed \BBBB rate, reduced \FCFs by directly administering formal loans with centrally set rates via village committees, and potentially helped relax \CLCs by increasing the number of formal loans available. As a result of these policies, the proportion of households that used formal \BBBB increased by up to \FIPNdiffRnd\xspace percentage points, and the formal borrowing interest rates decreased by up to 8 percentage points. The local informal \ITR also dropped by up to 14 percentage points. I introduce the Thaksin policy shift in my model as possible changes in centrally set formal interest rates, in formal \FCFs, and in formal \BBBB \CLCs.

The model is estimated using simulated maximum likelihood. I develop solution and estimation algorithms that capture the differential effects of variations in formal and informal access costs and constraints on asset choices and distributions. The estimation allows for the identification of unobserved \FCFs and \CLCs, and household preferences and productivities. I allow the parameters that characterize the credit market to vary over time and estimate their changes based on changes in credit choice category participation probabilities. I find that the \FCFs to access the formal credit market dropped from about 9\% of average annual \ICM to about 4.6\%, and the \CLCs were also significantly relaxed, more than doubling how much each household could borrow \NLCLY given the same amount of \PYS. Along with the lowering of the centrally set formal \BBBB interest rates, these changes resulted in significant improvements in households' access to formal \BBBB.

In terms of welfare comparisons, the estimated model suggests that there were winners and losers. The majority of rural households are better off under Thaksin's policies, although the estimated steady state gains are close to zero in consumption-equivalent variation for most households.\footnote{While it is difficult to compare across models that consider different equilibrium objects, \textcite{buera_macroeconomics_2012, townsend_welfare_2010} estimate welfare gains on the order of 5\% to 15\% from financial deepening and consider equilibrium changes in wage and aggregate interest rates.} Steady state welfare effects are heterogeneous across households. High productivity households could gain up to about 5\%. These households were previously constrained by the \CLCs or by the \FCFs from investing more in their household farms or businesses. However, low productivity households could lose up to 1\% in consumption-equivalent variation. These households have relatively higher preferences for safe savings over taking out loans to finance their low expected return risky investments. Their gains from lower borrowing costs are outweighed by diminished savings opportunities due partly to the drop in locally determined equilibrium informal interest rates.

I conduct counterfactual experiments to decompose the relative contributions of each of the three formal \BBBB access dimensions in explaining the effects of the Thaksin policies.
The redistributive welfare effects are largely driven by the relaxation of the collateral constraint, which by itself accounts for \IRNdiffCLShareRnd\% of the reduction in the informal \ITR.
Reductions in formal \BBBB \FCFs had a large impact on participation shares, and by themselves these reductions account for \FIPNafterFCShareRnd\% of the increase in the fraction of households using formal credit options. Interestingly, the reduction in the centrally set formal \BBBB \ITRs had limited effects on aggregate participation and the informal market \ITRs. The three measures of access have different impacts because they shift the average costs of formal loans differently for different types of households.

\paragraph{Related Literature}

The analysis in this paper contributes to several strands of the literature. First, there has been substantial research on the impacts of greater financial access on developing economies \autocite{greenwood_financial_1990, lloyd-ellis_enterprise_2000, gine_evaluation_2004, kaboski_structural_2011, NBERw20821, buera_macroeconomics_2012}. Despite the importance of informal financial arrangements \autocite{udry_risk_1994, townsend_risk_1994}, these dynamic models of financial deepening---formal credit market expansion---generally do not explicitly consider informal financial options. Additionally, studies that test the fit of informal risk-sharing models to data do not model formal options explicitly \autocite{alem_evaluation_2014, karaivanov_dynamic_2014, KinnanThai}. In this paper, I model risky entrepreneurs' choices over formal and informal credit market options in an exogenous incomplete markets setting. While there are different ways for rural households to transfer financial resources, \textcite{karaivanov_dynamic_2014} find that a model with exogenously incomplete borrowing and savings options fit consumption and investment data in rural Thai villages better than constrained efficient credit/insurance models. The model in effect augments equilibrium models of risky entrepreneurs (see review: \textcite{quadrini_entrepreneurship_2009}) with additional exogenous borrowing and savings options. Similar to \textcite{kaboski_structural_2011}, I treat villages as small open economies where formal prices are exogenously determined, but I extend the framework to explicitly consider informal choices and equilibrium interest rates determined within a local informal credit market. My approach here focuses on the \textit{micro-equilibrium} effects of formal credit market expansion on village credit markets. This complements the work of \textcite{buera_macroeconomics_2012, BrezaKinnanNBERw24329}, which focuses on the macro equilibrium effects of large microfinance roll-outs on prices, including interest rates and wages, on the aggregate economy.

Second, this paper contributes to work that studies the interaction between formal and informal credit markets. Due to the expansion of development banking, there has been significant interest in the interaction between formal and informal credit markets \autocite{HoffStiglitzWBER, BesleyJEP}. Studies have analyzed the impact of differential access costs and contract enforcement on the sorting between formal and informal borrowing options \autocite{gine_access_2011, KARAIVANOV2018100}; the interaction between suppliers of informal credits and formal banks \autocite{sagrario_floro_vertical_1997, madestam_informal_2014}; and the effects of formal credit expansion on informal credit market interest rates \autocite{mookherjee_theory_2016, DEMONT201621}. To pin down the analytical structures, these papers generally rely on non-dynamic models and generate sorting conditional on wealth and types. This paper incorporates some core features of formal and informal credit markets interaction into an empirical dynamic equilibrium framework with endogenous asset distributions. Credit market participation now includes borrowing and savings in both markets, and the supply side of informal credit is endogenized through the dynamics of \SSSS.

Third, there is a significant and growing empirical literature that analyzes separate dimensions of credit market policies. Studies have found that formal borrowing and savings choices are elastic to subsidies or shifts in the interest rates of borrowing \autocite{karlan_credit_2008, dehejia_interest_2012} and saving \autocite{schaner_persistent_2018}. The expansion of formal borrowing or savings opportunities, which partly reduces the fixed/transaction costs of access, have had modest but heterogeneous effects on households in previously under-exposed areas \autocite{banerjee_six_2015, dupas_banking_2018}. Offering loans to borrowers who would otherwise be ineligible due to a possible lack of collaterals \autocite{banerjee_firms_2014-1, augsburg_impacts_2015} has also had mixed results. To allow for the coexistence of various formal and informal credit market participation categories, I incorporate into the model formal borrowing and savings interest rates, fixed costs, and collateral constraints. As a result, this paper provides a potential framework for studying the non-separable effects of these important dimensions of formal credit market policies.

Fourth, there is a literature that studies how the provision of formal insurance could crowd-out informal insurance \autocite{attanasio_consumption_2000, krueger_public_2011, ChandrasekharKinnanLarreguy}. These papers find that more formal insurance provisions might worsen informal conditions and lead to welfare losses. In this paper, I study formal and informal market interactions in the context of exogenously incomplete credit markets. The welfare gains and losses in this paper arise out of a competitive local-informal credit market's interaction with an exogenous formal credit market with centrally set rates. This contrasts with the insurance literature, where welfare losses arise in the context of an endogenously incomplete market structure that is constrained by limited commitment and other frictions.

The structure of the paper is as follows. Section \ref{sec:modelmodel} develops the model. In Section \ref{sec:Data-and-Background}, I describe the data and background. Section \ref{sec:estiesti} describes the estimation results and counterfactuals. I offer the conclusion in Section \ref{sec:concludeconclude}. Additional details for the solution and estimation methods are in Appendix Sections \ref{sec:soluestialgo}.

\section{The Model\label{sec:modelmodel}}
For a village economy, there is a continuum of infinitely lived household firms that are heterogeneous in their \PDT \(\A\), \PYS \(\K\), and \FNT \(\B\). \(\A\) is fixed over time and represents a household's heterogeneous ability to earn \ICM. \(\K\) is the input for the household firm. \(\B\) is the sum of principal and interests due from or owed to both \NLC and \LLC sources. Comparison of expected productivity and interest rates endogenously determine households' asset holdings. In each period, households face two kinds of shocks: a \SKE \(\SHK\) and a vector of \USK \(\SHKU\) for alternative \NLC and \LLC credit choice categories. Given these shocks, households choose assets within joint \NLC and \LLC credit choice categories.
 \subsection{Preference}
\label{subsec:preference}

Households are risk averse with respect to consumption within each period, and utility over consumption is separable over time. Individual expected utility over lifetime sequence of consumption \(\CC_{t}\), is \(\E\left[\sum_{t=0}^{\infty}\PBETA^{t}\utt\left(\CC_{t}\right)\right]\), where within each period, utility from consumption is \(\frac{\left(\CC_{t}\right)^{1-\PCRRA}}{1-\PCRRA}\). \(\PBETA\) is the discount factor, \(\PCRRA\) is the coefficient of relative risk aversion. The expectation is over realized values of \SKEs \(\SHK\) and \USK \(\SHKU\).\footnote{\(\SHKU\) primarily captures stochastic factors unrelated to intertemporal consumption and investments that impact households' willingness to save and borrow. For example, given realized shocks, households might have differing levels of non-pecuniary values associated with borrowing from and lending to other households that are in search of lending opportunities or that are in need of loans.}
 \subsection{Technology}
\label{subsec:production}
At the beginning of each period, for each household, the household \PDT \(\A\), normally distributed \SKE \(\SHK\), and \PYS \(\K\), which is chosen previously by the household based on expected productivity, jointly determine the \ICM of the household in the current period:\footnote{The model focuses on financial (safe asset) and \PYS  (risky asset) choices. Labor supply for the household firm is inelastic and captured by \(\A\).}
\begin{equation}
  \label{eq:ProductionFunction}
  \Y=\exp\left(\A+\SHK\right)\cdot \K^{\PALPHA}
  \thinspace.
\end{equation}
 \subsection{Credit Markets \label{subsec:Credit-Markets}}
\label{subsec:credit}
\newcommand{\modelcreditfoottwo}{For households with low \PDT is effectively a cash-under-mattress option.}
\newcommand{\modelcreditfootthree}{Pecuniary components of the \FCFs include transportation costs, fees, and opportunity costs among others. Non-pecuniary components of \FCFs include aversion or preference toward different credit market activities. There is evidence that transaction costs for formal banking access are significant in many developing countries \autocite{beck_banking_2008}, and there is also evidence that reducing the fixed costs for financial access \autocite{brune_facilitating_2015, dupas_effect_2017} and bank branch expansions \autocite{burgess_rural_2005,bruhn_real_2014} could be welfare improving.}

\newcommand{\modelcreditfootfour}{Policy makers have long used borrowing and savings interest rates offered by development banks as policy instruments \autocite{buttari_subsidized_1995}. Recent studies have found that borrowers are sensitive to formal borrowing interest rate changes and subsidies in South Africa \autocite{karlan_credit_2008} and Bangladesh \autocite{dehejia_interest_2012}. It has also long been observed that higher formal savings rates can mobilize savings \autocite{ravallion_estimates_1986}, and recent evidence from Kenya shows that savings interest rate subsidies could have persistent positive effects on income and assets \autocite{schaner_persistent_2018}.}

\newcommand{\modelcreditfootfive}{
Microfinance institutions generally distinguish themselves from traditional development banks by lowering the collateral requirements \autocite{cull_microfinance_2009}.
In recent occupational-choices models with entrepreneurial investment \autocite{NBERw20821,buera_macroeconomics_2012}, current period static capital investments are constrained by prior dynamic savings choices---collateral is savings that banks can confiscate. In this paper, risky capital decisions are made prior to the realization of productivity shocks, and
\(\PKPPA\) determines the proportion of risky physical capital that banks are willing to finance. Here, collateral is physical capital that can be used to pay off loans.
}

Villages are small open economies with respect to external-\NLC financial choices with centrally set interest rates. Villages also contain an endogenous internal-\LLC credit market with a locally determined interest rate. Following \textcite{gine_access_2011}, financial access in \NLC and \LLC credit markets is impacted by \FCF, \ITRs, and \CLCs. Households choose among seven credit alternatives: \PYS only; borrowing from the \LLC credit market; borrowing from the \NLC market; lending in the \LLC credit market; saving in the \NLC market; borrowing jointly from the \NLC and \LLC markets; or concurrently borrowing from the \NLC market and lending in the \LLC market.

\subsubsection{Fixed Costs}

There are four \FCFs: $\left\{\FXCjfb,\FXCjib,\FXCjfs,\FXCjil\right\}$. In each period, households pay the \FCFs associated with the chosen credit category. These \FCFs represent the pecuniary and non-pecuniary costs associated with \NLC and \LLC credit market activities,\footnote{Pecuniary components of the \FCFs can include the costs incurred at home (e.g., bookkeeping), costs incurred on the way to the bank (e.g., transportation or search costs), and costs incurred at the bank (e.g., fees). Non-pecuniary components of the \FCFs can include the opportunity cost of time as well as general preference aversion towards different credit market activities.
} and the formal \FCFs are potential policy instruments.

Reducing $\FXCjfb$ and $\FXCjfs$, the \FCFs of accessing formal savings and borrowing opportunities, can play an important role in deepening financial access \autocite{townsend_aer_1983, greenwood_financial_1990, GRANDA2019302, NBERw20821}. There is also strong empirical evidence that the fixed costs for formal banking access are significant in many developing countries \autocite{beck_banking_2008, dupas_banking_2018}.

The fixed costs for informal lending, $\FXCjil$, potentially captures the per-lender level costs of set-up and monitoring \autocite{TOWNSEND1979265, williamson_costly_1986, greenwood_financial_1990}.
While the literature has considered both aggregate as well as per-dollar-of-loan intermediation costs \autocite{greenwood_financial_1990, becsi_ping_wynne}, I focus on aggregate costs, which can be the dominant factor for small-scale lending activities.

\subsubsection{Interest Rates}

There are three \ITRs: the \NLC \BBBB \ITR \(\Rjfb\), the \NLC \SSSS \ITR \(\Rjfs\), and the \LLC \ITR \(\RI\). The two centrally set \NLC \ITRs are exogenous, and the \LLC \ITR is competitively determined by local demand and supply. Even though the locally determined \LLC \ITR \(\RI\) is the same for borrowers and lenders, the combination of \LLC interest rate \(\RI\) and \(\FXCjil\) leads to informal interest earning differentials.\footnote{A common interest rate differential arises if the cost of lending is incurred at the per-dollar-of-loan level \autocite{NBERw20821}. In that setting, the average and marginal rates are the same. In a setting with per-lender aggregate fixed costs, the average interest rate differentials differ by the lender depending on the lending level.}

The large roll-outs of micro-finance programs could potentially impact prices at the aggregate level \autocite{buera_macroeconomics_2012, BrezaKinnanNBERw24329}. Here, however, I take the interest rates offered by formal lending and deposit-taking institutions as centrally set policy rates and focus on the effects of microfinance on micro equilibrium interest rates.

\subsubsection{Collateral Constraint}

\CAPCAP\NLC banks allow households to borrow up to \(\PKPPA\) fraction of their \PYS choice \(\Kp\). This means a fraction of \RKI could be financed by \NLC borrowing. Informal borrowing is not constrained by \(\PKPPA\). In the absence of possibly binding borrowing constraints, households can borrow up to the natural borrowing constraint \autocite{aiyagari_uninsured_1994}. The natural borrowing constraint as well as the collateral constraint are visualized in Appendix \FIGRMjfcr.

The formal borrowing constraint can be interpreted as a collateral constraint that secures repayments given enforcement frictions \autocite{BKSannualReview}.\footnote{\modelcreditfootfive} In contrast, full enforcement (no strategic default) is assumed for informal loans. In addition to the collateral interpretation, \(\PKPPA\) might also capture rationing given potential limits to the amount of loans that are made available from formal lenders to households in each village \autocite{kaboski_structural_2011}.

\subsubsection{Coexistence of Formal and Informal Choices}

The four fixed costs, three interest rates, and one collateral constraint allow for the coexistence of the seven credit market participation categories.
Fixed costs can account for borrowers choosing loans with higher interest rates and savers choosing savings opportunities with lower savings rates. The collateral constraint allows households to jointly choose formal and informal borrowing and limits the amount of arbitrage from borrowing formally and lending informally.

In this setting, for a development banker, there are five levers of formal credit market policies: subsidizing centrally determined \(\Rjfb\) or \(\Rjfs\), reducing \(\FXCjfb\) or \(\FXCjfs\), and relaxing \(\PKPPA\). The model provides a framework for analyzing the effects of these policies jointly, taking into consideration the informal credit market and the conditions of the village economy.
 \subsection{Recursive Formulation of the Household's Problem}
\label{subsec:recursive}
Households maximize their lifetime utility by choosing sequences of consumption, \PYS, and \NLC and \LLC \FNT positions, subject to a sequence of budget constraints and \NLC \BBBB limits. Let \(\COH\) denote \CHOHH (alternatively wealth).
\(\COH\) is the sum of \FNT, depreciated \PYS, and \(\Y\) which is realized given \SKE \(\SHK\):
\begin{equation}
\FCOH =
\exp\left(\A + \SHK\right)
\cdot \K^{\PALPHA} +
\left(1-\PDELTA\right) \K +
\B
\thinspace.
\label{eq:CashOnHand}
\end{equation}
 \(\COH\) is the resources available for consumption and new asset choices.

At the beginning of a period, the state-space \(\STATEsetEle\) of the household is summarized by \PDT \(\A\), \PYS level \(\K\), \FNT level \(\B\), \SKE \(\SHK\), and a vector of credit type-specific utility shocks \(\SHKU\). Together, \(\STATEsetEle=\left(\STATES\right)\).  \(\VakbephiStates\) is the maximum over the value of the seven credit category alternatives discussed previously:
\begin{equation}
\VakbephiStates=\max_{\JinOtS }
\left\{
\VakbejujOptStates{1},...,\VakbejujOptStates{7}
\right\}
\thinspace.
\label{eq:MaxOf7}
\end{equation}
 The optimal credit category choice is \(\fo\left(\STATES\right)\in\left\{\OtS\right\}\). \(\VakbejujOptStates{j}\) is the value for each credit category \(\J\) given optimal continuous assets and consumption choices. \CAPCAP\USK \(\SHKUj\) is i.i.d. extreme value and influences the relative values of the seven credit choice categories given optimal continuous choices.

Equation \eqref{eq:ValVj} is the generic form of \(\VakbejujOptStates{j}\). Given the period budget constraint, a household chooses \(c\), \RKI \(\Kp\), and next period financial position \(\Bp\):

\begin{equation}
\VakbejujOptStates{j}
=
\max_{
\begin{array}{c}
  \CC,
  \Kp \in \KpSET{j},\\
  \BIp \in \BIpSET{j},
  \BFp \in \BFpSET{j}
\end{array}}
\left\{
  \begin{array}{c}
    \utt \left(\CC\right) + \SHKUj + \\
    +
    \PBETA
      \EVJEVp{\Bp}
  \end{array}
\right\}
\thinspace,
\label{eq:ValVj}
\end{equation}
\begin{equation}
\mbox{s.t.:\hspace{2.5mm}}
\CC = \FCOH - \Kp- \FXCj -
\left\{
\begin{array}{c}
\frac{\BFp \INDI \left\{\BFp>0\right\}}{1+\Rjfs} +
\frac{\BFp \INDI \left\{\BFp \le 0\right\}}{1+\Rjfb} +
\frac{\BIp}{1+\RI}
\end{array}
\right\}
\thinspace,
\label{eq:Budget}
\end{equation}
\[
\Bp = \BIp + \BFp
\thinspace.
\]
 The budget constraint for each category \(\J\) includes the \FCFs \(\FXCj\) and the associated \ITRs. $\KpSET{j}$, $\BIpSET{j}$ and $\BFpSET{j}$ are the constraint sets for asset choices in each participation category. The continuation value is a function of the state space elements \(\left(\STATESnext\right)\). Next period financial position \(\Bp\) represents the net amount of principal plus interest owed and earned from \BBBB and saving undertaken by the household in the current period. Full equation for each \(\left\{ \Vakbejuj \right\}_{\JinOtS}\) is shown in Appendix Section \ref{subsec:equavaljall}.

\FIGTYPEONE{fig_main}{\FIGTSckb}{\FIGPSckb}{1.0}{\FIGDSckb}

Nesting the safe and risky asset choice problem of Equation \eqref{eq:ValVj} within the discrete participation categories of Equation \eqref{eq:MaxOf7} allows for the analysis of key types of formal and informal participation possibilities. Despite the nesting, asset choices have expected patterns. In \FIGRSckb, the y-axis shows average continuous choice solutions to Equation \eqref{eq:ValVj} integrated over the discrete probabilities induced by Equation \eqref{eq:ValVj}. As \CZH increases, the overall fraction of \CZH invested in risky and safe assets increases, and the \RKI share of overall savings decreases. Lower \CZH households finance a portion of their \RKI through \BBBB.
Households with low \CZH and high \PDT externally finance the majority of their \RKI by \BBBB.
Households with high \CZH and low \PDT consume the smallest proportion of their wealth and invest more in safe savings.
These heterogeneous borrowing and savings needs of households are important in sorting households into demanders and suppliers of credit on the informal market with locally determined interest rates. Improvements in borrowing and savings access can benefit households differently.

 \subsection{Stationary Competitive Equilibrium}
\label{subsec:stationary}

\label{subsec:equaequi}

Given centrally set
\NLC \ITRs for \SSSS \(\Rjfs\) and \BBBB \(\Rjfb\),
\CLC \(\PKPPA\),
and a vector of \FCFs \(\FXC\),
a Recursive Competitive Equilibrium consists of the values and policy function for the household
\(\Vakbephi: \STATEset \rightarrow \REAL\),
\(\fo:\STATEset \rightarrow \left\{\OtS\right\}\),
\(\left\{\fc, \fk, \fbi, \fbf : \STATEset \rightarrow \REAL \right\}_{\JinOtS}\),
locally determined \LLC \ITR \(\RI\), as well as stationary measure \(\MEASURE\),
such that:
\begin{enumerate}
 \item Given \(\Rjfs\), \(\Rjfb\), \(\RI\), \(\PKPPA\), \(\FXC\), the value function \(\Vakbephi\) solves Equation \eqref{eq:MaxOf7} and the value functions \(\left\{\Vakbejuj\right\}_{\JinOtS}\) solves Equation \eqref{eq:ValVj} for each $\JinOtS$. \(\fo\) and \(\left\{ \fc, \fk, \fbi, \fbf \right\}_{\JinOtS}\) are the associated policy functions.
 \item \CAPCAP\LLC credit market clears:
 \begin{equation}
   \label{eq:modelequiequiequi}
  \ITG
  \left\{
  \SUMA_{\JinOtS}
  \fp \left(\A, \COH, \SHKUdist \right)
  \cdot
  \fbi \left(\A, \COH \right)
  \right\}
  \ITGD\MEASURE\left(\ITGD\A, \ITGD\COH\right) = 0
  \thinspace,
 \end{equation}
 where \(\fp \left(\A, \COH , \SHKUdist \right)\) denotes the probability of a household with \(\left(\A, \FCOH \right)\) choosing credit category \(\J\), given the distribution over the \USK \(\SHKUdist\).
\end{enumerate}
 In this stationary equilibrium, the measure \(\MEASURE\) over the state space is invariant with respect to the Markov process induced by \(\SHK\), \(\SHKU\), the distribution of \(\A\), and the policy functions.\footnote{
Households facing negative productivity shocks draw down their savings or resort to borrowing to finance consumption and maintain physical capital investments. Conversely, positive productivity shocks can lead to increases in savings and physical capital investments. The introduction of discrete credit category shocks generates greater uncertainties. Given credit category participation probabilities, I generate overall wealth transition matrices that integrate over credit-category-specific wealth transition matrices. While I do not offer a formal proof, within the range of parameter values explored during estimation, the numerical methods discussed in
Appendix Section \ref{subsec:steady} is able to find the model-induced Markovian wealth transition matrices and corresponding stationary distributions.
}

While the locally determined \LLC credit market clears, the \NLC market with centrally set rates does not have to. Positive excess supply of \NLC credit means that \NLC institutions are mobilizing rural savings. Village income would come from within village production as well as from returns to formal savings. Negative excess supply of \NLC credit means that \NLC institutions are injecting credit into the village economy. Risky investments in the village would be partly financed by outside banks that receive a portion of village outputs as interest payments.

For policy experiments, welfare in the aggregate village economy is measured by a social welfare function in the steady state\footnote{While it is important to consider welfare along the transition path, this paper focuses on steady state welfare comparisons. This is a limitation of the approach here. Estimating the structural model while accounting for the transition path would require solving for equilibrium transition paths at each set of parameters that the estimation algorithm searches through. For computational tractability, estimation is conducted under steady state assumptions. To preserve the consistency between estimation and counterfactual analysis, I compute welfare changes under steady state assumptions as well.} that is Utilitarian with equal weights assigned to all households.
 \subsection{Solving the Dynamic Programming Problem}
Solving heterogeneous-agent equilibrium models with multiple continuous and discrete states and constrained continuous and discrete choices is potentially time-consuming.
Furthermore, given that the model has four fixed costs, one collateral constraint, and three interest rates, the solution algorithms need to accurately capture how these eight values determine the feasible choice set as well as the average prices of borrowing and saving in formal and informal credit markets. I adopt a set of solution algorithms to reduce iteration and increase accuracy within each of the five standard steps involved in solving this class of model.\footnote{The five general parts are: 1, optimization of the choice problem; 2, iteration over the value function; 3, simulation of the steady state distribution; 4, integration over types; 5, finding equilibrium prices. Speed improvements for the first three steps are achieved through vectorization, and for the final two steps are achieved by using more parallel processors.}

The algorithm produces conditional asset distributions, which are used in constructing the estimation likelihood described in Section \ref{sec:likelihood}. To arrive at these conditional asset distributions, I first solve for optimal choices given current value function approximation using an iterative grid search algorithm as described in Appendix Section \ref{subsec:solu}. Then, I approximate the expected future value using multi-linear splines over \(\Kp\) and \(\Bp\) as described in Appendix Section \ref{subsec:vfi}. In Appendix Section \ref{subsec:steady}, I derive the distribution of \CZH tomorrow based on \CZH in the current period as a transition matrix. I project to solve for the steady state asset distributions. Finally, I solve for the equilibrium \ITR using a multi-section algorithm where I also integrate over different productivity types as described in Appendix Sections \ref{subsec:typeinter} and \ref{subsec:equi}.

\section{Data and Background\label{sec:Data-and-Background}}
\label{par:thaivillages}
I estimate the model with Thai village data. Thai villages have traditionally had strong \LLC credit markets \autocite{siamwalla_thai_1990}. The Thai government has also subsidized the expansion of state development banks led by the Bank for Agriculture and Agricultural Cooperatives (BAAC) \autocite{maurer_agricultural_2000}.

After the election of Thaksin Shinawatra in 2001, the government introduced programs to improve rural \BBBB access. The Million Baht Village Fund (MBF) program was the most prominent program and provided every village with one million baht in credits at centrally set rates \autocite{boonperm_does_2013}. The total funds amounted to 1.5 percent of GDP \autocite{kaboski_townsend_aejapplied_2012}. The funds were transferred to villages via the BAAC, but the power to approve loans was in the hands of local committees. In addition to the MBF, the government gave BAAC a greater mandate to expand its lending programs, and also created additional programs through other agencies.\footnote{The Government \CAPCAP\SSSS Bank was given additional resources. There were also lending programs aimed at supporting small businesses, funding the purchase of low-priced household goods, and financing schooling costs \autocite{phongpaichit_thaksin:_2004, menkhoff_village_2011, boonperm_does_2013}.}

I use the 1999 to 2009 waves of the Townsend Thai Monthly data, which is a panel for about 650 households in 16 villages \autocite{samphantharak_townsend_2009}.\footnote{There were 684 households in 1999. Summary statistics and estimation rely on 606 households for whom there is information for all years. 304 households are from the Northeast.} The villages are split between the wealthier Central and the more impoverished Northeast regions. Households consist of multiple generations and operate businesses and farms of varying scales. In contrast to the Townsend Thai Annual Survey, which has 64 villages and allows for analysis using village-level variations of MBF credit-per-household \autocite{kaboski_structural_2011, kaboski_townsend_aejapplied_2012}, the monthly survey has more complete data on formal and informal financial transactions.
 \subsection{Formal and Informal Channels}
\subsubsection{Borrowing and Savings Alternatives}
\label{par:formalinformal}

Table \ref{tab:BORROWERparticipate} presents the relative popularities of the main lenders: MBF and BAAC are external-formal lenders with centrally set rates; Village Coops, friends/neighbors, and moneylenders are internal-informal lenders with locally determined rates.\footnote{Village Coops include Production Cooperative Groups and Village Agricultural Cooperatives. Commercial banks are external-formal lenders, but I do not consider them in the analysis since they only account for 0.3 percent of the total number of loans.} Village coops are by definition specific to the village or township. 72 percent of loans from friends/neighbors and money-lenders are from within the village or township (tambon), and 93 percent are from within the same province (changwat). For the analysis in this paper, I aggregate across villages in the two wealthier central provinces and the two poorer northeastern provinces separately. The analysis assumes that local conditions are similar within each region.

In the Northeast, 78\% percent of households borrowed at least once from the MBF, 62\% from the BAAC, 36\% from Village Coops, 83\% from friends and neighbors, and 38\% from moneylenders. In the Central villages, \LLC participation rates are roughly half of those in the Northeast. On average, \NLC borrowers borrowed \NLCLY in half of the years between 1999 and 2009, and informal borrowers borrowed from friends and neighbors in four of these years.

Table \ref{tab:LENDERparticipate} also presents the relative popularities of the main channels for savings: commercial banks, the BAAC, and the GSB are external-formal deposit takers; saving at Village Coops and lending to other households constitute internal-informal savings. In the Northeast, 94\% of households saved with either the BAAC or the GSB, 37\% saved with commercial banks, 64\% saved with Village Coops, and 67\% lent directly to other households. In the Central villages, the use of commercial banks for \SSSnoS was twice as frequent as in the Northeast, and individual lending took place half as often. On average, formal depositors made deposits in 7.8 out of the 11 years, and household lenders lent out in 2.6 out of the 11 years.
 \subsubsection{Formal and Informal Credit Categories}
\label{par:Credit-Choice-Categories}

I annualize household credit market participation. Some households used one option annually: \BBBB or \SSSnoS \NLCLY, \BBBB or \SSSnoS \LLCLY, or no \BBBB or \SSSnoS. Some households borrowed from both \NLC and \LLC sources, and others borrowed from \NLC sources and lent \LLCLY in the same year.\footnote{Joint credit choices have long been observed in village economies \autocite{hoff_moneylenders_1998, sagrario_floro_vertical_1997}. \textcite{bell_rationing_1997} discuss joint \NLC and \LLC \BBBB.} Households are classified as choosing one of these seven participation categories, which allow for substitutability and complementarity between \NLC and \LLC options.\footnote{Households with other combinations of activities are grouped into these seven categories based on the level of \BBBB and savings they undertake \NLCLY and \LLCLY.}

Thaksin's policies started in the second half of 2001, but came into full force in 2002. Given potential heterogeneities in borrowing and savings frictions as well as household productivity and wealth distributions across villages, external-formal interventions could have differential local-informal effects depending on whether the complementarity or substitutability of informal and formal options dominate. Tables \ref{tab:NECEcreditSahres4m} and \ref{tab:NECEcreditShares7m} show the average annual household participation shares in the seven credit market categories between 1999--2001 and between 2002--2009. Overall, the \LLC household participation shares are higher in the Northeast villages, and the \SSSnoS shares are higher in the Central villages. There is a large shift after the implementation of Thaksin's policies. After 2001, Northeast households that only borrowed or saved \NLCLY increased by 23 percentage points, and those that only borrowed or lent \LLCLY decreased by 21 percentage points. In the Central villages, these shares increased by 7 and decreased by 2 percentage points, respectively. At the same time, the proportion of households using both \NLC and \LLC credit markets increased after 2001, shifting from 6.7 percent to 18 percent of households in the Central villages, for example.
 \subsection{Formal and Informal Loan Characteristics}
\subsubsection{Interest Rates}
\label{par:Interest-Rate}

\newcommand{\DATAINTFOOTONE}{BAAC and MBF rates, which are centrally set and tend to be subsidized, have limited variations within areas and periods, but \LLC \ITRs have higher mean as well as variance. While the definition for informality differs across studies, local-informal loans have been found to carry high rates \autocite{conning_chapter_2007}. For example, \textcite{banerjee_credit_2017} find this to be the case in Hyderabad, India. At the same time, a significant proportion of loans to relatives and friends often have low reported rates. \textcite{KARAIVANOV2018100} show that these ostensibly low-cost loans might have high implicit social-tie-based "shadow costs" that could make them more expensive, in expectation, than formal loans.}

\newcommand{\DATAINTFOOTTWO}{International capital markets and national policies could shift \NLC interest rates. The BAAC, for example, uses a "transfer-pricing" system that allows branches to borrow from or deposit at BAAC's headquarters at policy rates, which might shift to subsidize lending or mobilize \SSSS \autocite{maurer_agricultural_2000}.}

\newcommand{\DATAINTFOOTTHR}{Using the Annual survey, \textcite{kaboski_townsend_aejapplied_2012} find that the level of household MBF borrowing does not have significant impacts on average credit-interest rates. This is consistent with MBF borrowers relying on MBF and BAAC loans with similar rates.}

I compute annual \ITRs taking into consideration pecuniary repayments as well as repayments in kind.
Table \ref{tab:intRatereg3yr2m2} and Figure \ref{fig:Interest1} show distributions of annualized average village \ITRs for \NLC and \LLC categories. I find that loans with locally determined rates, informal loans in the context of this paper, were higher than loans with centrally set rates, formal loans in the context of this paper.\footnote{\DATAINTFOOTONE} Additionally, \LLC \BBBB \ITRs were higher than \NLC \SSSS \ITRs. Both \NLC and \LLC \ITRs decreased significantly over time.\footnote{\DATAINTFOOTTHR} The average \LLC interest rate in the Northeast was 28\% before 2002, compared to 15\% for \NLC \BBBB and 3.3\% for \NLC saving. The corresponding rates in the Central region were 18\%, 13\%, and 3.2\%, respectively. After the policy shift, the \LLC interest rates decreased by 14 percentage points to 13.9\% and 8 percentage points to 9.4\% in the Central and Northeast regions, respectively. Accompanying these shifts, the \NLC \BBBB interest rates decreased by 8 percentage points in both the Northeast and Central regions to approximately 6\%, and the \NLC saving interest rates in both regions also decreased by approximately 2 percentage points to about 1\%.
 \subsubsection{Borrowings and Savings Amounts}
\label{par:Amount-of-Borrowings}

Using annualized sums of \NLC or \LLC borrowing activities and the net-flow of \NLC or \LLC \SSSS activities, Appendix Figure \ref{fig:SaveLoanSize} shows that \NLC loan sizes tend to be larger than \LLC loan sizes and \LLC savings/lending tend to be larger than \NLC savings. The median \LLC individual lending size is 17,500 baht. Although some households have very large amounts of \NLC \SSSS, aggregating across all, the median net flow of \NLC \SSSS is approximately 8,000 baht per year. The median \NLC and \LLC \BBBB loan sizes are 27,000 and 15,500 baht, respectively.
These compare to a mean annual income across survey years of around 125,000 Baht and a median annual income of around 65,000 Baht among the sampled households.

Disaggregating across lenders, Appendix Figure \ref{fig:LoanSize} shows that BAAC loans tend to be larger in size, and loan sizes from those that are categorized as village moneylenders and friends are similar. Additionally, aggregating the loan volumes of key lenders, Appendix Table \ref{tab:borrowintensive} shows that there was a significant increase in MBF loan volume shares and a corresponding sharp drop in informal lender loan volume shares after the Thaksin policy shift. \subsubsection{Fixed Costs}
\label{par:Fixed-Cost}
The survey captures aspects of the pecuniary \FCFs for borrowing. Appendix Table \ref{tab:FCBorrDetailType} and Figure \ref{fig:FixedCost} show significant reductions in loan-specific fees and transportation costs after 2001 for \NLC \BBBB, mainly due to the lower reported costs for MBF loans. MBF loans had fees and transportation costs similar to informal loans, which is likely due to their village committee-based administration. Reported BAAC costs also decreased over time, which might have been driven by improvements in the BAAC itself or reductions in transportation and communication costs. BAAC costs, however, were still twice as high as the average costs reported for informal loans. Strikingly, reported costs for the few commercial bank loans were, on average, seven times larger than the costs for BAAC loans and almost 35 times larger than the costs for MBF or \LLC loans.
 \subsubsection{Collateral and Repayment}
\label{par:Repayment}

Across all years, 21.0 percent of all BAAC loans have collateral requirements. Before 2002, this proportion was 32.3 percent. After the policy shift, 15.7 percent of BAAC loans have collateral requirements. For MBF loans, less than 0.1 percent have collateral requirements. This evidence indicates that the overall collateral requirements for these formal loans with centrally set interest rates are relaxed after the policy shift. In contrast, 6.9 percent of loans from non-relatives and 4.5 percent of loans from relatives are reported as having collateral requirements. Over time, the share of informal loans requiring collateral did not change significantly.

For repayments, Appendix Tables \ref{tab:Repay_ontime} and \ref{tab:Repay_notFull} show that 97 percent of \NLC loans and 95 percent of \LLC loan principals were repaid.\footnote{Every month, repayment amounts of each loan are tracked until full repayment. I compute repayment rates three months after the initially stated due date for each loan.} Additionally, households that do not fully pay back their loans still pay a significant proportion of their debts.\footnote{Using the Townsend Thai Annual data, \textcite{kaboski_structural_2011} find that repayment rates for MBF loans are 97 percent, and for other loans are between 77 to 88 percent. The repayment of interest and principal seems to be differentiated for informal loans. Repayment rates considering principals only is high, but there are variations in initially stated interests and actual interests paid.}
Relatively high repayment rates could be due to strong informal enforcement mechanisms, localized monitoring of MBF loans, or the presence of collateral constraints for BAAC loans.

The combination of limited collateral and high repayment for locally determined informal loans motivates the full-enforcement assumption for informal loans in the model. For formal loans with centrally set rates, the relatively higher collateral requirements and their reductions over time motivate the modeling of formal loans with a collateral constraint that could vary over time.

\section{Estimation and Counterfactuals\label{sec:estiesti}}
\subsection{Parameters}
\label{sec:parameters}
I interpret the Thaksin policy shifts as changes in the formal \ITRs, the formal \FCFs, and the formal \BBBB \CLCs.\footnote{Large-scale expansions of credit market access might often involve shifting multiple dimensions of credit market access jointly. For example, through significant interest rate subsidies, the Reserve Bank of India helped to finance the expansion of branch banking, with formal borrowing and savings services, to 30,000 rural villages between 1969 and 1990 \autocite{burgess_rural_2005}. This Indian policy could also be interpreted as shifting formal savings and borrowing interest rates, fixed costs, and collateral constraints jointly.} The focus of estimation is to identify the two unobserved dimensions of credit access, namely \FCFs and \CLCs, before and after the policy shift, as well as parameters for household preference and the production function.

To estimate the model, I divide the data into a pre-Thaksin period and a post-Thaksin period. For each period, I estimate separate values for \FCFs and \CLCs for villages in the Northeast and Central regions of Thailand. The main parameters that need to be estimated are:
\[
 \Theta_{\ETR}=
 \left(
 \underset{\mbox{Preference}}{
  \underbrace{
  \PBETA_{\EstiRegion},
  \PCRRA_{\EstiRegion},
  \PPHISIGMASub{_\EstiRegion}
}
 },
 \thinspace
 \thinspace
 \thinspace
\thinspace
 \thinspace
 \underset{\mbox{Production}}{
  \underbrace{\PALPHA_{\EstiRegion}, \PDELTA_{\EstiRegion}, \PAMUSub{\ETR}, \PASIGMASub{_{\EstiRegion}}, \EESIGMASub{_{\EstiRegion}}}
 },
 \underset{\mbox{Credit Market}}{
  \underbrace{\FXC\ETR,\PKPPA\ETR}
 }
 \right)
 \thinspace,
\]
where $\EstiRegion$ is region, and $\EstiTime$ represents the two periods. Preference and production function parameters differ across regions except for the mean of \(\A\), which is time- and region-specific to allow for productivity shifts. Credit market parameters are different across periods and regions.\footnote{Formal credit market parameters shift over time due to policy changes. In addition to that, I also allow the informal savings/lending fixed costs to vary over time. The growth of formal alternatives reduces households' reliance on the informal credit market for consumption smoothing and investment financing, potentially weakening their commitments to informal arrangements \autocite{attanasio_consumption_2000, ChandrasekharKinnanLarreguy}. This could increase the information acquisition and monitoring costs for local lenders, which is captured by the informal savings/lending \FCF.} I assume that households do not predict the changes in credit market access.

The identification for the credit access parameters $\psi\ETR\mbox{ and }\PKPPA\ETR$ comes from the fact that changes in \FCFs and \CLCs have separate impacts on households' credit choice probabilities. If the \FCFs for formal \SSSS increase, it will tend to reduce the probability of households choosing formal \SSSS. If the \FCFs for formal \BBBB increase, that will tend to reduce the probability of formal \BBBB along with the probability of choosing the two joint formal and informal credit choice categories. As the \CLC relaxes, the relative probability of choosing the joint options would decrease.

In addition to the credit access parameters, the identification of utility parameters $\PBETA, \PCRRA \text{, and }\PPHISIGMA$ come from households' preferences for \BBBB versus \SSSS and the equilibrium interest rates, which differ across regions. The identification of production function parameters comes from the relationship between \RKI and \OTP.\footnote{Both permanent productivity types and a persistent shock process can capture persistence in the income process, but they are in practice difficult to jointly identify given the data. For estimation, I set shock persistence to zero.}

Finally, I solve and estimate the model under steady state assumptions.
Potentially, given the vectors of $\Theta_{\ETR}$ for two periods in the same region, one could solve for policy functions and prices along the transition path, and construct transition path-specific likelihoods. This would, however, compound the computational burdens of the estimation procedure. Additionally, parameters might become not identifiable if there is indeterminacy in mapping the microdata to positions along the transition path. Given these concerns, for estimation, I assume that the observed data capture steady state choices.\footnote{On the one hand, by averaging over choices during the transition and at the steady state, the estimation procedure might underestimate the magnitudes of the improvements in borrowing conditions. On the other hand, conditional on parameter estimates, welfare analysis might overstate gains by assuming that consumption functions shift to the steady state immediately. Jointly, the effects of imposing the steady state assumption on estimation, which is a limitation of the approach in this paper, are ambiguous.}
 \subsection{Maximum Likelihood}
\label{sec:likelihood}

\newcommand{\likelihoodFootOne}{In standard partial equilibrium life cycle models with discrete choices and continuous asset states and choices that are measured with errors, the econometrician simulates the life-cycle a large number of times given sequences of shocks. The differences between these paths and the actual path of observed individual choices are components of the likelihood function \autocite[e.g.,][]{keane_effect_2001, imai_intertemporal_2004}. In these models, the asset distribution is partially endogenous, given initial exogenous asset distributions. In the problem here, the asset distribution is fully endogenous. Given the non-iterative procedure for deriving conditional asset distribution described in Appendix Section \ref{subsec:steady}, \(\PDF \left( \A \condi \COH \right)\) can be obtained without simulating histories. Consequently, evaluating Equation \eqref{eq:likefour} does not involve sequential simulations but is obtained from a set of matrix multiplications and additions.}
\newcommand{\likelihoodFootTwo}{
 Given that \(\SHKU\) is assumed to be i.i.d extreme value,  \(\FPcohs\) follows standard multinomial-logit formulations. The probability of choosing one of the credit categories is:
 \begin{equation}
  \begin{split}\label{eq:multilogit}
   \PROB
   \left(
   \OO\HT = \J
   \condi
   \A_\HH,\K\HT,\B\HT,\SHK\HT
\right)
   =
   \prod_{\J=1}^{\JJ}
   \left(
   \frac{
    \exp
    \left(
      \Vakbeitj
      \left(
        \A_\HH, \K\HT, \B\HT, \SHK\HT
      \right)
      /
      \PPHISIGMA
    \right)
   }{
    \sum_{l=1}^{\JJ}
    \exp
    \left(
      \Vakbeitl
      \left(
        \A_\HH,\K\HT,\B\HT,\SHK\HT
      \right)
      /
      \PPHISIGMA
    \right)
   }
   \right)^{\INDI\left\{ \OO\HT=\J\right\} }
   \thinspace.
  \end{split}
 \end{equation}
Since $\Vakbeitj$ is not linear-in-parameters, $\PPHISIGMA$ allows for conventional logit scaling normalization \autocite[Chapter 3]{train_2009}. Additionally, while $\phi_{jit}$ is uncorrelated across $j$, the categories share common shock $\epsilon_{it}$.
}
\newcommand{\likelihoodFootThree}{\textcite{karaivanov_dynamic_2014} estimate, among other models, a partial equilibrium dynamic borrowing and savings model with maximum likelihood also using these rural Thai data. They build a likelihood function based on policy functions tracing out the path of choices without steady state assumptions given fixed interest rates.}

For estimation, in addition to the discrete credit market participation information described earlier, I use income, asset, and consumption data from household balance sheets compiled by \textcite{samphantharak_townsend_2009}. Income is household revenue from all sources minus costs. Physical assets include land, business and agricultural assets, livestock, and household assets. Following \textcite{karaivanov_dynamic_2014}, I exclude from physical capital land, livestock, and durable household assets.\footnote{\textcite{Pawasutipaisit11} describe key asset variables and find that while overall wealth increased at 2.7 percent in nominal terms every year, 44 percent of households experienced negative growth in net wealth between 1999 and 2005. For the data periods here, real consumption in the Northeast increased, on average, from 47.1 thousand baht to 55.5 thousand baht from the before-policy to the after-policy period. 1st, 5th, 15th, and 25th percentile real consumption decreased from 12.6, 18.5, 25.5, and 29.4 thousand baht to 11.0, 17.0, 24.5, and 29.3 thousand baht, respectively. Median and third-quartile consumption increased from 39.1 and 51.2 to 42.8 and 66.9 thousand baht. (Nominal consumption percentiles are uniformly higher except for the 1st percentile.)} I deflate prices by 2005 northeast rural price levels using different rural price indexes for each region.

I estimate the model using simulated maximum likelihood with measurement errors. I simulate the likelihood by integrating over the steady state joint \CZH \(\COH\) and \PDT \(\A\) distributions derived in Appendix Sections \ref{subsec:steady} and \ref{subsec:typeinter}.
The likelihood is determined by the individual \(\estiobsi\) and time \(\estitimei\) specific continuous choices and states observed with error \(\left(\KpMSR\EIT, \BpMSR\EIT, \CCMSR\EIT, \COHMSR\EIT \left( \KMSR\EIT, \BMSR\EIT, \YMSR\EIT \right)\right)\), the observed discrete choices \(\J\EIT\), the policy functions \(\FKcohsESTI\), \(\FBcohsESTI\), \(\FCcohsESTI\), \(\FPcohsESTI\), and the region \(\EstiRegion\) and period \(\EstiTime\) specific joint \(\PDF \left( \COH,\A \psep \ESTIALLETR \right)\) steady state distributions.\footnote{\likelihoodFootThree}

The probability of observing \(\left(\KpMSR, \BpMSR, \CCMSR, \J\right)\) given \(\left( \COHMSR \left( \KMSR, \BMSR , \YMSR \right)\right)\) is:
\begin{equation}
 \label{eq:likeone}
 \PDF
 \left(\KpMSR, \BpMSR, \CCMSR, \J \condi \COHMSR \left( \KMSR, \BMSR , \YMSR \right) \right)
 =
 \ITG_{\COH}
 \ITG_{\A}
 \PDF \left( \KpMSR , \BpMSR , \CCMSR, \J \condi \A , \COH \right)
 \PDF \left( \A \condi \COH \right)
 \PDF \left( \COH \condi \COHMSR \right)
 \ITGD \A
 \ITGD \COH
 \thinspace,
\end{equation}
where the first term in the integral is the product of lognormal measurement errors and the discrete credit choice category probability:
\(
\PDF \left( \KpMSR , \BpMSR , \CCMSR, \J \condi \A , \COH \right)
=
\KMsrErr\cdot\BMsrErr\cdot\CCMsrErr
\cdot
\FPcohs
\)
, with \(\KMsrErr=\KMsrErr\left( \log(\KpMSR) - \log(\FKcohs) \right)\), and similarly for \(\BMsrErr\) and \(\CCMsrErr\).\footnote{\likelihoodFootTwo} The log ratio measurement error assumes that observed and model-generated \(\B\) choices have the same signs, which implies that discrete choice \(\J\) is not observed with measurement error. Additionally, I obtain \(\PDF \left( \COH \condi \A \right)\) while solving the model following the procedure in Appendix Section \ref{subsec:typeinter}, hence \(\PDF \left( \A \condi \COH \right)\) can be replaced using Bayes' rule:
\(
 \label{eq:likethree}
 \PDF \left( \A \condi \COH \right)
 =    \frac{
  \PDF \left( \COH \condi \A \right)
  \PDF \left( \A \right)
 }{
  \ITG
  \PDF \left( \COH \condi \Ahat \right)
  \PDF \left( \Ahat \right)
  \ITGD \Ahat
 }
\)
. Following Appendix Sections \ref{subsec:steady} and \ref{subsec:typeinter}, the integrals in Equation \eqref{eq:likeone} are approximated using mid-point Riemann-sum for \(\COH\) and quadrature for \(\A\). Finally, \(\COHMSR\) is determined by measurement errors for \(\KMSR\), \(\BMSR\), and \(\YMSR\).  Given these, the likelihood is:
\begin{equation}
  \label{eq:likefour}
 \ell_{\EstiRegion\EstiTime} =
 \PMULT_{\estiobsi=1}^{\estiobs}
 \PMULT_{\estitimei=1}^{\estitime}
 \left\{
 \ITG_{\COH}
 \ITG_{\A}
 \PDF \left( \KpMSR\EIT , \BpMSR\EIT , \CCMSR\EIT, \J\EIT \condi \A , \COH \psep  \ESTIALLETR \right)
 \PDF \left( \A \condi \COH  \psep \ESTIALLETR \right)
 \PDF \left( \COH \condi \COHMSR\EIT \right)
 \ITGD \A
 \ITGD \COH
 \right\}
 .
\end{equation}
I estimate the model using the log of \(
\PMULT_{\EstiRegion \in \left\{1,2\right\},\EstiTime\in \left\{1,2\right\}}
  \left(
    \ell_{\EstiRegion\EstiTime}
  \right)
\) as the objective function.\footnote{\likelihoodFootOne}

It is important to note that both measurement errors on the continuous choices as well as \USK \(\SHKU\) are needed to assure that the joint choice probabilities can be computed during the estimation process. Within bounds of the state space, \USK assure positive choice probabilities for each credit market participation category. Given the discrete choice, the measurement errors on the asset choices assure positive probabilities of observing any asset choices.

 \subsection{Estimation Routine}
\label{subsec:summalgo}
Dynamic heterogeneous-agent equilibrium models with discrete and continuous choices are time consuming to estimate. Searching for the correct local minimum often requires restarts to calibrate estimation step-size and tolerance levels. I develop a global optimization routine that minimizes the use of loops and iteration by parallel computing. The algorithm has four steps that take advantage of scalable parallel computing resources and is described more fully in Appendix Section \ref{subsec:nestedestimation}. In brief, the process is as follows: 1. I evaluate the model concurrently at a large set of randomly drawn model parameters; 2. I approximate the estimation objective function using polynomials; 3. I estimate an approximated version of the model by evaluating the objective function using polynomial regression coefficients; and 4. I estimate the model using the actual model objective function and use results from earlier steps to find initial parameters. The first three steps are fully parallelizable and could provide step four with parameter values that are potentially close to global minimizers. The procedure potentially reduces estimation time significantly. I discuss computational structures and the costs of running this algorithm on cloud computing services in Appendix Sections \ref{subsec:containerorchestration} and \ref{subsec:awsprice}.
 \subsection{Estimation Results and Interpretations \label{subsec:Estimation-Results}}
\label{subsec:estimates}

Parameter estimates are shown in Tables \ref{tab:costEsti} and \ref{tab:otherEstimates}. Fits are shown in Table \ref{tab:modelFit}. In Table \ref{tab:costEsti}, estimated fixed costs are shown as fractions of annual average income in the first period. The estimated model fits credit participation shares and asset choices generally well across regions and periods. The exception in the Northeast is that the model overpredicts the proportion of households borrowing informally only, and consequently the average share of informal borrowing in total borrowing.\footnote{This is perhaps because in the estimation---while other fixed costs are allowed to vary across periods---the informal borrowing fixed costs are fixed across periods. Additionally, given the likelihood-based estimation approach, it is possible that at parameters that maximize the log-likelihood, predicted moments do not match well with a subset of empirical counterparts. This is especially true in this model, given the relatively parsimonious parameter specifications. In the tradition of \textcite{keane_career_1997}, a strategy to improve fits is to incorporate additional exogenous observables and discrete unobserved types into the model. That could allow for greater heterogeneities in predicted choices conditional on states. That approach, however, would significantly increase the computational burdens for solving and estimating the model.
} The exception in the Central provinces is that the model overpredicts consumption.\footnote{In the Central provinces, the model predicts that consumption should be in excess of three-fourths of outputs, but in the data, it is only half. This might be related to the fact that Central households purchase more household durable goods than Northeast households, and durable household goods are excluded from estimation.}

The estimated formal \BBBB \FCFs fractions decreased in the Northeast region from \FBnNY\xspace to \FBnZY, a 49\% reduction. In the Central region, where changes in participation rates were more limited, formal \BBBB \FCFs fractions decreased from \FBcNY\xspace to \FBcZY. In comparison, using an earlier round of the Thai survey, \textcite{gine_access_2011} provides estimates for formal borrowing fixed costs that would be approximately 7\% of the earlier period borrowing fixed costs in the Northeast region. These reductions are driven by the introduction of MBF and other policies that brought formal lending decisions to local communities, reducing the fixed costs of what I categorize as external-formal lending. Reductions in these fixed costs help to explain increases in formal credit market participation rates over time. The \FCFs for formal \SSSS decreased as well, but their changes were less significant. The estimated formal \BBBB \FCFs fractions decreased in the Northeast region from \FSnNY\xspace to \FSnZY. In the Central region, formal \SSSS \FCFs fractions decreased from \FScNY\xspace to \FScZY, indicating the availability of low-cost deposit-taking services in these more affluent areas.

Table \ref{tab:costEsti} also presents \FCFs estimates for informal \BBBB and \SSSS. The informal \BBBB \FCFs fractions are \IBnNY\xspace in the Northeast region and \IBcNY\xspace in the Central region. Interestingly, in the Central region, to fit the low participation shares in informal borrowing, the fixed costs estimates for informal borrowing are higher than for formal borrowing.\footnote{Without preference shocks for credit participation categories, fixed costs for the lower interest rate formal borrowing option would need to generally be higher than the fixed costs for the higher interest rate informal borrowing option \autocite{gine_access_2011}. Higher informal costs could reflect the higher costs of finding credit suppliers when the informal market is small.} Additionally, the estimated \FCFs fractions for informal \SSSS are \ILnNY\xspace for the Northeast region in the earlier period and \ILnZY\xspace in the later period, and are \ILcNY\xspace for the Central region in the earlier period and \ILcZY\xspace in the later period. Households with low \SSSS needs do not lend \LLCLY unless they receive a large positive shock.\footnote{When positive shocks induce households to lend informally, households might be lending small amounts informally with negative average returns. It is often observed that informal savings/lending might earn negative returns. For example, \textcite{banerjee_credit_2017} explicitly consider a fixed negative savings interest rate.} Informal fixed costs could be partly attributed to local preferences toward informal borrowing and savings activities, and they could also partly represent credit search costs for borrowers and intermediation costs for savers/lenders.

\SUBTRACT{\FSnNY}{\FSnZY}{\FSnDec}
\DIVIDE{\FSnDec}{\FSnNY}{\FSnDecRatio}
\MULTIPLY{\FSnDecRatio}{100}{\FSnDecRatio}
\ROUND[0]{\FSnDecRatio}{\FSnDecRatioRnd}

\SUBTRACT{\ILnZY}{\ILnNY}{\ILnInc}
\DIVIDE{\ILnInc}{\ILnNY}{\FSnIncRatio}
\MULTIPLY{\FSnIncRatio}{100}{\FSnIncRatio}
\ROUND[0]{\FSnIncRatio}{\FSnIncRatioRnd}

An important result here is that while borrowing opportunities improved in the Northeast, savings opportunities did not. The formal savings fixed costs decreased by \FSnDecRatioRnd\xspace percent, improving formal savings opportunities, but the informal savings fixed costs increased by \FSnIncRatioRnd\xspace percent.
Given these results and the drops in both the formal savings interest rates as well as the informal interest rates, the return to savings decreased in the Northeast.

In addition to changes in the \FCFs parameters, Table \ref{tab:costEsti} also shows that in the Northeast region, the \CLC fraction increased from \KAPPAnN\xspace to \KAPPAnZ, and in the Central region, the \CLC fraction increased from \KAPPAcN\xspace to \KAPPAcZ. In comparison, modeling MBF as a shift in borrowing constraints---specified as a fraction of a household's permanent income---\textcite{kaboski_structural_2011} find that constraints under MBF relaxed from about 8 percent of permanent income to about 28 percent, on average.\footnote{The borrowing constraints fractions in this paper are multiplied by physical capital, and hence not directly comparable to the \textcite{kaboski_structural_2011} results. Both results show that relaxing borrowing constraints was a key Thaksin policy channel, but it seems that the borrowing constraints were tighter initially and experienced greater relaxation under \textcite{kaboski_structural_2011}'s framework. In the set-up here, shifts in fixed costs and interest rates can change the fraction of households constrained (in joint participation categories), holding borrowing constraints constant.} Relaxing the collateral constraints increases the fraction of risky investments that could be financed by formal borrowing and also increases the amount of consumption that could be financed by formal borrowing given the same level of risky investments.

The changes in formal borrowing conditions were likely brought about by the MBF program. As discussed earlier, the program allowed households to borrow from village committees. Pecuniary borrowing costs could have decreased: for example, there were reductions in observed formal \BBBB fees and transportation costs, as discussed in Section \ref{par:Fixed-Cost}. It should be noted that the estimated fixed costs are much higher than the reported fees and transportation costs, indicating that the reported costs might not fully capture various pecuniary and non-pecuniary costs associated with formal banking access.
For example, non-pecuniary borrowing costs could also have changed: communicating with village committees about credit needs might be less burdensome than less familiar BAAC managers. Additionally, the MBF did not have explicit collateral requirements, as noted in the data section. MBF also made large quantities of additional funds available for formal lending through these committees with local knowledge.

\subsection{Welfare Consequences of Thaksin\textquoteright s Policies\label{subsec:Welfare-Consequences-of}}
\label{subsec:thaksinwelfare}

\begin{figure}[!htbp]
\vspace*{-3mm}
\makebox[\textwidth][c]{
\fcolorbox{white}{white}{
\centering
\begin{minipage}{1.0\textwidth}
\begin{center}
\caption{\label{fig:wel11}Welfare Change in Northeast from 1999-2001 to 2002-2009}
\begin{tikzpicture}
  \node (img)  {
  \includegraphics[clip, trim=1.37cm 0.72cm 0cm 0cm, scale=0.70]{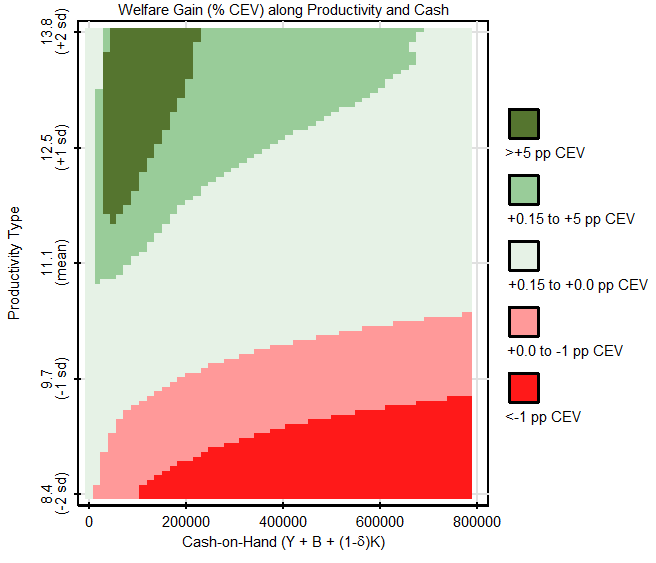}};
  \node[below=of img, node distance=0cm, yshift=1cm,font=\color{black}] {\hspace{-30mm} \CAPCAP\CZH};
  \node[left=of img, node distance=0cm, rotate=90, anchor=center,yshift=-0.7cm,font=\color{black}] {Productivity Type};
 \end{tikzpicture}
\end{center}
\small \emph{Notes:} This figure shows the steady state welfare impacts of Thaksin's policies that lowered formal borrowing interest rates and fixed costs, and relaxed formal borrowing collateral constraints. I compare \emph{steady state} welfare for households facing credit market access parameters from the \emph{1999-2001} and the \emph{2002-2009} periods. I present welfare changes in terms of \emph{consumption equivalent variation} (\textbf{CEV}) percentage points changes.  Credit market access parameters are observed or estimated.
\end{minipage}
}}

\end{figure}

\newcommand{\FOOTthaksinwelfareOne}{Formal \BBBB \ITRs could have moved down due to overall changes in international \ITRs as well as government policies that shifted them.}

Thaksin's policies expanded formal borrowing access, especially in the Northeast. However, returns to savings generally diminished in the Northeast, as discussed in Section \ref{subsec:estimates}. Northeast welfare effects are shown in Figure \ref{fig:wel11}. All estimated and observed parameters from the two periods are used to calculate welfare changes in terms of consumption equivalent variation (CEV) gains for households with different productivity and \CZH levels. To be consistent with the estimation procedure, I conduct steady state welfare analysis. The welfare analysis does not consider the costs incurred to the government for financing Thaksin's policies.

Figure \ref{fig:wel11} shows welfare effects for subsets of households. At the top left, high \PDT and low \CZH households gain on the order of slightly more than 5\% CEV. These households have high \BBBB needs and can now borrow formally more cheaply given both the drop in the \ITR and \FCF.\footnote{These households also face future states in which they would have high savings needs. The worsening of savings conditions is detrimental to them. But overall, they benefit more from the improvements in borrowing conditions.} The second group of households includes those who have more than average \PDT and higher levels of \CZH. These households gain on the order of 0.15\% to 5\% CEV. In this group, some of the gains are due to reductions in the formal borrowing \ITR. The presence of informal options weakens welfare gains here: the gains of switching from informal borrowing to a lower-cost formal borrowing option is less than the gains of moving from autarchy to borrowing.

The third group consists of a large proportion of households with average levels of productivity and \CZH, shown in the central area of Figure \ref{fig:wel11}. These households have modest current period borrowing needs and benefit less from the lower formal borrowing interest rates or relaxed collateral constraints. Looking forward, they face future states in which they have higher borrowing or higher savings needs. Overall, these households experience slightly positive but limited gains from the policy changes.

The last two groups have higher \CZH and lower productivity levels. They are shown in the bottom right of Figure \ref{fig:wel11}. These households suffer welfare losses of up to 1\% CEV due to the worsening of savings conditions. They have low expected returns for investments, and hence also lower borrowing needs compared to more productive households. Gains from improvements in borrowing conditions are outweighed by losses due to the lower formal and informal savings interest rates as well as the higher informal lending fixed costs. The welfare losses are smaller for households with slightly higher productivity types and lower levels of wealth.

Welfare effects for the last two groups could be different if informal markets are not considered. Specifically, in \textcite{kaboski_structural_2011}, poor and low productivity households benefit from microfinance consumption loans. Here, the same households already had access to low fixed costs informal consumption loans and switching to microfinance consumption loans led to limited gains.
Additionally, conditional on productivity type, higher wealth households in this setting can lose from improvements in borrowing conditions: productive but low wealth households rely less on higher wealth informal lenders as financiers for their high return risky capital investments.
 \subsection{Impacts of Shifting Dimensions of Formal Access}
\label{subsec:threedimsprob}

In this section, I conduct counterfactual policy experiments in which I shift each formal borrowing policy parameter from its value in the 2002-2009 period back to its value in the 1999-2001 period while holding other parameters at their 2002-2009 levels. As in Section \ref{subsec:thaksinwelfare}, I focus on the Northeast region only. The policy effects are nonlinear and non-separable. Hence, their individual impacts do not sum up to their joint effects. The policy experiments investigate the incremental effects of policy shifts along each policy dimension individually, conditional on the empirical estimates of other policy parameters.\footnote{\textcite{gine_access_2011} also analyzes the effects of shifting these policy dimensions in magnitudes unrelated to Thaksin's policies and in the context of a static partial equilibrium model. In both static partial equilibrium and dynamic general equilibrium settings, reductions in borrowing interest rates, drops in borrowing fixed costs, and the relaxation of collateral constraints lead to greater formal borrowing participation. The addition of dynamics means that effects are due to both changes in policy functions as well as distributions. The addition of equilibrium means that changes in formal borrowing conditions might shift informal borrowing and savings conditions as well. The overall dynamic equilibrium model allows for welfare analysis.}

\subsubsection{Fixed Costs}

In this particular policy setting, I find that fixed costs changes can be effective in changing credit market participation shares but have limited equilibrium effects. As discussed previously, lowering the fixed costs reduces the average costs of borrowing, especially for households with limited borrowing needs or households that are constrained in how much they can borrow, and leads to higher formal borrowing participation shares. Specifically, shifting the \NLC \BBBB \FCFs parameter back up to its previously higher level reduces the proportion of households using only the \NLC credit market from \FPNafter\% to \FPNafterFC\%. This means that the lowering of the \NLC \BBBB \FCFs by itself accounts for \FPNafterFCShareRnd\% of the total change in \NLC only participation shares and increases the first-period \NLC only participation shares by \FPNafterFCPropIncRnd\%; if households that use \NLC and \LLC markets jointly are also considered, these two percentages would be \FIPNafterFCShareRnd\% and \FIPNafterFCPropIncRnd\%, respectively.

Setting the \NLC \BBBB \FCFs at their previous levels also increases the informal \ITR from \IRNafter\% to \IRNafterFC\%. The change is modest because the impacts from reduced demand for informal borrowing only would be attenuated by increased demand for informal loans from households who borrow jointly. Limited equilibrium effects mean that the changes in fixed costs were not the key factor driving the redistributive welfare consequences presented in Figure \ref{fig:wel11}.

\subsubsection{Interest Rates}

Interestingly, the large change in the formal \BBBB \ITR from \FBRNbefore\% to \FBRNafter\% has relatively small impacts on participation rates. This is because given formal \BBBB \FCFs and \CLC, changing formal \BBBB \ITRs mainly impacts the average cost of formal loans for households with higher borrowing needs---low wealth but very productive households. However, these households would have already been \BBBB \NLCLY, even at higher \NLC \ITRs. Given other parameter values in 2002-2009, increasing the \NLC \BBBB \ITR back to its previous level would have less than a 3 percentage point impact on participation rates and would only increase the informal \ITR by \IRNafterRRdiff\%. Given the relatively small impacts on equilibrium interest rates, similar to fixed costs, interest rate changes had limited redistributive welfare consequences in this policy setting.

\subsubsection{Collateral Constraints}
Relaxing the formal \BBBB \CLC can allow \NLCLY constrained households to borrow more and at lower average costs. Given the estimated parameters, this brings about a significant change in the demand for informal loans as previously constrained households who borrowed \LLCLY switch to formal \BBBB. As a result, tightening the \CLC back to its previous level would increase the informal \ITR from \IRNafter\% to \IRNafterCL\%. The difference  is equal to a 24\% reduction in the pre-policy period informal interest rate. The result also indicates that the lowering of the \CLC by itself can account for \IRNdiffCLShareRnd\% of the total change in the \LLC interest rate.

The impacts of shifting the collateral constraint on aggregate participation rates, however, are limited to up to three percentage points of change.

\begin{figure}[!htbp]
\vspace*{-3mm}
\makebox[\textwidth][c]{
\fcolorbox{white}{white}{
\centering
\begin{minipage}{1.0\textwidth}
\begin{center}
\caption{\label{fig:pegecoll}Welfare Change in Northeast with $\gamma$ Variations between 1999-2001 and 2002-2009}
\includegraphics[scale=1.0]{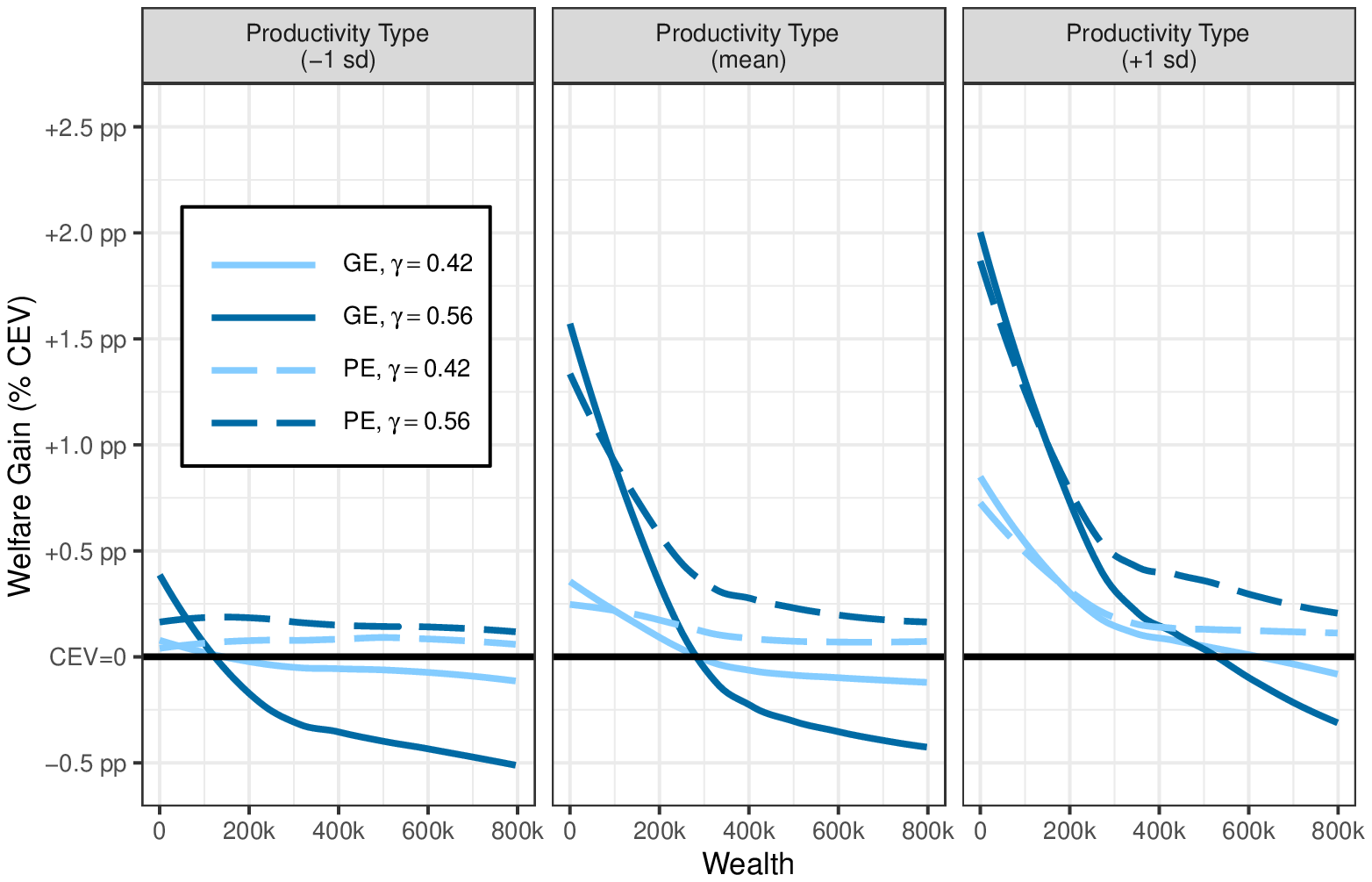}
\end{center}
\small \emph{Notes:} This figure shows the welfare impacts of the estimated collateral constraints relaxation component of Thaksin's policies. I compare \emph{steady state} partial equilibrium and general equilibrium welfare for households facing the collateral constraint parameter from the \emph{1999-2001} and the \emph{2002-2009} periods, holding all other parameters at \emph{2002-2009} levels. I present welfare changes in terms of \emph{consumption equivalent variation} (\textbf{CEV}) percentage point changes.
\end{minipage}
}}
\end{figure}

Given the significant impacts on the informal interest rate, the relaxation of collateral constraints is the key contributor to the redistributive welfare consequences shown in Figure \ref{fig:wel11}. To analyze this further, in Figure \ref{fig:pegecoll} I present steady state partial and general equilibrium welfare changes driven by the shift of the collateral constraints $\gamma$. The y-axis shows percentage point changes in CEV, the x-axis indicates wealth levels, and three subplots present results for lower (-1 standard deviation), mean, and higher (+1 standard deviation) productivity types, respectively. Dashed lines show partial equilibrium results, and solid lines show general equilibrium results. CEV changes from relaxing $\gamma$ from 0.28, the estimated value during the 1999-2001 period, to 0.56, the estimated value from the 2002-2009 period, are shown in dark blue. For comparison, CEV changes from relaxing $\gamma$ from 0.28 to 0.42 only are shown in lighter blue.

First, the partial equilibrium CEV changes are all positive. Under the partial equilibrium analysis here, as the collateral constraints relax, the value from choosing formal borrowing improves, but the values associated with alternative credit market choices do not worsen. The government finances a costly program, and the model predicts that all households would benefit, given their current and future expected needs for financing consumption and investments.

Second, the positive welfare impacts under partial equilibrium are heterogeneous across households and can be very small. For lower productivity households, CEV gains do not exceed 0.25\% across wealth levels despite doubling $\gamma$ from 0.28 to 0.56. For mean and higher productivity type households, those with lower wealth gain up to 1.9\% CEV. The gains decrease rapidly but stay positive as wealth increases. Similar to the results from Figure \ref{fig:wel11}, gains are limited by the prior availability of informal loans.

Third, allowing for equilibrium informal interest rate responses lead to greater gains for some households. Across productivity types, at lower wealth levels, the general equilibrium CEV gains are larger than the partial equilibrium gains. Households that are in need of loans benefit both from the relaxed formal borrowing constraint and the drop in the informal interest rate. However, the general equilibrium gains in excess of the partial equilibrium gains are very limited, not exceeding 0.2\% CEV in Figure \ref{fig:pegecoll}. This is because the improvements in formal borrowing conditions reduce households' reliance on informal borrowing and reduce the benefits of a lower informal borrowing interest rate.

Fourth, the reduction in informal equilibrium interest rates leads to welfare losses. For households with higher wealth, the small gains under partial equilibrium are outweighed by losses in the expected values from the informal savings alternatives due to lower informal interest rates. Losses are greater for lower productivity type households and can amount to up to 0.5\% CEV. Lower productivity and higher wealth households, as shown earlier in \FIGRSckb, invest comparatively more in lending and savings and less in risky capital investments. They benefit from lending to lower wealth but more productive households in their villages. Improvements in formal borrowing access, given the parameters here, reduce the number of high productivity local household-firms that have to rely on local financing and are detrimental to the welfare of lower productivity and higher wealth households in the villages.

Overall, Figure \ref{fig:pegecoll} shows two key sets of crossing points that determine the redistributive welfare consequences of relaxing collateral constraints when informal markets are considered. Each productivity type has a welfare gain crossing point along the wealth axis where the general equilibrium CEV gains fall below partial equilibrium gains. Additionally, each productivity type's general equilibrium CEV curve has an x-intercept marking the wealth level where CEV gains become losses. Both crossing points are increasing in productivity type, and the second comes after the first given strictly positive welfare gains under partial equilibrium.

Since collateral constraints are the key driver of equilibrium interest rate changes, the results from Figure \ref{fig:pegecoll} match up with the aggregate welfare consequences from Figure \ref{fig:wel11} when all Thaksin policies are jointly considered.
I should note that at alternative combinations of fixed costs and interest rates, the level of gains or losses from relaxing $\gamma$ might be magnified or dampened compared to the results in Figure \ref{fig:pegecoll}. Given the potential complementarity as well as substitutability between formal and informal options that the model allows for, under some parameter combinations, a shift in collateral constraints might not have significant impacts on the informal interest rates. The quantitative welfare consequences of microfinance policies have to be evaluated on a case-by-case basis.

\section{Conclusion\label{sec:concludeconclude}}
In recent decades, \NLC financial services have expanded significantly in developing countries. This paper evaluates the impacts of improving access to the \NLC credit market on rural households, taking into consideration the impacts of changing \NLC credit market conditions on the \LLC credit market.

I built a risky entrepreneur model assuming that villages are small open economies with respect to formal credit market options that have centrally set rates, but households can also borrow and save in an equilibrium local credit market with the locally determined rate. The model allows for evaluating the impacts of \NLC credit market expansions through interest rate subsidies, access fixed costs reductions, and collateral constraint relaxations. Policy evaluations take into consideration equilibrium responses of the informal credit market. In the empirical section of the paper, I explored detailed data on \NLC and \LLC credit market interactions from Thai villages. I connected the model with the Thai micro-data by estimating the model using Simulated Maximum Likelihood. The likelihood function incorporates choice probabilities given endogenous asset distributions determined by model parameters.

Using the estimated model, in the case of these Thai villages, I showed that
there are redistributive consequences of microfinance policies through general equilibrium effects on informal interest rates.
Without equilibrium considerations, all households can experience welfare gains from improvements in formal borrowing conditions. These gains are, however, limited because informal borrowing alternatives were already available.
In equilibrium, policies that improve \NLC \BBBB conditions can drive down \LLC \ITRs and could hurt less productive and higher wealth households who face diminished opportunities for \SSSS.
In particular, I find that in the context of the Thaksin policies, relaxing the \CLCs was the key driver of equilibrium interest rate changes and redistributive steady state welfare effects.

In solving and estimating this model, I developed solution and estimation algorithms that minimize the use of loops and iteration by parallel computing. The solution algorithm captures the impacts of variations in the four fixed costs, three interest rates, and one collateral constraint on household choices and equilibrium outcomes.
The global estimation algorithm chooses initial values for estimation by first exploring the estimation objective function through polynomial approximations. The solution and estimation algorithms can reduce estimation time by taking advantage of scalable cloud computing resources.

Currently, while many microfinance institutions offer both borrowing and savings services, there remain many development financial institutions, including many development banks, that are mainly oriented toward expanding borrowing access \autocite{luna-martinez_2017_2018}. My counterfactual simulations suggest that when households rely on local informal credit markets to meet differential needs for borrowing and saving, efforts that mainly improve formal access for borrowing could lead to more limited informal savings opportunities for households and their subsequent welfare losses.

\pagebreak

\begingroup
\setstretch{1.0}
\setlength\bibitemsep{3pt}
\printbibliography[title=References]
\endgroup
\pagebreak

\section*{Tables and Additional Figures}
{\setstretch{1.3}

\begin{figure}[H]
\vspace*{-3mm}
\makebox[\textwidth][c]{
\fcolorbox{white}{white}{
\centering
\begin{minipage}{0.87\textwidth}
\begin{table}[H]\centering
\def\sym#1{\ifmmode^{#1}\else\(^{#1}\)\fi}
\caption{\label{tab:BORROWERparticipate} Channels for Borrowing from 1999 to 2009}
\begin{tabular}{l*{2}{c}}
\toprule
                    &\multicolumn{1}{c}{Percentage of }&\multicolumn{1}{c}{Number of Years}\\
                    &     Households Who          &      with New Loans      \\
                    &    Have Borrowed &      (out of 11)      \\
\midrule
\rowgroup{\textbf{Northeast} Region (poorer)}&                   &                   \\
Million Baht Fund {\footnotesize\emph{formal}}   &    78.0          &     7.13          \\
BAAC {\footnotesize\emph{formal}}               &    61.6          &     5.58          \\
Village Coop {\footnotesize\emph{informal}}       &    35.9          &     3.00          \\
Friends and Neighbors {\footnotesize\emph{informal}}&    83.3          &     3.99          \\
Village Moneylenders {\footnotesize\emph{informal}}&    37.5          &     1.93          \\
\midrule
\rowgroup{\textbf{Central} Region (richer)}&                   &                   \\
Million Baht Fund {\footnotesize\emph{formal}}  &    74.5          &     6.73          \\
BAAC {\footnotesize\emph{formal}}               &    40.8          &     5.70          \\
Village Coop {\footnotesize\emph{informal}}        &    16.9          &     3.56          \\
Friends and Neighbors {\footnotesize\emph{informal}} &    48.5          &     2.27          \\
Village Moneylenders {\footnotesize\emph{informal}} &    12.9          &     1.50          \\
\bottomrule
\end{tabular}
\end{table}

 \small \textcolor{blue}{\emph{Notes:}} To effectively model the variety of borrowing choices shown in Table \ref{tab:BORROWERparticipate}, I group lenders into formal and informal categories. The Million Baht Fund (MBF) and the Bank for Agriculture and Agricultural Cooperatives (BAAC) are formal lenders with centrally set interest rates. Friends, neighbors and village moneylenders are local-informal lenders. Village Coop includes Production Cooperative Groups (PCG) and Village Agricultural Cooperatives, which are local semi-formal organizations that mostly intermediate credit among households within villages. This paper considers Village Coops as falling within the informal credit market with locally determined interest rates. See Section \ref{par:formalinformal} for detail.
\end{minipage}
}}
\end{figure}

\begin{figure}[H]
\vspace*{-3mm}
\makebox[\textwidth][c]{
\fcolorbox{white}{white}{
\centering
\begin{minipage}{0.87\textwidth}
\begin{table}[H]\centering
\def\sym#1{\ifmmode^{#1}\else\(^{#1}\)\fi}
\caption{\label{tab:LENDERparticipate} Channels for Saving from 1999 to 2009}
\begin{tabular}{l*{2}{c}}
\toprule
                    &\multicolumn{1}{c}{Percentage of }&\multicolumn{1}{c}{Number of Years}\\
                    &     Households Who        &      with New Deposits      \\
                    &     Have Saved        &      (out of 11)      \\
\midrule
\rowgroup{\textbf{Northeast} Region (poorer)}&                   &                   \\
Commercial {\footnotesize\emph{formal}}         &    37.4          &     4.56          \\
BAAC and GSB {\footnotesize\emph{formal}}       &    94.1          &     7.82          \\
Village Coop {\footnotesize\emph{informal}}       &    64.1          &     6.33          \\
Individual lending {\footnotesize\emph{informal}} &    66.6          &     2.57          \\
\midrule
\rowgroup{\textbf{Central} Region (richer)}&                   &                   \\
Commercial {\footnotesize\emph{formal}}         &    77.9          &     5.25          \\
BAAC and GSB {\footnotesize\emph{formal}}       &    85.6          &     6.41          \\
Village Coop {\footnotesize\emph{informal}}       &    75.2          &     7.54          \\
Individual lending {\footnotesize\emph{informal}} &    33.1          &     2.23          \\
\bottomrule
\end{tabular}
\end{table}
 \small \textcolor{blue}{\emph{Notes:}} To effectively model the variety of savings choices shown in Table \ref{tab:LENDERparticipate}, I group savings channels into formal and informal categories. Households could save formally at commercial banks, the BAAC, and the Government Savings Bank (GSB). Households can also save informally by lending to other individuals directly. When they save in village coops, I consider that households are saving informally. See Notes for Table \ref{tab:BORROWERparticipate} for what village coops are. See Section \ref{par:formalinformal} for detail.
\end{minipage}
}}
\end{figure}

\begin{figure}[H]
\vspace*{-3mm}
\makebox[\textwidth][c]{
\fcolorbox{white}{white}{
\centering
\begin{minipage}{1.0\textwidth}
\begin{table}[H]\centering
\def\sym#1{\ifmmode^{#1}\else\(^{#1}\)\fi}
\caption{\label{tab:NECEcreditSahres4m} Credit Category Participation Shares}
\begin{tabular}{l*{4}{c}}
\toprule
&\multicolumn{2}{c}{\textbf{Northeast} Region (poorer)}&\multicolumn{2}{c}{\textbf{Central} Region (richer)}\\
\cmidrule(l{12pt}r{12pt}){2-3} \cmidrule(l{12pt}r{12pt}){4-5}
                    & 1999-2001& 2002-2009& 1999-2001& 2002-2009\\
                    &    \footnotesize{percent}&    \footnotesize{percent}&    \footnotesize{percent}&    \footnotesize{percent}\\
\midrule
Formal and Informal     &     31.5&     40.3&      6.7&     18.1\\
Formal Only         &     21.1&     44.0&     51.5&     58.3\\
Informal Only       &     32.0&     11.1&     12.1&     9.9\\
Neither             &     15.5&     5.6&     29.7&     13.8\\
\bottomrule
\end{tabular}
\end{table}
 \begin{table}[H]\centering
\def\sym#1{\ifmmode^{#1}\else\(^{#1}\)\fi}
\caption{\label{tab:NECEcreditShares7m} Credit Category Shares Across Seven Participation Types}
\begin{tabular}{l*{4}{c}}
\toprule
&\multicolumn{2}{c}{\textbf{Northeast} Region (poorer)}&\multicolumn{2}{c}{\textbf{Central} Region (richer)}\\
\cmidrule(l{12pt}r{12pt}){2-3} \cmidrule(l{12pt}r{12pt}){4-5}
& 1999-2001& 2002-2009& 1999-2001& 2002-2009\\
&    \footnotesize{percent}&    \footnotesize{percent}&    \footnotesize{percent}&    \footnotesize{percent}\\
\midrule
Formal Borrowing       &       \DataNEoneFBprob    & \DataNEtwoFBprob   & \DataCEoneFBprob   & \DataCEtwoFBprob   \\
Formal Saving          &       \DataNEoneFSprob    & \DataNEtwoFSprob   & \DataCEoneFSprob   & \DataCEtwoFSprob   \\
Informal Borrowing     &       \DataNEoneIBprob    & \DataNEtwoIBprob   & \DataCEoneIBprob   & \DataCEtwoIBprob   \\
Informal Saving        &       \DataNEoneISprob    & \DataNEtwoISprob   & \DataCEoneISprob   & \DataCEtwoISprob   \\
Formal+Informal Borrowing&     \DataNEoneFBIBprob & \DataNEtwoFBIBprob & \DataCEoneFBIBprob & \DataCEtwoFBIBprob \\
Formal Borrow+Informal Saving& \DataNEoneFBISprob & \DataNEtwoFBISprob & \DataCEoneFBISprob & \DataCEtwoFBISprob \\
No Credit Transactions&        \DataNEoneNONEprob & \DataNEtwoNONEprob & \DataCEoneNONEprob & \DataCEtwoNONEprob \\
\bottomrule
\end{tabular}
\end{table}
 \small \textcolor{blue}{\emph{Notes:}} Households' credit market choices within a calendar year can be categorized as falling within  one of the seven credit participation categories listed in Table \ref{tab:NECEcreditShares7m}. This table shows that after 2001, due to policies that promoted formal borrowing such as the Million Baht Fund Program (see Section \ref{sec:Data-and-Background}), the proportion of households participating in the formal credit market increased significantly, especially in the Northeast region. See Section \ref{par:Credit-Choice-Categories} for detail.
\end{minipage}
}}
\end{figure}

\begin{figure}[H]
\vspace*{-3mm}
\makebox[\textwidth][c]{
\fcolorbox{white}{white}{
\centering
\begin{minipage}{1.0\textwidth}
\caption{\label{fig:Interest1}Village/Year Real Interest Rates for Different Lenders}
\vspace*{-1mm}
\includegraphics[scale=1]{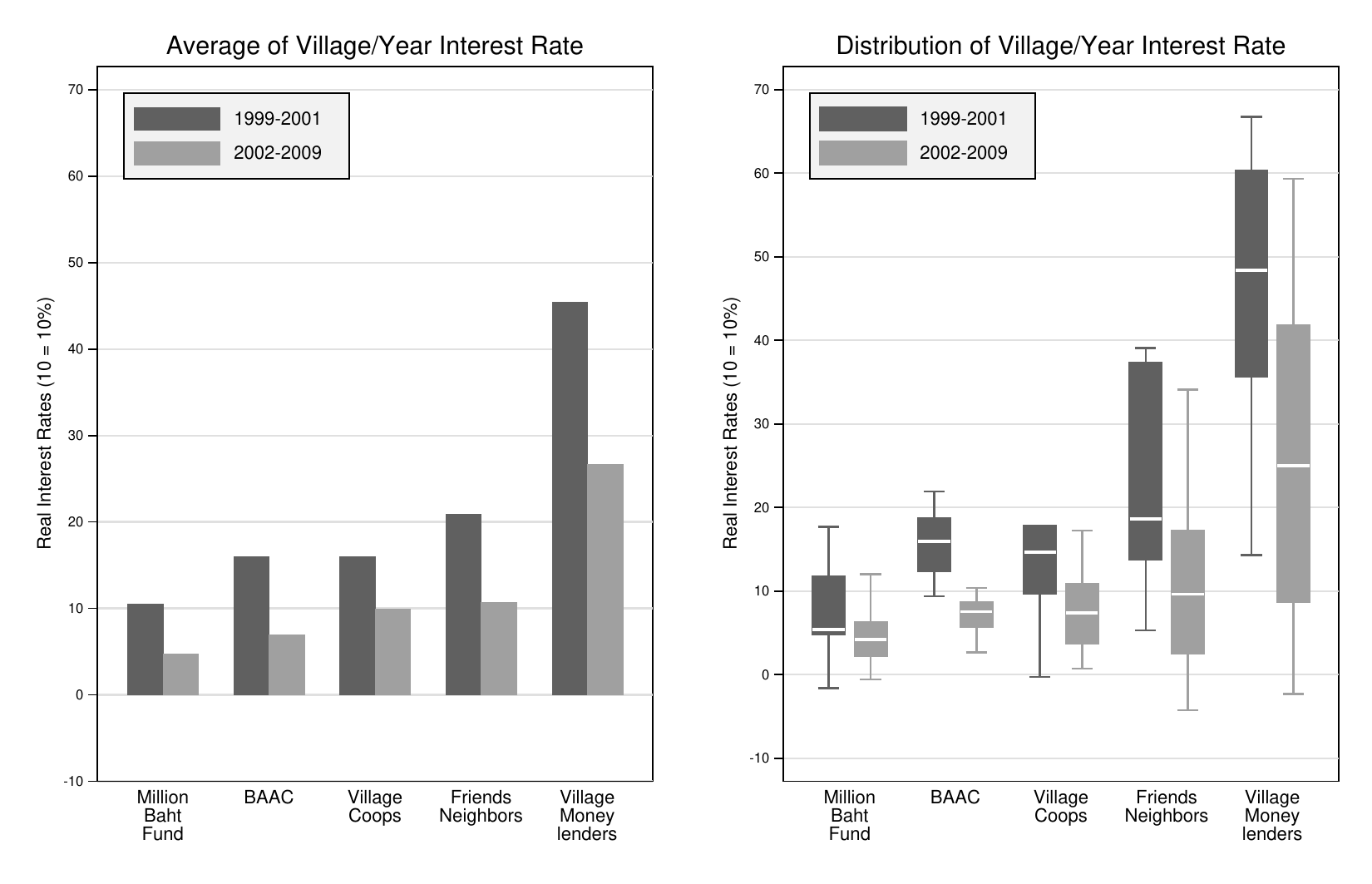}
\vspace*{-1mm}
\small \textcolor{blue}{\emph{Notes:}} Village/Year Interest rates are average interest rates for each lender type for each village in each calendar year. See Section \ref{par:Interest-Rate} for detail. \\
\begin{table}[H]\centering
\def\sym#1{\ifmmode^{#1}\else\(^{#1}\)\fi}
\caption{\label{tab:intRatereg3yr2m2}Annual Average Real Interest Rate (percent)}
\begin{tabular}{l*{2}{c}}
\toprule
                    &\multicolumn{1}{c}{1999-2001}&\multicolumn{1}{c}{2002-2009}\\
\midrule
\rowgroup{\textbf{Northeast} Region (poorer)}  &                   &                   \\
\emph{Informal} Interest   &    27.9          &    13.9          \\
\emph{Formal} Borrowing Interest&    14.8          &     6.1          \\
\emph{Formal} Saving Interest&     3.3          &     1.1          \\
\midrule
\rowgroup{\textbf{Central} Region (richer)}   &                   &                   \\
\emph{Informal} Interest   &    17.6          &     9.4          \\
\emph{Formal} Borrowing Interest&    13.0          &     5.4          \\
\emph{Formal} Saving Interest&     3.2          &     1.0          \\
\bottomrule
\end{tabular}
\end{table}
 \small \textcolor{blue}{\emph{Notes:}} Table \ref{tab:intRatereg3yr2m2} and Figure \ref{fig:Interest1} show that formal and informal interest rates decreased after 2001.
\end{minipage}
}}
\end{figure}

 }
{\setstretch{1.15}
\renewcommand{\FBcZY}{0.033} \renewcommand{\FBcNY}{0.044} \renewcommand{\FBnZY}{0.046} \renewcommand{\FBnNY}{0.090} \renewcommand{\FScZY}{0.001} \renewcommand{\FScNY}{0.011} \renewcommand{\FSnZY}{0.071} \renewcommand{\FSnNY}{0.083} \renewcommand{\IBcZY}{0.124} \renewcommand{\IBcNY}{0.124} \renewcommand{\IBnZY}{0.053} \renewcommand{\IBnNY}{0.053} \renewcommand{\ILcZY}{0.156} \renewcommand{\ILcNY}{0.151} \renewcommand{\ILnZY}{0.186} \renewcommand{\ILnNY}{0.134}  \renewcommand{\FBcZsY}{0.0053} \renewcommand{\FBcNsY}{0.044} \renewcommand{\FBnZsY}{0} \renewcommand{\FBnNsY}{0.008} \renewcommand{\FScZsY}{0.00025} \renewcommand{\FScNsY}{0.037} \renewcommand{\FSnZsY}{0.0054} \renewcommand{\FSnNsY}{0.0045} \renewcommand{\IBcZsY}{0.098} \renewcommand{\IBcNsY}{0.098} \renewcommand{\IBnZsY}{0.0044} \renewcommand{\IBnNsY}{0.0044} \renewcommand{\ILcZsY}{0.042} \renewcommand{\ILcNsY}{0.056} \renewcommand{\ILnZsY}{0.011} \renewcommand{\ILnNsY}{0.034}  \renewcommand{\FBnZsY}{0.02}  \renewcommand{\FBcZs}{0.0079} \renewcommand{\FBcNs}{0.015} \renewcommand{\FBnZs}{0} \renewcommand{\FBnNs}{0.095} \renewcommand{\FScZs}{0.0002} \renewcommand{\FScNs}{0.0049} \renewcommand{\FSnZs}{0.029} \renewcommand{\FSnNs}{0.00012} \renewcommand{\IBcZs}{0.021} \renewcommand{\IBcNs}{0.021} \renewcommand{\IBnZs}{0.039} \renewcommand{\IBnNs}{0.039} \renewcommand{\ILcZs}{0.021} \renewcommand{\ILcNs}{0.052} \renewcommand{\ILnZs}{7.2e-05} \renewcommand{\ILnNs}{0.29} \renewcommand{\AMUcZs}{0.21} \renewcommand{\AMUcNs}{0.22} \renewcommand{\AMUnZs}{7.2e-05} \renewcommand{\AMUnNs}{0.045} \renewcommand{\ASDcs}{0.02} \renewcommand{\ASDns}{0} \renewcommand{\KDEPRECIATIONcs}{0.054} \renewcommand{\KDEPRECIATIONns}{0.0048} \renewcommand{\ALPHAKcs}{0.058} \renewcommand{\ALPHAKns}{0.01} \renewcommand{\BETAcs}{0} \renewcommand{\BETAns}{0} \renewcommand{\KAPPAcZs}{0.07} \renewcommand{\KAPPAcNs}{0.14} \renewcommand{\KAPPAnZs}{0.016} \renewcommand{\KAPPAnNs}{0.062} \renewcommand{\LOGITSDSCALEcs}{0.39} \renewcommand{\LOGITSDSCALEns}{0.014} \renewcommand{\RHOcs}{0.11} \renewcommand{\RHOns}{0.0052} \renewcommand{\STDEPSEcs}{0.45} \renewcommand{\STDEPSEns}{0.52} \renewcommand{\STDEPScs}{0.24} \renewcommand{\STDEPSns}{0.052}  \renewcommand{\FBcZs}{0.019} \renewcommand{\FBcNs}{0.03} \renewcommand{\FBnZs}{4.3e-18} \renewcommand{\FBnNs}{0.35} \renewcommand{\FScZs}{8.1e-05} \renewcommand{\FScNs}{0.0083} \renewcommand{\FSnZs}{0.061} \renewcommand{\FSnNs}{0.036} \renewcommand{\IBcZs}{0.064} \renewcommand{\IBcNs}{0.064} \renewcommand{\IBnZs}{0.14} \renewcommand{\IBnNs}{0.14} \renewcommand{\ILcZs}{0.058} \renewcommand{\ILcNs}{0.073} \renewcommand{\ILnZs}{0.01} \renewcommand{\ILnNs}{0.51} \renewcommand{\AMUnZs}{0.12} \renewcommand{\AMUnNs}{0.17}  \renewcommand{\FBcZ}{0.050} \renewcommand{\FBcN}{0.087} \renewcommand{\FBnZ}{0.403} \renewcommand{\FBnN}{1.350} \renewcommand{\FScZ}{0.000} \renewcommand{\FScN}{0.024} \renewcommand{\FSnZ}{1.303} \renewcommand{\FSnN}{1.305} \renewcommand{\IBcZ}{0.363} \renewcommand{\IBcN}{0.363} \renewcommand{\IBnZ}{0.491} \renewcommand{\IBnN}{0.491} \renewcommand{\ILcZ}{0.518} \renewcommand{\ILcN}{0.498} \renewcommand{\ILnZ}{3.041} \renewcommand{\ILnN}{2.319} \renewcommand{\AMUcZ}{0.58} \renewcommand{\AMUcN}{0.34} \renewcommand{\AMUnZ}{0.28} \renewcommand{\AMUnN}{0.18} \renewcommand{\ALPHAKn}{0.32}  \renewcommand{\FBcZ}{0.008} \renewcommand{\FBcZs}{0.0072} \renewcommand{\FBcN}{0.037} \renewcommand{\FBcNs}{0.015} \renewcommand{\FBnZ}{0.000} \renewcommand{\FBnZs}{0} \renewcommand{\FBnN}{0.589} \renewcommand{\FBnNs}{0.073} \renewcommand{\FScZ}{0.000} \renewcommand{\FScZs}{0.00024} \renewcommand{\FScN}{0.019} \renewcommand{\FScNs}{0.0042} \renewcommand{\FSnZ}{1.486} \renewcommand{\FSnZs}{0.023} \renewcommand{\FSnN}{1.500} \renewcommand{\FSnNs}{0.00011} \renewcommand{\IBcZ}{0.145} \renewcommand{\IBcZs}{0.014} \renewcommand{\IBcN}{0.145} \renewcommand{\IBcNs}{0.014} \renewcommand{\IBnZ}{0.435} \renewcommand{\IBnZs}{0.039} \renewcommand{\IBnN}{0.435} \renewcommand{\IBnNs}{0.039} \renewcommand{\ILcZ}{0.573} \renewcommand{\ILcZs}{0.017} \renewcommand{\ILcN}{0.508} \renewcommand{\ILcNs}{0.051} \renewcommand{\ILnZ}{3.500} \renewcommand{\ILnZs}{0} \renewcommand{\ILnN}{2.948} \renewcommand{\ILnNs}{0.27} \renewcommand{\KAPPAcZ}{0.13} \renewcommand{\KAPPAcZs}{0.052} \renewcommand{\KAPPAcN}{0.3} \renewcommand{\KAPPAcNs}{0.15} \renewcommand{\KAPPAnZ}{0.56} \renewcommand{\KAPPAnZs}{0.015} \renewcommand{\KAPPAnN}{0.13} \renewcommand{\KAPPAnNs}{0.047}  \renewcommand{\FBcZ}{0.139} \renewcommand{\FBcZs}{0.15} \renewcommand{\FBcN}{0.183} \renewcommand{\FBcNs}{0.16} \renewcommand{\FBnZ}{1.287} \renewcommand{\FBnZs}{1.2} \renewcommand{\FBnN}{1.482} \renewcommand{\FBnNs}{0.91} \renewcommand{\FScZ}{0.012} \renewcommand{\FScZs}{0.045} \renewcommand{\FScN}{0.024} \renewcommand{\FScNs}{0.037} \renewcommand{\FSnZ}{1.126} \renewcommand{\FSnZs}{0.48} \renewcommand{\FSnN}{1.232} \renewcommand{\FSnNs}{0.48} \renewcommand{\IBcZ}{0.393} \renewcommand{\IBcZs}{0.16} \renewcommand{\IBcN}{0.393} \renewcommand{\IBcNs}{0.16} \renewcommand{\IBnZ}{0.990} \renewcommand{\IBnZs}{1} \renewcommand{\IBnN}{0.990} \renewcommand{\IBnNs}{1} \renewcommand{\ILcZ}{0.490} \renewcommand{\ILcZs}{0.11} \renewcommand{\ILcN}{0.469} \renewcommand{\ILcNs}{0.12} \renewcommand{\ILnZ}{2.489} \renewcommand{\ILnZs}{1.2} \renewcommand{\ILnN}{2.127} \renewcommand{\ILnNs}{0.94} \renewcommand{\ASDn}{0.12} \renewcommand{\ASDns}{0.1} \renewcommand{\ALPHAKc}{0.38} \renewcommand{\ALPHAKcs}{0.18} \renewcommand{\BETAc}{0.92} \renewcommand{\BETAcs}{0.04} \renewcommand{\BETAn}{0.91} \renewcommand{\BETAns}{0.038} \renewcommand{\KAPPAcZ}{0.41} \renewcommand{\KAPPAcZs}{0.23} \renewcommand{\KAPPAcN}{0.36} \renewcommand{\KAPPAcNs}{0.21} \renewcommand{\LOGITSDSCALEc}{1.7} \renewcommand{\LOGITSDSCALEcs}{0.33} \renewcommand{\LOGITSDSCALEn}{1.5} \renewcommand{\LOGITSDSCALEns}{0.45} \renewcommand{\RHOc}{1.3} \renewcommand{\RHOcs}{0.15} \renewcommand{\RHOn}{1.2} \renewcommand{\RHOns}{0.17}  \renewcommand{\FBcZ}{0.025} \renewcommand{\FBcN}{0.064} \renewcommand{\FBnZ}{-0.000} \renewcommand{\FBnN}{0.935} \renewcommand{\FScZ}{0.000} \renewcommand{\FScN}{0.022} \renewcommand{\FSnZ}{1.473} \renewcommand{\FSnN}{1.497} \renewcommand{\IBcZ}{0.212} \renewcommand{\IBcN}{0.212} \renewcommand{\IBnZ}{0.432} \renewcommand{\IBnN}{0.432} \renewcommand{\ILcZ}{0.527} \renewcommand{\ILcN}{0.481} \renewcommand{\ILnZ}{3.499} \renewcommand{\ILnN}{2.644} \renewcommand{\ASDc}{0.13} \renewcommand{\KDEPRECIATIONc}{0.071} \renewcommand{\KDEPRECIATIONn}{0.056} \renewcommand{\BETAn}{0.88} \renewcommand{\STDEPSc}{0.55} \renewcommand{\STDEPSn}{0.37}  \renewcommand{\FBcZ}{0.017} \renewcommand{\FBcN}{0.050} \renewcommand{\FBnZ}{-0.000} \renewcommand{\FBnN}{0.789} \renewcommand{\FScZ}{0.000} \renewcommand{\FScN}{0.019} \renewcommand{\FSnZ}{1.480} \renewcommand{\FSnN}{1.500} \renewcommand{\IBcZ}{0.183} \renewcommand{\IBcN}{0.183} \renewcommand{\IBnZ}{0.439} \renewcommand{\IBnN}{0.439} \renewcommand{\ILcZ}{0.549} \renewcommand{\ILcN}{0.492} \renewcommand{\ILnZ}{3.500} \renewcommand{\ILnN}{2.764} \renewcommand{\KAPPAnZ}{0.56} \renewcommand{\KAPPAnN}{0.28}

\renewcommand{\ALPHAKc}{0.34} \renewcommand{\ALPHAKcs}{0.03} \renewcommand{\ALPHAKn}{0.15} \renewcommand{\ALPHAKns}{0.02}

\renewcommand{\AMUcZ}{7.94} \renewcommand{\AMUcZs}{0.44} \renewcommand{\AMUcN}{7.87} \renewcommand{\AMUcNs}{0.4} \renewcommand{\AMUnZ}{8.96} \renewcommand{\AMUnZs}{0.46} \renewcommand{\AMUnN}{8.75} \renewcommand{\AMUnNs}{0.42} \renewcommand{\ASDc}{1.18} \renewcommand{\ASDcs}{0.025} \renewcommand{\ASDn}{0.84} \renewcommand{\ASDns}{0.025}

\renewcommand{\STDEPSc}{0.97} \renewcommand{\STDEPScs}{0.041} \renewcommand{\STDEPSn}{0.87} \renewcommand{\STDEPSns}{0.019}
\FIGESTIMATESFY{Parameter Estimates}

\renewcommand{\IBFBcZ}{0.240} \renewcommand{\IBFBcZa}{0.130} \renewcommand{\IBFBcZe}{0.341} \renewcommand{\IBFBcN}{0.183} \renewcommand{\IBFBcNa}{0.083} \renewcommand{\IBFBcNe}{0.286}  \renewcommand{\IBFBnZ}{0.511} \renewcommand{\IBFBnZa}{0.372} \renewcommand{\IBFBnZe}{0.707} \renewcommand{\IBFBnN}{0.474} \renewcommand{\IBFBnNa}{0.289} \renewcommand{\IBFBnNe}{0.651}  \renewcommand{\IBBFnZ}{0.274} \renewcommand{\IBBFnZe}{0.338} \renewcommand{\IBBFnZa}{0.231} \renewcommand{\IBBFnN}{0.345} \renewcommand{\IBBFnNe}{0.412} \renewcommand{\IBBFnNa}{0.171}  \renewcommand{\IBBFcZ}{0.100} \renewcommand{\IBBFcZe}{0.277} \renewcommand{\IBBFcZa}{0.040} \renewcommand{\IBBFcN}{0.107} \renewcommand{\IBBFcNe}{0.322} \renewcommand{\IBBFcNa}{0.055}  \renewcommand{\KKnZ}{84.34} \renewcommand{\KKnZa}{70.95} \renewcommand{\KKnZe}{91.24} \renewcommand{\KKnN}{76.10} \renewcommand{\KKnNa}{66.50} \renewcommand{\KKnNe}{85.93}  \renewcommand{\KKcZ}{97.34} \renewcommand{\KKcZa}{87.97} \renewcommand{\KKcZe}{120.41} \renewcommand{\KKcN}{77.71} \renewcommand{\KKcNa}{69.52} \renewcommand{\KKcNe}{99.56}  \renewcommand{\CCcZ}{148.49} \renewcommand{\CCcZa}{128.54} \renewcommand{\CCcZe}{173.04} \renewcommand{\CCcN}{119.78} \renewcommand{\CCcNa}{103.72} \renewcommand{\CCcNe}{138.62} \renewcommand{\YYcZ}{198.16} \renewcommand{\YYcZa}{163.24} \renewcommand{\YYcZe}{252.16} \renewcommand{\YYcN}{181.20} \renewcommand{\YYcNa}{138.12} \renewcommand{\YYcNe}{215.68}  \renewcommand{\CCnZ}{54.09} \renewcommand{\CCnZs}{28.34} \renewcommand{\CCnZa}{49.24} \renewcommand{\CCnZe}{60.78} \renewcommand{\CCnN}{48.98} \renewcommand{\CCnNs}{22.31} \renewcommand{\CCnNa}{37.80} \renewcommand{\CCnNe}{57.92}  \renewcommand{\YYnZ}{64.97} \renewcommand{\YYnZa}{38.13} \renewcommand{\YYnZe}{88.54} \renewcommand{\YYnN}{52.15} \renewcommand{\YYnNa}{30.49} \renewcommand{\YYnNe}{73.97}  \renewcommand{\PBBcZ}{0.11} \renewcommand{\PBBcZs}{0.002} \renewcommand{\PBBcZa}{0.104} \renewcommand{\PBBcZb}{0.105} \renewcommand{\PBBcZc}{0.106} \renewcommand{\PBBcZd}{0.107} \renewcommand{\PBBcZe}{0.109} \renewcommand{\PBBcZz}{0.11} \renewcommand{\PBBcN}{0.06} \renewcommand{\PBBcNs}{0.006} \renewcommand{\PBBcNa}{0.057} \renewcommand{\PBBcNb}{0.060} \renewcommand{\PBBcNc}{0.062} \renewcommand{\PBBcNd}{0.070} \renewcommand{\PBBcNe}{0.070} \renewcommand{\PBBcNz}{0.06} \renewcommand{\PBBnZ}{0.28} \renewcommand{\PBBnZs}{0.001} \renewcommand{\PBBnZa}{0.280} \renewcommand{\PBBnZb}{0.281} \renewcommand{\PBBnZc}{0.281} \renewcommand{\PBBnZd}{0.282} \renewcommand{\PBBnZe}{0.282} \renewcommand{\PBBnZz}{0.28} \renewcommand{\PBBnN}{0.22} \renewcommand{\PBBnNs}{0.002} \renewcommand{\PBBnNa}{0.215} \renewcommand{\PBBnNb}{0.215} \renewcommand{\PBBnNc}{0.215} \renewcommand{\PBBnNd}{0.218} \renewcommand{\PBBnNe}{0.218} \renewcommand{\PBBnNz}{0.22} \renewcommand{\PBScZ}{0.042} \renewcommand{\PBScZs}{0.005} \renewcommand{\PBScZa}{0.037} \renewcommand{\PBScZb}{0.038} \renewcommand{\PBScZc}{0.040} \renewcommand{\PBScZd}{0.046} \renewcommand{\PBScZe}{0.050} \renewcommand{\PBScZz}{0.042} \renewcommand{\PBScN}{0.021} \renewcommand{\PBScNs}{0.006} \renewcommand{\PBScNa}{0.015} \renewcommand{\PBScNb}{0.017} \renewcommand{\PBScNc}{0.020} \renewcommand{\PBScNd}{0.026} \renewcommand{\PBScNe}{0.028} \renewcommand{\PBScNz}{0.021} \renewcommand{\PBSnZ}{0.11} \renewcommand{\PBSnZs}{0.001} \renewcommand{\PBSnZa}{0.110} \renewcommand{\PBSnZb}{0.111} \renewcommand{\PBSnZc}{0.111} \renewcommand{\PBSnZd}{0.112} \renewcommand{\PBSnZe}{0.112} \renewcommand{\PBSnZz}{0.11} \renewcommand{\PBSnN}{0.08} \renewcommand{\PBSnNs}{0.001} \renewcommand{\PBSnNa}{0.083} \renewcommand{\PBSnNb}{0.084} \renewcommand{\PBSnNc}{0.085} \renewcommand{\PBSnNd}{0.085} \renewcommand{\PBSnNe}{0.086} \renewcommand{\PBSnNz}{0.08} \renewcommand{\PFBcZ}{0.25} \renewcommand{\PFBcZs}{0.008} \renewcommand{\PFBcZa}{0.234} \renewcommand{\PFBcZb}{0.242} \renewcommand{\PFBcZc}{0.248} \renewcommand{\PFBcZd}{0.252} \renewcommand{\PFBcZe}{0.254} \renewcommand{\PFBcZz}{0.25} \renewcommand{\PFBcN}{0.21} \renewcommand{\PFBcNs}{0.003} \renewcommand{\PFBcNa}{0.208} \renewcommand{\PFBcNb}{0.210} \renewcommand{\PFBcNc}{0.210} \renewcommand{\PFBcNd}{0.212} \renewcommand{\PFBcNe}{0.214} \renewcommand{\PFBcNz}{0.21} \renewcommand{\PFBnZ}{0.25} \renewcommand{\PFBnZs}{0.003} \renewcommand{\PFBnZa}{0.247} \renewcommand{\PFBnZb}{0.250} \renewcommand{\PFBnZc}{0.251} \renewcommand{\PFBnZd}{0.254} \renewcommand{\PFBnZe}{0.254} \renewcommand{\PFBnZz}{0.25} \renewcommand{\PFBnN}{0.17} \renewcommand{\PFBnNs}{0.002} \renewcommand{\PFBnNa}{0.163} \renewcommand{\PFBnNb}{0.165} \renewcommand{\PFBnNc}{0.166} \renewcommand{\PFBnNd}{0.167} \renewcommand{\PFBnNe}{0.168} \renewcommand{\PFBnNz}{0.17} \renewcommand{\PFScZ}{0.33} \renewcommand{\PFScZs}{0.003} \renewcommand{\PFScZa}{0.331} \renewcommand{\PFScZb}{0.333} \renewcommand{\PFScZc}{0.334} \renewcommand{\PFScZd}{0.336} \renewcommand{\PFScZe}{0.338} \renewcommand{\PFScZz}{0.33} \renewcommand{\PFScN}{0.30} \renewcommand{\PFScNs}{0.002} \renewcommand{\PFScNa}{0.297} \renewcommand{\PFScNb}{0.299} \renewcommand{\PFScNc}{0.299} \renewcommand{\PFScNd}{0.301} \renewcommand{\PFScNe}{0.303} \renewcommand{\PFScNz}{0.30} \renewcommand{\PFSnZ}{0.12} \renewcommand{\PFSnZs}{0.001} \renewcommand{\PFSnZa}{0.118} \renewcommand{\PFSnZb}{0.118} \renewcommand{\PFSnZc}{0.119} \renewcommand{\PFSnZd}{0.119} \renewcommand{\PFSnZe}{0.120} \renewcommand{\PFSnZz}{0.12} \renewcommand{\PFSnN}{0.11} \renewcommand{\PFSnNs}{0.003} \renewcommand{\PFSnNa}{0.101} \renewcommand{\PFSnNb}{0.104} \renewcommand{\PFSnNc}{0.105} \renewcommand{\PFSnNd}{0.106} \renewcommand{\PFSnNe}{0.108} \renewcommand{\PFSnNz}{0.11} \renewcommand{\PIBcZ}{0.08} \renewcommand{\PIBcZs}{0.003} \renewcommand{\PIBcZa}{0.081} \renewcommand{\PIBcZb}{0.081} \renewcommand{\PIBcZc}{0.082} \renewcommand{\PIBcZd}{0.084} \renewcommand{\PIBcZe}{0.087} \renewcommand{\PIBcZz}{0.08} \renewcommand{\PIBcN}{0.08} \renewcommand{\PIBcNs}{0.002} \renewcommand{\PIBcNa}{0.081} \renewcommand{\PIBcNb}{0.084} \renewcommand{\PIBcNc}{0.084} \renewcommand{\PIBcNd}{0.086} \renewcommand{\PIBcNe}{0.087} \renewcommand{\PIBcNz}{0.08} \renewcommand{\PIBnZ}{0.14} \renewcommand{\PIBnZs}{0.002} \renewcommand{\PIBnZa}{0.143} \renewcommand{\PIBnZb}{0.143} \renewcommand{\PIBnZc}{0.146} \renewcommand{\PIBnZd}{0.146} \renewcommand{\PIBnZe}{0.146} \renewcommand{\PIBnZz}{0.14} \renewcommand{\PIBnN}{0.24} \renewcommand{\PIBnNs}{0.001} \renewcommand{\PIBnNa}{0.238} \renewcommand{\PIBnNb}{0.239} \renewcommand{\PIBnNc}{0.240} \renewcommand{\PIBnNd}{0.241} \renewcommand{\PIBnNe}{0.241} \renewcommand{\PIBnNz}{0.24} \renewcommand{\PIScZ}{0.035} \renewcommand{\PIScZs}{0.005} \renewcommand{\PIScZa}{0.029} \renewcommand{\PIScZb}{0.032} \renewcommand{\PIScZc}{0.034} \renewcommand{\PIScZd}{0.038} \renewcommand{\PIScZe}{0.041} \renewcommand{\PIScZz}{0.035} \renewcommand{\PIScN}{0.028} \renewcommand{\PIScNs}{0.008} \renewcommand{\PIScNa}{0.020} \renewcommand{\PIScNb}{0.024} \renewcommand{\PIScNc}{0.025} \renewcommand{\PIScNd}{0.032} \renewcommand{\PIScNe}{0.040} \renewcommand{\PIScNz}{0.028} \renewcommand{\PISnZ}{0.07} \renewcommand{\PISnZs}{0.000} \renewcommand{\PISnZa}{0.067} \renewcommand{\PISnZb}{0.068} \renewcommand{\PISnZc}{0.068} \renewcommand{\PISnZd}{0.068} \renewcommand{\PISnZe}{0.068} \renewcommand{\PISnZz}{0.07} \renewcommand{\PISnN}{0.08} \renewcommand{\PISnNs}{0.003} \renewcommand{\PISnNa}{0.074} \renewcommand{\PISnNb}{0.076} \renewcommand{\PISnNc}{0.079} \renewcommand{\PISnNd}{0.080} \renewcommand{\PISnNe}{0.082} \renewcommand{\PISnNz}{0.08}  \renewcommand{\PBBcZa}{0.057} \renewcommand{\PBBcZe}{0.104} \renewcommand{\PBBcNa}{0.037} \renewcommand{\PBBcNe}{0.075} \renewcommand{\PBBnZa}{0.192} \renewcommand{\PBBnZe}{0.291} \renewcommand{\PBBnNa}{0.180} \renewcommand{\PBBnNe}{0.227} \renewcommand{\PBScZa}{0.025} \renewcommand{\PBScZe}{0.055} \renewcommand{\PBScNa}{0.006} \renewcommand{\PBScNe}{0.043} \renewcommand{\PBSnZa}{0.078} \renewcommand{\PBSnZe}{0.144} \renewcommand{\PBSnNa}{0.074} \renewcommand{\PBSnNe}{0.127} \renewcommand{\PFBcZa}{0.225} \renewcommand{\PFBcZe}{0.257} \renewcommand{\PFBcNa}{0.198} \renewcommand{\PFBcNe}{0.223} \renewcommand{\PFBnZa}{0.207} \renewcommand{\PFBnZe}{0.274} \renewcommand{\PFBnNa}{0.128} \renewcommand{\PFBnNe}{0.179} \renewcommand{\PFScZa}{0.297} \renewcommand{\PFScZe}{0.335} \renewcommand{\PFScNa}{0.289} \renewcommand{\PFScNe}{0.309} \renewcommand{\PFSnZa}{0.102} \renewcommand{\PFSnZe}{0.150} \renewcommand{\PFSnNa}{0.062} \renewcommand{\PFSnNe}{0.122} \renewcommand{\PIBcZa}{0.060} \renewcommand{\PIBcZe}{0.096} \renewcommand{\PIBcNa}{0.065} \renewcommand{\PIBcNe}{0.098} \renewcommand{\PIBnZa}{0.087} \renewcommand{\PIBnZe}{0.170} \renewcommand{\PIBnNa}{0.187} \renewcommand{\PIBnNe}{0.252} \renewcommand{\PIScZa}{0.021} \renewcommand{\PIScZe}{0.050} \renewcommand{\PIScNa}{0.014} \renewcommand{\PIScNe}{0.044} \renewcommand{\PISnZa}{0.017} \renewcommand{\PISnZe}{0.090} \renewcommand{\PISnNa}{0.049} \renewcommand{\PISnNe}{0.102}

\renewcommand{\PIBcZz}{0.061} \renewcommand{\PIBcNz}{0.092} \renewcommand{\PIBnZz}{0.062} \renewcommand{\PIBnNz}{0.25} \renewcommand{\PIScZz}{0.038} \renewcommand{\PIScNz}{0.03} \renewcommand{\PISnZz}{0.049} \renewcommand{\PISnNz}{0.07}

\renewcommand{\PFBcZz}{0.24} \renewcommand{\PFBcNz}{0.22} \renewcommand{\PFBnZz}{0.29} \renewcommand{\PFBnNz}{0.15} \renewcommand{\PFScZz}{0.34} \renewcommand{\PFScNz}{0.30} \renewcommand{\PFSnZz}{0.14} \renewcommand{\PFSnNz}{0.062}

\renewcommand{\PBBcZz}{0.14} \renewcommand{\PBBcNz}{0.05} \renewcommand{\PBBnZz}{0.33} \renewcommand{\PBBnNz}{0.24} \renewcommand{\PBScZz}{0.038} \renewcommand{\PBScNz}{0.017} \renewcommand{\PBSnZz}{0.078} \renewcommand{\PBSnNz}{0.074}

\renewcommand{\YYcNz}{115.9} \renewcommand{\YYcZz}{134.5} \renewcommand{\YYnNz}{31.8} \renewcommand{\YYnZz}{42.2}

\renewcommand{\KKcNz}{32.2} \renewcommand{\KKcZz}{44.7} \renewcommand{\KKnNz}{23.6} \renewcommand{\KKnZz}{39.0}

\renewcommand{\CCcNz}{64.5} \renewcommand{\CCcZz}{74.2} \renewcommand{\CCnNz}{39.1} \renewcommand{\CCnZz}{42.8}

\renewcommand{\YYcZz}{206.3} \renewcommand{\YYcNz}{185.5} \renewcommand{\YYnZz}{75.1} \renewcommand{\YYnNz}{49.0}

\renewcommand{\KKcZz}{97.1} \renewcommand{\KKcNz}{79.5} \renewcommand{\KKnZz}{82.0} \renewcommand{\KKnNz}{76.5}

\renewcommand{\CCcZz}{90.2} \renewcommand{\CCcNz}{79.8} \renewcommand{\CCnZz}{55.5} \renewcommand{\CCnNz}{47.1}

\renewcommand{\IBFBcZz}{0.239} \renewcommand{\IBFBcNz}{0.192} \renewcommand{\IBFBnZz}{0.516} \renewcommand{\IBFBnNz}{0.440}

\renewcommand{\IBBFcZz}{0.089} \renewcommand{\IBBFcNz}{0.109} \renewcommand{\IBBFnZz}{0.168} \renewcommand{\IBBFnNz}{0.266}

\FITPRBKY{Model Fits}
 }
\clearpage
\pagebreak

\appendix

\setlength{\footnotemargin}{5.80mm}
\begingroup
\doublespacing
\centering
\Large\vspace*{-2.5em}ONLINE APPENDIX \\[0.20em]
\Large\vspace*{-0.5em}\begin{singlespace}\href{\PAPERDOIURL}{\PAPERTITLE}\end{singlespace}
\large\AUTHORWANG\\[0.25em]
\endgroup

\vspace*{-2.0em}
\section{Value Functions for Discrete Choices (Online Appendix)\label{subsec:equavaljall}}
\renewcommand{\thefigure}{A.\arabic{figure}}
\setcounter{figure}{0}
\renewcommand{\thetable}{A.\arabic{table}}
\setcounter{table}{0}
\renewcommand{\theequation}{A.\arabic{equation}}
\setcounter{equation}{0}
\renewcommand{\thefootnote}{A.\arabic{footnote}}
\setcounter{footnote}{0}

\vspace*{-1em}
\begin{figure}[!htp]
\makebox[\textwidth][c]{
  \fcolorbox{white}{white}{
   \centering
   \begin{minipage}{1.15\textwidth}

    \begin{equation}
     \VakbejujOptStates{1}=
     \max_{
      \begin{array}{c}
       \EVJKset, \\
       \BIp=0, \BFp=0 \\
      \end{array}}
     \left\{
     \begin{array}{c}
      \utt \left( \FCOH - \Kp \right)
      + \SHKUjopt{1} \\
      + \PBETA \EVJEVp{0}
     \end{array}
     \right\} \label{eq:v1}
    \end{equation}

    \begin{equation}
     \VakbejujOptStates{2}=
     \max_{\begin{array}{c}
       \EVJKset,\BIp=0, \\
       \EVJFSset
      \end{array}}
     \left\{
     \begin{array}{c}
      \UtCBK{\BFp}{\Rjfs}{\FXCjfs}{2} \\
      + \PBETA \EVJEVp{\left(\BFp\right)}
     \end{array}
     \right\}
     \label{eq:v2}
    \end{equation}

    \begin{equation}
     \VakbejujOptStates{3}=
     \max_{\begin{array}{c}
       \EVJKset,\BFp=0, \\
       \EVJISset
      \end{array}}
     \left\{
     \begin{array}{c}
      \UtCBK{\BIp}{\RI}{\FXCjil}{3} \\
      + \PBETA \EVJEVp{\left(\BIp\right)}
     \end{array}
     \right\} \label{eq:v3}
    \end{equation}

    \begin{equation}
     \VakbejujOptStates{4}=
     \max_{\begin{array}{c}
       \EVJKset,\BIp=0, \\
       \EVJFBset
      \end{array}}
     \left\{
     \begin{array}{c}
      \UtCBK{\BFp}{\Rjfb}{\FXCjfb}{4} \\
      + \PBETA \EVJEVp{\left(\BFp\right)}
     \end{array}
     \right\} \label{eq:v4}
    \end{equation}

    \begin{equation}
     \VakbejujOptStates{5}=
     \max_{
      \begin{array}{c}
       \EVJKset,\BFp=0, \\
       \EVJIBset
      \end{array}}
     \left\{
     \begin{array}{c}
      \UtCBK{\BIp}{\RI}{\FXCjib}{5} \\
      + \PBETA \EVJEVp{\left(\BIp\right)}
     \end{array}
     \right\} \label{eq:v5}
    \end{equation}

    \begin{equation}
     \VakbejujOptStates{6}=
     \max_{
      \begin{array}{c}
       \EVJKset,\EVJIBset, \\
       \EVJFBset
      \end{array}}
     \left\{
     \begin{array}{c}
      \UtCBKjoint{\BIp}{\RI}{\BFp}{\Rjfb}{\FXCjib}{\FXCjfb}{6} \\
      + \PBETA \EVJEVp{\left(\BFp+\BIp\right)}
     \end{array}
     \right\} \label{eq:v6}
    \end{equation}

    \begin{equation}
     \VakbejujOptStates{7}=
     \max_{
      \begin{array}{c}
       \EVJKset,\EVJISset, \\
       \EVJFBset
      \end{array}}
     \left\{
     \begin{array}{c}
      \UtCBKjoint{\BIp}{\RI}{\BFp}{\Rjfb}{\FXCjil}{\FXCjfb}{7} \\
      + \PBETA \EVJEVp{\left(\BFp+\BIp\right)}
     \end{array}
     \right\} \label{eq:v7}
    \end{equation}
\end{minipage}}}
\end{figure}
 \pagebreak

\section{Additional Data Details (Online Appendix)}
\renewcommand{\thefigure}{B.\arabic{figure}}
\setcounter{figure}{0}
\renewcommand{\thetable}{B.\arabic{table}}
\setcounter{table}{0}
\renewcommand{\theequation}{B.\arabic{equation}}
\setcounter{equation}{0}
\renewcommand{\thefootnote}{B.\arabic{footnote}}
\setcounter{footnote}{0}

\begin{figure}[H]
\subsection{Data: Village Locations}
\vspace*{-3mm}
\makebox[\textwidth][c]{
\fcolorbox{white}{white}{
\centering
\begin{minipage}{1.0\textwidth}
\caption{\label{fig:map}Townsend Thai Monthly Survey Village Locations}
\begin{center}
\includegraphics[scale=0.7]{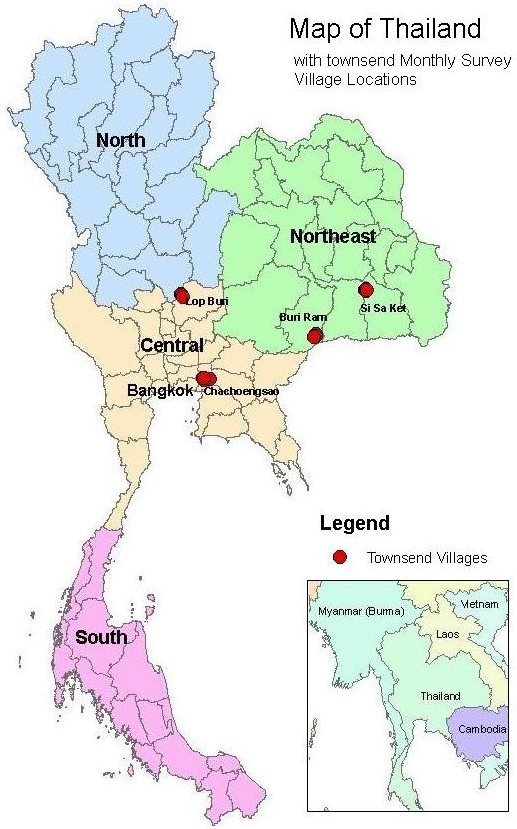}
\end{center}
\small \textcolor{blue}{\emph{Notes:}} This paper uses panel data from 16 Thai villages collected between 1999 and 2009. The panel includes detailed credit market transaction information for more than 650 households. 8 of the 16 villages are located in the poorer Northeast region, and 8 villages are located in the wealthier Central region of Thailand. See Section \ref{sec:Data-and-Background} for details.
\end{minipage}
}}
\end{figure}

\begin{figure}[H]
\subsection{Data: Volume Shares by Lender}
\vspace*{-3mm}
\makebox[\textwidth][c]{
\fcolorbox{white}{white}{
\centering
\begin{minipage}{0.87\textwidth}
\begin{table}[H]\centering
\def\sym#1{\ifmmode^{#1}\else\(^{#1}\)\fi}
\caption{\label{tab:borrowintensive} Share of Volume of Loans by Lender Type}
\begin{tabular}{l*{3}{c}}
\toprule
&\multicolumn{1}{c}{Overall}
&\multicolumn{1}{c}{1999-2001}
&\multicolumn{1}{c}{2002-2009}\\
\midrule
\rowgroup{\textbf{Northeast} Region (poorer)}&                   &            &            \\
Million Baht Fund {\footnotesize\emph{formal}}      &    39.0  &    10.8  &    48.2  \\
BAAC {\footnotesize\emph{formal}}                   &    36.1  &    50.3  &    30.9  \\
Village Coop {\footnotesize\emph{informal}}         &     2.7  &     1.1  &     2.6  \\
Friends and Neighbors {\footnotesize\emph{informal}}&    18.3  &    29.2  &    15.7  \\
Village Moneylenders {\footnotesize\emph{informal}} &     4.0  &     8.4  &     2.6  \\
\midrule
\rowgroup{\textbf{Central} Region (richer)}&                    &           &           \\
Million Baht Fund {\footnotesize\emph{formal}}      &    17.5  &     4.5  &   24.4  \\
BAAC {\footnotesize\emph{formal}}                   &    49.9  &    53.2  &   46.8  \\
Village Coop {\footnotesize\emph{informal}}         &    11.2  &    17.5  &    8.6  \\
Friends and Neighbors {\footnotesize\emph{informal}}&    20.0  &    22.4  &   19.1  \\
Village Moneylenders {\footnotesize\emph{informal}} &     1.5  &     2.4  &    1.1  \\
\bottomrule
\end{tabular}
\end{table}
\vspace*{-1em}
\small \textcolor{blue}{\emph{Notes:}}
Across the years and regions, for savings, there were 60.5, 39.8, and 8.7 million Baht of deposits made at commercial banks, the BAAC/GSB, and village coops, respectively. Surveyed households lent out 46.2 million Baht of loans. Across the years and regions, for borrowing, there were 95.1, 146.0, and 25.9 million Baht of loans borrowed from MBF, the BAAC, and village coops, respectively. Surveyed households borrowed 64.3 million Baht of loans from friends and neighbors and 7.1 million Baht of loans from money-lenders. While these statistics can be skewed by outliers, especially when studied annually, they provide a rough picture of the relative aggregate significance of the key channels for borrowing and saving along intensive margins.
In Table \ref{tab:borrowintensive}, I aggregate the total volume of loans (in Baht) from the five key types of lenders shown in the table. The shares are computed by dividing each lender's volume of loans by the total across lenders. The overall column shows shares for all years aggregated together. The columns for the year groups show proportions based on total lending within each period. Table \ref{tab:borrowintensive} shows a rise in the proportion of loans from MBF over time and reductions in aggregate loan volume shares from informal lenders.
\end{minipage}
}}
\end{figure}

\begin{figure}[H]
\subsection{Data: Borrowing and Savings Amounts}
\makebox[\textwidth][c]{
\fcolorbox{white}{white}{
\centering
\begin{minipage}{0.9\textwidth}
\caption{\label{fig:SaveLoanSize}Distribution of borrowing and savings Amount}
\includegraphics[scale=0.9]{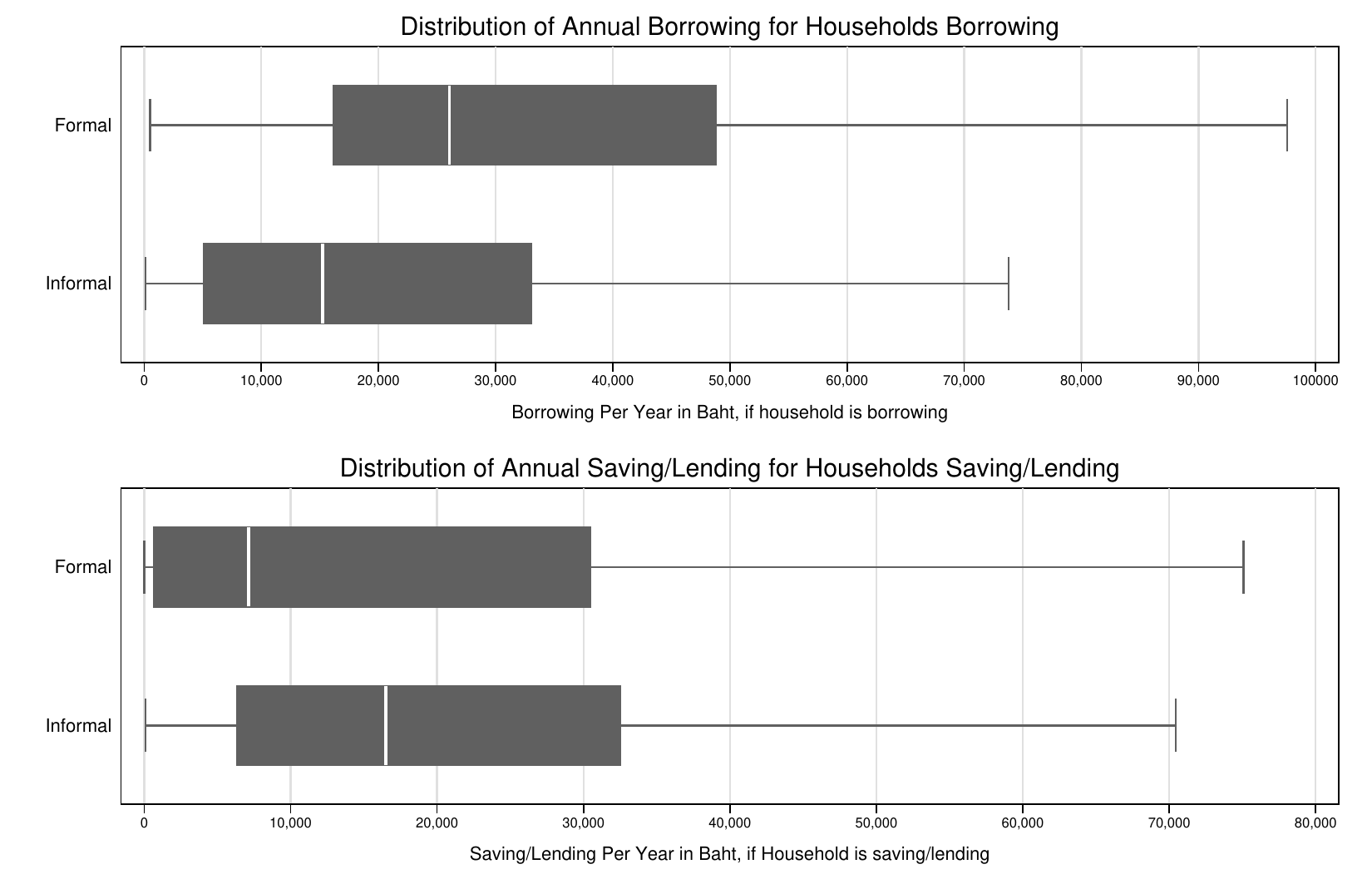}
\caption{\label{fig:LoanSize}Distribution of Borrowing Amount by Lenders}
\includegraphics[scale=0.9]{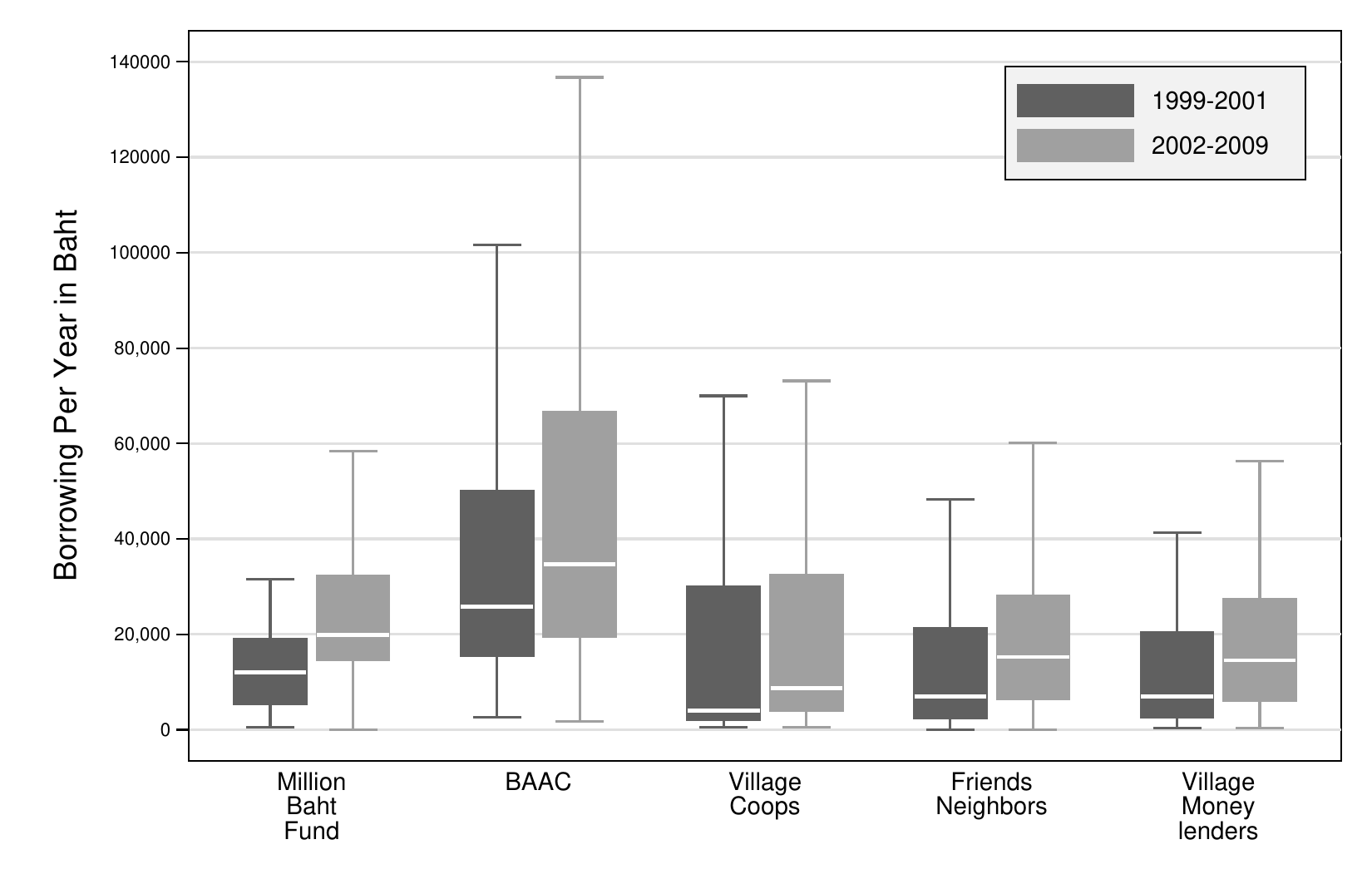}
\small \textcolor{blue}{\emph{Notes:}} Amount of borrowing and savings represents annualized sums for each lending and saving categories for each household. (The Million Baht Fund had a limited number of transactions close to the end of 2001.) See Section \ref{par:Amount-of-Borrowings} for detail.
\end{minipage}
}}
\end{figure}

\begin{figure}[H]
\subsection{Data: Reported Fees and Transport Costs for Formal Borrowing}
\makebox[\textwidth][c]{
\fcolorbox{white}{white}{
\centering
\begin{minipage}{1.0\textwidth}
\caption{\label{fig:FixedCost}Fees and Transport Cost for BAAC and Million Baht Fund}
\includegraphics[scale=1]{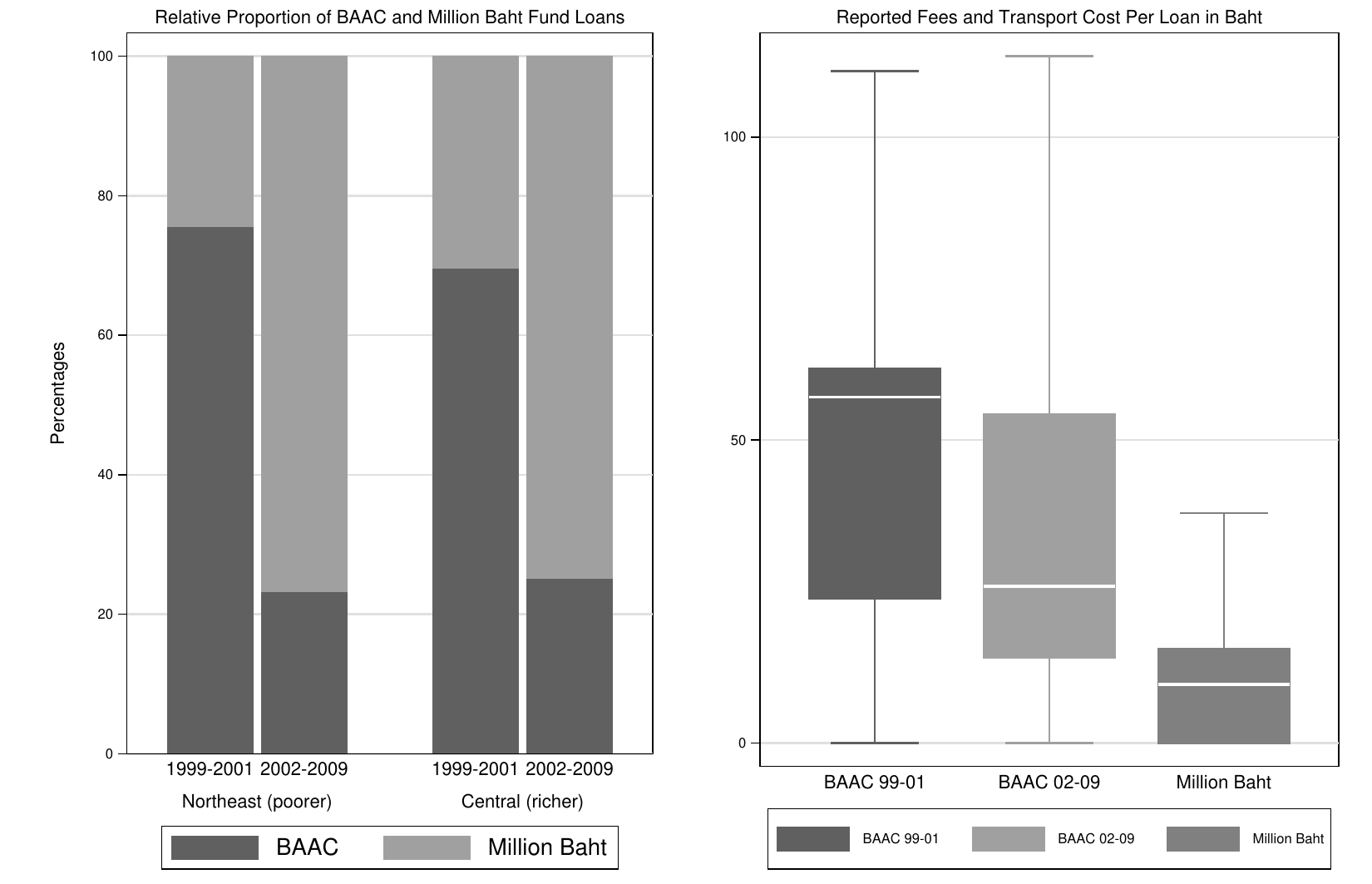}
\small \textcolor{blue}{\emph{Notes:}}
The Million Baht Fund is administered by village committees, and hence require fewer fees and transport costs to access. Formal borrowing fees and transport costs decrease significantly after 2001 as Million Baht Fund became the leading provider of formal loans. See Section \ref{par:Fixed-Cost} for detail.
\begin{table}[H]\centering
\def\sym#1{\ifmmode^{#1}\else\(^{#1}\)\fi}
\caption{\label{tab:FCBorrDetailType} Fees and Transport Cost for Each Loan by Lender Types (Baht)}
\begin{tabular}{l*{3}{c}}
\toprule
&\multicolumn{1}{c}{Relatives}&\multicolumn{1}{c}{Neighbor}&\multicolumn{1}{c}{Moneylender}\\
&  mean(sd)         &  mean(sd)         &  mean(sd)         \\
\midrule
Fees and Transport Cost          &25.8 (133.6)         &4.0 (41.9)         &22.1 (38.3)         \\
\midrule
                    &\multicolumn{1}{c}{Commercial-Bank}&\multicolumn{1}{c}{BAAC}&\multicolumn{1}{c}{Million Baht Fund}\\
\midrule
Fees and Transport Cost          &379.7 (737.6)         &51.6 (92.1)         &11.3 (16.6)         \\

\bottomrule
\end{tabular}
\end{table}
 \small \textcolor{blue}{\emph{Notes:}} Commercial bank loans are rare, likely due to the high fixed costs associated with getting them. Fees and Transport costs are only a proportion of the total pecuniary fixed costs associated with borrowing. There are also possibly non-pecuniary costs of borrowing. See Section \ref{par:Fixed-Cost} for detail.
\end{minipage}
}}
\end{figure}

\begin{figure}[H]
\subsection{Data: Repayment Rate Based on Payment History}
\vspace*{-3mm}
\makebox[\textwidth][c]{
\fcolorbox{white}{white}{
\centering
\begin{minipage}{1.0\textwidth}
\begin{table}[H]\centering
\def\sym#1{\ifmmode^{#1}\else\(^{#1}\)\fi}
\caption{\label{tab:Repay_ontime} Repayment of Loans}
\begin{tabular}{l*{3}{c}}
\toprule
                    &Fully Repaid&Not Fully Paid\\
\midrule
\textbf{Formal Loans}      &                  &                   \\
Percentage of Loans             &     97.01 &      2.99\\
\midrule
\textbf{Informal Loans}    &                   &                   \\
Percentage of Loans             &     95.16 &      4.84\\
\bottomrule
\end{tabular}
\end{table}
 \small \textcolor{blue}{\emph{Notes:}} Every month, new loans that households take out are recorded and the month in which repayment should be completed is also recorded. Then every month, repayment of each loan is tracked until full repayment is made. Table \ref{tab:Repay_ontime} shows that most households repay their loans and that the repayment difference between formal and informal loans is small. Given these data, this paper does not model default. See Section \ref{par:Repayment} for detail.
\begin{table}[H]\centering
\def\sym#1{\ifmmode^{#1}\else\(^{#1}\)\fi}
\caption{\label{tab:Repay_notFull}Repayment Percentage for Loans That Are Not Fully Paid}
\begin{tabular}{l*{3}{c}}
\toprule
                    &\multicolumn{1}{c}{By Month 24}&\multicolumn{1}{c}{By Month 36}&\multicolumn{1}{c}{By Month 48}\\
\midrule
\textbf{Formal Loans}              &                   &                   &                   \\
Percentage of Principal Repaid   &30.3          &36.7         &39.1         \\
\midrule
\textbf{Informal Loans}            &                   &                   &                   \\
Percentage of Principal Repaid   &32.4          &39.9         &44.0         \\
\bottomrule
\end{tabular}
\end{table}
 \small \textcolor{blue}{\emph{Notes:}} For the 2.99 percent of formal loans that are not fully repaid, and 4.84 percent of informal loans that are not fully repaid, Table \ref{tab:Repay_notFull} shows that a significant proportion of the principal on these loans is repaid within two years of the start of the loans. Given this information and data, as shown in Table \ref{tab:Repay_ontime}, the repayment gap between formal and informal loans is very small; hence, this paper does not model default. See Section \ref{par:Repayment} for detail.
\end{minipage}
}}
\end{figure}
\pagebreak

\section{Solution and Estimation Algorithm\label{sec:soluestialgo}  (Online Appendix)}
\renewcommand{\thefigure}{C.\arabic{figure}}
\setcounter{figure}{0}
\renewcommand{\thetable}{C.\arabic{table}}
\setcounter{table}{0}
\renewcommand{\theequation}{C.\arabic{equation}}
\setcounter{equation}{0}
\renewcommand{\thefootnote}{C.\arabic{footnote}}
\setcounter{footnote}{0}

The model is computationally intensive to estimate. Building on standard solution and estimation concepts, I develop a global optimization routine that minimizes the use of loops and iteration by parallel computing. The routine solves and estimates the heterogeneous agent equilibrium model with discrete and constrained continuous choices.\footnote{There has been significant progress made in using endogenous grid methods to solve and estimate models with discrete and continuous choices \autocite{fella_generalized_2014, IskhahovJorgensenRustSchjerning17}. Some of the steps described here can potentially be combined with endogenous grid or other methods when the model structure allows.} The resulting algorithm is resource intensive concurrently but saves time if concurrency is possible. The algorithm is also designed to accurately capture the impacts of differences in fixed costs, interest rates and collateral constraints on the choice sets, the average costs (returns) of loans (savings), and the conditional asset distributions facing households across the credit market options.

Typically, university computing clusters have a maximum set of computing resources that are shared, which limits resources that are available to individual researchers. Cloud computing services such as Amazon Web Services (AWS) offers from the econometrician's perspective potentially unlimited concurrent resources. On AWS's managed containerized services, the cost of running 1 CPU for 10000 minutes is identical to running 10000 CPUs for 1 minute. The idea here is to take advantage of scalable on-demand concurrent computing resources by modifying some standard solution and estimation steps. If the estimation task can be parallelized explicitly or implicitly, then the researcher has more control over how much time estimation should take, given their budget.

In terms of solution algorithms, I reduce the iterative steps required for joint continuous and discrete optimization through a "zooming-in" algorithm (Section \ref{subsec:solu}), I adopt a non-simulation based procedure for deriving steady state asset distributions (Section \ref{subsec:steady}), and I solve for equilibrium interest rates using a multi-section algorithm (Section \ref{subsec:equi}).\footnote{Implementation examples for using aspects of the solution algorithms can be found at the \href{http://fanwangecon.github.io/MEconTools/}{MEconTools} repository.}

For estimation, I add additional initialization steps to standard estimation routines. These initial steps help me find initialization parameters that are potentially closer to the minimizers of the estimation objective function. I do this by simulating the model at a large set of parameter values given data, and then conducting an initial round of multi-start estimation using a set of polynomials that approximate components of the estimation objective functions. I describe the estimation algorithm in Section \ref{subsec:nestedestimation} and its implementation on AWS in Sections \ref{subsec:containerorchestration}, \ref{subsec:containerorchestration}, \ref{subsec:awsbatchqueue}, and \ref{subsec:awsprice}.\footnote{Examples for how to use AWS web services can be found at the \href{https://fanwangecon.github.io/Py4Econ}{Py4Econ} repository.}
 \subsection{Solution Algorithm\label{sec:solualgo}}
\subsubsection{Iterative Optimization on Asset Share Grids}
\paragraph{Triangular Asset Choice Sets}
\label{subsec:jfcr}
\FIGTYPEONE{fig_online_appendix}{\FIGTMjfcr}{\FIGPMjfcr}{0.70}{\FIGDMjfcr}

\CAPCAP\ITRs, \FCF and collateral constraints differences across credit market options determine the choice set for \RKI and the cost of financing \RKI. Using \(\widehat{\Bp}\) to denote principal borrowed for here, the average cost of \BBBB, \(\frac{\widehat{\Bp} \PRj + \FXCj}{\widehat{\Bp}}\), is increasing in \(\FXCj\) and decreasing in \(\widehat{\Bp}\). With \YMINZERO equal to zero, the natural \BBBB constraint restricts interest and principal due next period to be less than the depreciated \RKI. This means a portion of \RKI must be financed by existing \CZH. The marginal effect of additional \CZH on the maximum \RKI level is constant and equal to \(1+\frac{1-\PDELTA}{\PRj+\PDELTA}\). This ratio is the \BBBB multiplier. Lower \(\PRj\) expands the choice set by increasing the multiplier. Lower \(\FXCj\) also expands the choice set by increasing usable \CZH. The \CLC restricts the choice set if it is tighter than the natural borrowing constraint. Graphically, \ITRs, \FCF, and collateral constraints shift vertices of an obtuse triangle that represents the choice set, as shown in \FIGRMjfcr. Formal and informal borrowing choices form separate obtuse triangle choice sets. Given combinations of wealth, informal interest rates, and informal borrowing fixed costs, borrowing choices are constrained as shown by \FIGRMjfcr.\footnote{If exogenous quantity constraints are imposed as well, graphically, they would be vertical lines cutting potentially through the obtuse triangles shown in \FIGRMjfcr.}

For \SSSS, the average return to \SSSS is decreasing in \(\FXCj\) and increasing in the amount saved. An increase in \(\FXCj\) reduces one to one the maximum \RKI, and increases in the \SSSS \ITR reduce next period's \CZH and subsequent \RKI choice sets.
 \paragraph{Bounds on Continuous Choices}
\newcommand{\solufootone}{Assuming that \(\totk=\totb\), the maximum bounds of \(\Bp\) for the \(\zoom=I^{\text{th}}\) iteration are:
	\begin{align}
	\begin{split}
	\Bphhjmaxzoomopt{I^{\text{th}}} =&
	\frac{\Bphhjmax - \Bphhjmin}{\totb^{I-1}}
	\cdot
	\min \left\{ \SUMA_{s = 1}^{I-1} \totb^{s - 1} \cdot \discboptiopt{ (I - s)^{\text{th}} } + 1, \totb^{I-1} \right\}
	+ \Bphhjmin \\
	\Bphhjminzoomopt{I^{\text{th}}} =&
	\frac{\Bphhjmax - \Bphhjmin}{\totb^{I-1}}
	\cdot
	\max \left\{ \SUMA_{s = 1}^{I-1} \totb^{s - 1}\cdot\discboptiopt{(I-s)^{\text{th}}} - 1, 0 \right\}
	+ \Bphhjmin \\
	\end{split}
\end{align}}
\newcommand{\solufoottwo}{Equation \eqref{equ:solulessthan} applies to constrained maximization problems that are possibly non-differentiable but concave. With sufficient grid points in each iteration, the method should also generally find optimal choices for non-concave problems. If the non-concaveness is due to fixed costs or other clearly defined parameters---as is the case for this model (Equation \eqref{equ:solumaindiscrete})---the problem could be restated as multiple concave continuous-choice problems nested within discrete-choice problems.}

\label{subsec:solu}
Following previous notations, I rewrite the continuous choice problem over safe asset \(\Bp\) and risky asset \(\Kp\) for each household \(i\) as:
\begin{align}
 \begin{split}
   \label{equ:solumaindiscrete}
  \max_{
   \mmsbeg
   \J \in \left\{ 1,...,7 \right\} \\
   \mmsend
  }
  {\left\{
   \max_{
    \mmsbeg
    \Bp \in \left( \Bphhjmin, \MINBORRj \right] \\
    \mmsend
   }
   {\left\{
    \max_{
     \mmsbeg
     \Kp \in \left( \Kphhjbmin, \Kphhjbmax \right) \\
     \mmsend
    }
    \uallj \left( \A, \fcoh, \SHKUj ; \Bp, \Kp \right)
    \right\}}
   \right\}}
 \end{split}
\end{align}
where, given the \(\Bp\) choice, the minimum and maximum for the \(\Kp\) choice are:
\begin{align}
  \label{equ:soluctskmax}
  \Kphhjbmax =&
  \left( \COHi - \FXCj \right) - \frac{\Bp}{1+\Rj} \\
  \label{equ:soluctskmin}
  \Kphhjbmin =&
  \begin{cases}
   \max \left\{- \dfrac{\YMIN}{\left( 1 - \PDELTA  \right)\cdot\Rj } - \dfrac{\Bp}{1-\PDELTA}, 0\right\}
    & \text{if } j \text{ is borrow} \\
   0
    & \text{if } j \text{ is save}   \\
  \end{cases}
\end{align}
To maintain positive consumption in the current period, \(\Kp \) must be less than \(\Kphhjbmax\), and to maintain positive consumption in all states in the next period, \(\Kp\) must be greater than \(\Kphhjbmin\). For \BBBB, geometrically, plotting \( \Kp \) along the y-axis, and \( \Bp \) along the x-axis, Equation \eqref{equ:soluctskmax} and Equation \eqref{equ:soluctskmin} represent two intersecting lines where the intersection point is in the region of positive \( \Kp\) and negative \( \Bp \) as shown in \FIGRMjfcr. Individual \(i\) can only borrow less than the \(\Bp\) value at the intersecting point, which is:
\begin{align}
   \label{equ:soluctsbmin}
   \Bphhjmin = & -\left( \COHi - \FXCj + \frac{\YMIN}{\left( 1 - \PDELTA \right)\cdot\Rj } \right)
   \cdot
   \frac{ \left( 1 + \Rj \right) \cdot \left( 1 - \PDELTA \right) }{ \PDELTA + \Rj}
   \thinspace.
\end{align}
Household \(i\)'s \(\Kp\) choice must be less than the \(\Kp\) value associated with \(\Bphhjmin\) at the intersecting point.

\paragraph{Discretization using Asset Share Indexes}

The geometry of the problem allows me to solve for the minimum and maximum asset choices as shown above. I now transform the problem in Equation \eqref{equ:solumaindiscrete}, which is continuous, to the percentage-grid asset choice problem in Equation \eqref{equ:solumainshare}, which has identical scales for all households. This homogenization of the choice grid allows for efficient vectorization.\footnote{Utility maximization requires two matrix operations only, and utility evaluation and maximization can fully take advantage of single instruction parallelization without additional code.}

To approximate the exact optimal choice, I solve the discretized percentage asset choice problem \( \zoom \) times. This is a type of iterative grid search method, which has long been used in the literature, starting perhaps with \textcite{imrohoroglu_numerical_1993}. After each iteration, an index operation is used to find the closest set of surrounding share indexes in each direction from the current optimal index. I solve this model here with \(\totk=50\), \(\totb=50\) and \( \zoom=3 \). Specifically, using \( \zoom \) as a superscript to indicate the current index search iteration, the iterative percentage maximization problem can be written as:
\begin{align}
 \begin{split}
   \label{equ:solumainshare}
  (\disckopti, \discbopti) = \argmax_{
   \begin{array}{c}
    \disckzoom \in {0,...,\totk} \\
    \discbzoom \in {0,...,\totb} \\
   \end{array}
  }
  \uallj^{\zoom} \left( \A, \fcoh, \SHKUj ; \Bijnbzoom, \Kijnbnkzoom \right)
 \end{split}
\end{align}
Choices \(\discbzoom \), \( \disckzoom\) are indexes for sequences \(\left\{ \Bijnbzoom \right\}^{\totb}_{\discb = 0}\) and \(\left\{ \Kijnbnkzoom \right\}^{\totk}_{\disck = 0}\), where \(\Bijnbzoom =
\Bphhjmaxzoom \cdot \frac{\discb}{\totb}
+ \Bphhjminzoom \cdot \left( 1 - \frac{\discb}{\totb} \right)\) and \(
\Kijnbnkzoom =
\Kphhjnbmaxzoom \cdot \frac{\disck}{\totk}
+ \Kphhjnbminzoom \cdot \left( 1 - \frac{\disck}{\totk} \right)
\).

By the \(\zoom=I-1\) iteration, the difference between the optimal choice \(\Bphhjopti\) and the approximate choice \(\Bijnbzoomopti{I-1}\) is less than a fraction determined by \(\totb\) and \( \zoom\):\footnote{\solufoottwo}\footnote{\solufootone}
\begin{align}
 \begin{split}
   \label{equ:solulessthan}
  \lvert \Bphhjopti - \Bijnbzoomopti{I-1} \rvert  \le \left(\Bphhjmax - \Bphhjmin \right) \cdot \frac{2}{\totb^{I-1}}
 \end{split}
\end{align}
If \(b_{\text{maxgap}} = \max_{i,j} \lvert \Bphhjmax - \Bphhjmin \rvert\) among all \(i\) in the state space solution grid and all \(j \in J\), then with \(\totb=50\) and \( \zoom=3 \):
\(\lvert \Bphhjopti - \Bijnbzoomopti{3} \rvert \le \frac{b_{\text{maxgap}}}{62500} \text{ , } \forall i \text{ and } \forall j\).
 \subsubsection{Value Function Iteration and Expectation}

\label{subsec:vfi}
To solve the dynamic stochastic programming problem, within a value function iteration step and after optimization, I integrate over the value of the continuation value function in Equation \eqref{eq:ValVj}:
\begin{align*}
E_{\SHK,\SHKU} \left( \Vakbephi \RAWONE \right) &
= \ITG
  \ITG\max_{\J \in \JJ}
    \Vakbejuj
    \RAWTWO
  \ITGD \CDF \left(\SHKU \right)
  \ITGD \CDF \left(\SHK \right)
  \thinspace,
\end{align*}
where \(\left\{ \SHKUj \right\}_{j\in\left\{ 1,...7\right\} }\) are assumed i.i.d and extreme-value Type I distributed. Hence:
\begin{equation}
  \EVakbe \left(\STATESj\right) =
  \ITG\max_{\J \in \JJ}
    \Vakbejuj
    \RAWTWO
  \ITGD \CDF \left(\SHKU \right)
  =
  \log
  \left(
    \sum_{\J \in \JJ}
    \exp \left(
      \Vakbej
      \RAWTHREE/\sigma_{\phi}
         \right)
  \right)
  \thinspace.
  \label{eq:integrate1}
\end{equation}

Using the optimization routine described earlier, I solve \(\EVakbe\left(\STATESj\right)\) at all combinations of a grid of \(\K\), a grid of \(\B\), and three \(\SHK\) grid points \(\SHK \in \left\{\SHK-3\EESIGMA, 0, \SHK+3\EESIGMA\right\}\). I obtain an \(\A\) specific interpolant \(\EVacohk \INTPONES \) based on multi-linear spline. Using \(\EVacohk\) in combination with $D$ $\SHK$ draws for each \(\INTPTWOS\),\footnote{To avoid extrapolation, I restrict the \(\SHK\) draws to be within the \(\SHK\) points where I solved \(\EVakbe\).} I approximate this expectation:
\[
E_{\SHK}\EVakbe\left(\STATESj\right)
\approx
\frac{1}{D} \sum_d \EVacohk
\left(
  \left\{ \B +\left(1-\PDELTA\right) \K + \exp\left(\SHK_D + \A \right)\cdot \K^{\PALPHA}
  \right\},
  \K
\right)
\]
\[
\textit{Interpolants, 1st: } \EVacohk \INTPONES \textit{ 2nd: } \EVakb \INTPTWOS
\]
With these simulated averages, I estimate a second \(\A\) specific interpolant \(\EVakb \INTPTWOS\) based on multi-linear spline. This second interpolant captures the differential marginal effects on expected future utility from investments in safe and risky assets. I substitute the ${\EVakb}$ for the integral over expected value function in Equation \eqref{eq:ValVj}. Value function iteration continues until convergence. The method here is similar in structure to the value function approximation method in \textcite{imai_intertemporal_2004}.\footnote{\textcite{imai_intertemporal_2004} present a model with both continuous state space and continuous choice space. In that model, individuals make human capital and financial asset choices.}

It should be noted that the multinomial logit assumption for credit category shocks significantly reduces the integration burdens here. As an alternative, multinomial normal errors would accommodate correlations among discrete category shocks with a seven by seven variance-covariance matrix for the seven credit alternatives. Estimating the covariance matrix would substantially add to estimation costs. In the context of this paper, the gains of allowing for multinomial normal errors might be limited. First, the dominant drivers for credit market participation are asset positions and productivity types, which are captured by the modeled component of the choice-specific value functions. Second, credit categories share the same productivity shocks, which means that overall uncertainty is not uncorrelated across credit category alternatives.

An additional issue to note is that due to the multinomial shocks, CEV changes can not be computed analytically following standard procedures under CRRA preference assumptions \autocite{conesa_social_1999}. For the welfare analysis, holding policy functions constant, I recompute value functions for each element of a dense grid of consumption percentage changes. The resulting array of values for each state-space element is compared against value function outputs under policy shifts to find the percentage consumption changes that minimize the differences. \subsubsection{Steady State Distribution Conditional on Type}

\paragraph{Conditional \( \COHp \) distribution}
\label{subsec:steady}

\newcommand{\COHpinverse}{ \LOGG \left(  \dfrac{\COHp - \Wphhjmin}{ \EXPP(\A) \cdot \FKcohs^{\PALPHA}  }  \right)    }

The stationary asset distribution is determined by the conditional transition \(\PDF \left( \COHp \condi \COH \right)\), which equals to:
\begin{align}
\begin{split}
\PDF \left( \COHp \condi \COH \right)
=
\ITG_\A
\SUMA_\J
\FPcohs
\PDF \left( \COHp \condi \A, \COH, j \right)
\PDF \left( \A \right)
\ITGD \A
\thinspace\thinspace.
\\
\end{split}
\end{align}
Given the structure of the problem, \(\PDF \left( \COHp \condi \A, \COH, j \right)\) is:
\begin{align}
\begin{split}
\PDF \left( \COHp \condi \A, \COH, j \right) =
\begin{cases}
0                                                         & \text{ if } \COHp \le \Wphhjmin \\
\PDF_\SHK \left( \SHK^{\opti}\left( \COHp \right) \right) & \text{ otherwise }              \\
\end{cases}
\thinspace\thinspace,
\end{split}
\end{align}
where:
\begin{align}
\begin{split}
\label{equ:steadywphhjminshkopti}
\Wphhjmin &= \FKcohs \left( 1 - \PDELTA \right) + \FBcohs + \YMIN\\
\SHK^{\opti}\left( \COHp \right) &= \COHpinverse\\
\end{split}
\end{align}
Hence, the conditional probability that \( \COHp \) falls between \( \left[ \COHp_x, \COHp_y \right] \) is:
\begin{align}
\begin{split}
\label{equ:steadycohpcondicts}
\PROB \left( \COHp_y - \COHp_x \condi \A, \COH, \J \right)
=
\begin{cases}
0                                                                                                                       & \text{ if } \COHp_x \le \Wphhjmin \\
\CDF_\SHK \left( \SHK^{\opti}\left(\COHp_y\right)  \right) - \CDF_\SHK \left( \SHK^{\opti}\left(\COHp_x\right)  \right) & \text{ otherwise}                 \\
\end{cases}
\end{split}
\end{align}
The method described here follows the spirit of the inverse decision rule method \autocite{rios-rull_computation_1997}.

\paragraph{Discretizing Conditional \( \COHp \) distribution}

\newcommand{\COHpinversedischigh}{ \LOGG \left( \frac{\Wnwcdfpp - \Wphhjmin}{ \EXPP(\A) \cdot \FKcohs^{\PALPHA}  }  \right)    }
\newcommand{\COHpinversedisclow}{ \LOGG \left( \frac{\Wnwcdf - \Wphhjmin}{ \EXPP(\A) \cdot \FKcohs^{\PALPHA}  }  \right)    }

I discretize \( \COH \) along a grid with \( \totcoh \) grid points. The mid-points sequence with \(\totcoh-1\) elements is \(\left\{ \Wnw \right\}^{\totcoh-1}_{\disccoh = 1} \) where \(\Wnw =
\left( \frac{\Wmax - \Wmin}{\totcoh} \right)
\cdot
\left( \disccoh + 0.5 \right)
+
\Wmin \). The corresponding end-points sequence is  \(\left\{ \Wnwcdf \right\}^{\totcoh}_{\subdisccohcdf = 1} \) where \( \Wnwcdf =
\left( \frac{\Wmax - \Wmin}{\totcoh} \right)
\cdot
\left( \subdisccohcdf \right)
+
\Wmin\). Given these, I define:
\begin{align}
\begin{split}
\PROB \left( \Wnwp \condi \A, \Wnw, \J \right)
\coloneqq
\PROB \left( \Wnwcdfpp - \Wnwcdf \condi \A, \Wnw, \J \right)
\thinspace,
\end{split}
\end{align}
where the right-hand side is evaluated using Equation \eqref{equ:steadycohpcondicts}. The discretized conditional \CHOHH distribution for a particular productivity type \(\Adisc\) is:
\begin{align}
\begin{split}
\PROB \left( \Wnwp \condi \Adisc, \Wnw \right)
=
\SUMA_\J
\FPcohsn
\PROB \left( \Wnwp \condi \Adisc, \Wnw, \J \right)
\thinspace.
\\
\end{split}
\end{align}
Given the policy functions derived in Sections \ref{subsec:solu} and \ref{subsec:vfi}, computing this conditional probability for all current- and next-period grid points involves simply finding the cumulative normal probabilities over a matrix of quantile points. The procedure is vectorizable and almost instantaneous. When there are multiple persistent shocks, the potential dimensionality of this discretized full-states transition matrix increases exponentially. However, the computational size increases only linearly with the usage of sparse matrices. \footnote{For examples with multiple persistent shocks, see \href{https://fanwangecon.github.io/CodeDynaAsset/}{Dynamic Asset Code Repository}.}

\paragraph{Discretized marginal \( \COHp \) distribution}
\newcommand\inv[1]{#1\raisebox{1.15ex}{$\scriptscriptstyle-\!1$}}

I now solve for \( \PROB \left( \Wnw \condi \Adisc \right) \) using \( \PROB \left( \Wnwp \condi \Adisc, \Wnw \right) \) following standard first order Markov transition procedure. At this stage, either an eigenvector or projection approach could be taken. The projection approach is more stable generally, but some languages' eigenvector algorithm is more efficient for large sparse full-states transition matrices. For projection, suppose \( b \) is a \( 1 \times (\totcoh-1) \) vector, which is all zero except for the final element which equals \( 1 \), \( Q \) is the \( (\totcoh-1) \times (\totcoh-1) \) Markov transition matrix where each row corresponds to a different \(\Wnw\) and each cell corresponds to \( \PROB \left( \Wnwp \condi \Adisc, \Wnw \right) \) minus corresponding values in an identity matrix. Given these, I compute \(P = \left( b \cdot Q^{\intercal} \right) \cdot \inv{\left( Q \cdot Q^{\intercal} \right)}\), where each \(\disccoh\) element of matrix \( P \) is equal to the marginal distribution for \( \Wnw \) conditional on \(\Adisc\), \(P_{\disccoh} = \PROB \left( \Wnw \condi \Adisc \right) \). As the number of discrete points increases, the discretized marginal distribution approaches the continuous marginal distribution.
 \subsubsection{Integration over Productivity Types}
\label{subsec:typeinter}
Integration over types and solving for equilibrium \ITR both involve solving the model at different parameters. This is in contrast to the solution procedures described in Sections \ref{subsec:solu}, \ref{subsec:vfi}, and \ref{subsec:steady}, where computations share common instructions and benefit from vectorization-based parallelization. Here, I solve the model at different productivity types and interest rates in separate concurrent processes.

I assume that the productivity type is normal, and lognormal with respect to \OTP. I solve the model concurrently along a sequence of productivity types \(\left\{ \Adisc \right\}^{\totA}_{\discA = 1}\) where \(\Adisc = \QUADz \cdot \PASIGMA \cdot \sqrt{2} + \PAMU\). \( \left\{ \QUADz \right\}^{\totA}_{\discA = 1} \) are Hermite Quadrature points, with associated weights sequence \( \left\{ \QUADWGTz \right\}^{\totA}_{\discA = 1} \). Following the previous section's notations, the unconditional distribution of \(\COH\) is approximated by:

\begin{align}
 \begin{split}
   \label{eq:typeinterone}
  \ITG_{x=\Wnwcdf}^{y=\Wnwcdfk} \PDF \left( \COH \right) \ITGD \COH
  \approx &
  \SUMA_{\subdisccoh = \subdisccohcdf}^{(\subdisccohcdf+k-1)}
  \AQSUMA
  \frac{1}{\sqrt{\pi}} \cdot \QUADWGTz \cdot
  \PROB \left( \Wnw \condi \Adisc \right)
  \\
 \end{split}
\end{align}
 \subsubsection{Equilibrium Interest Rates}
\paragraph{Multi-section algorithm}
\label{subsec:equi}
\newcommand{\equifootone}{The initial minimum and maximum interest rates should be picked to generate more borrowing than savings and more savings than borrowing, respectively. Given that, Equation \eqref{eq:equimove} looks for the smallest borrowing and savings gap among rates that generate more borrowing than savings.}
To solve for the equilibrium \ITR, I solve the model in \( \eqit \) iterations and during each iteration concurrently at \( \eachSolveR \) different \ITR points. If \( \eachSolveR = 1\), this is equivalent to bisection. In each iteration, I find:\footnote{\equifootone}
\begin{align}
 \begin{split}
 \label{eq:equimove}
  \RIstariter
  =
  \argmax_{ \RI \in \left\{ \RnReqit \right\}_{\discReqit = 1}^{\totR^{\eqit}}
  }
  \left\{ \HATT{\E} \left(\BIp \psep \RI \right)
  \suchthat
   \HATT{\E} \left(\BIp \psep \RI  \right)
  < 0
  \right\}
  \thinspace,
  \\
 \end{split}
\end{align}
where the net difference in borrowing savings is:
\begin{align}
\begin{split}
\label{eq:solveequibclear}
\HATT{\E}_{\Adisc} \left( \BIp \psep \RnReqit \right)
= &
\sum_\disccoh
\PROB \left( \Wnw \condi \Adisc \psep \RnReqit \right)
\cdot
\left(
\sum_\J
\fp \left( \Adisc, \COHd \psep \RnReqit \right)
\cdot
\fbi \left(\Adisc, \COHd \psep \RnReqit \right)
\right)\\
\HATT{\E} \left(\BIp \psep \RnReqit \right)
= &
\AQSUMA \PROBA \cdot \HATT{\E}_{\Adisc} \left(  \BIp \psep \RnReqit \right)\\
\end{split}
\end{align}
and where the discrete constrained choice set for each \( \eqit \) evolves following:
\begin{align}
\begin{split}
\left\{ \RnReqitopt{1} \right\}_{\discReqitopt{1} = 1}^{\totR^{\eqitopt{1}} = \totR^{\eqit} + \eachSolveR}
= \left\{ \RnReqit \right\}_{\discReqit = 1}^{\totR^{\eqit}}
\cup
\left\{
\R_{\eachSolveReach}
\suchthat
\begin{array}{c}
\dfrac{\RIstariterpone - \RIstariter}{\eachSolveR + 1}
\cdot
\eachSolveReach
+
\RIstariter \\
\text{, } \eachSolveReach \in \INTEGER
\text{, } 1 \le \eachSolveReach \le \eachSolveR
\\
\end{array}
\right\}
\thinspace.
 \end{split}
\end{align}

If the equilibrium rate is between \(1.00\) and \(1.30\), setting \( \eachSolveR = 8 \), after \( \eqitoptequal{3} \) iterations, I arrive at an approximated equilibrium \ITR within \(0.053\) percentage points of the equilibrium \ITR:
$\Requiapp = \RI_{\discReqitopt{3}, \eachSolveR = 8} \in \left[ \Requi - 0.00053, \Requi + 0.00053 \right]$.

At some values of the fixed costs parameters, given grids on productivity and assets, in the absence of the credit category shocks, there might be no households that are willing to save or borrow within the bounds on interest rates. The credit category shocks help to improve the numerical stability of equilibrium solution and estimation by assuring non-zero aggregate borrowing and savings levels as the estimator explores possibly high fixed costs.

Overall, solving the model involves Monte-Carlo integration for expected value function, weighting the probabilities of each credit participation category \(j\), the analytical integration of \(\PROB \left( \COHp_y - \COHp_x \condi \A, \COH, \J \right)\), Riemann-sum mid-point integration to approximate \( \ITG_{x=\Wnwcdf}^{y=\Wnwcdfk} \PDF \left( \COH | \Adisc \right) \ITGD \COH \), and the Hermite Quadrature integration here.

\paragraph{Tradeoffs between \( \eachSolveR \) and \( \eqit \)}
\newcommand{\cwith}{0.76in}
\newcommand{\mcwith}{1.52in}

\begin{table}
 \caption{\label{tab:equiiter}Concurrency and iteration: solving equilibrium, if $r^{I,equi} \in [1.0, 1.3]$}
 \vspace{0.1in}
 \begin{tabular}{| >{\arraybackslash}m{\cwith} | >{\centering\arraybackslash}m{\cwith} | >{\centering\arraybackslash}m{\cwith} | >{\centering\arraybackslash}m{\cwith} | >{\centering\arraybackslash}m{\cwith} | >{\centering\arraybackslash}m{\cwith} | >{\centering\arraybackslash}m{\cwith} |}
  \toprule
  \multicolumn{1}{|C{\cwith}|}{} & \multicolumn{2}{C{\mcwith}|}{\makecell{ \( \eachSolveR =1 \) \\ 1 model evaluation                                                                                                                                                                            \\per iteration\\ (bisection) }} & \multicolumn{2}{C{\mcwith}|}{\makecell{ \( \eachSolveR = 4 \)\\ 4 concurrent \\ model evaluations\\ per iteration}} & \multicolumn{2}{C{\mcwith}|}{\makecell{ \( \eachSolveR = 8 \)\\ 8 concurrent \\ model evaluations\\ per iteration}} \\
  \toprule
  \emph{Iterations}              & Number of Model Evaluations                                & Accuracy ($r^I$ percentage points) & Number of Model Evaluations & Accuracy ($r^I$ percentage points) & Number of Model Evaluations & Accuracy ($r^I$ percentage points) \\
  \midrule
\(\eqitoptequal{1st}\)  & 3                                                          & 15                                 & 4                           & 10                                 & 8                           & 4.3                                \\
  \(\eqitoptequal{2nd}\)                            & 4                                                          & 7.5                                & 8                           & 2                                  & 16                          & 0.48                               \\
  \(\eqitoptequal{3rd}\)                            & 5                                                          & 3.75                               & 12                          & 0.3                                & 24                          & 0.053                              \\
  \(\eqitoptequal{6th}\)                            & 8                                                          & 0.47                               & 24                          & 0.00003                            & 48                          & 0.0000007                          \\
  \bottomrule
  \hline
 \end{tabular}
\end{table}

Table \ref{tab:equiiter} explains the trade-off between \( \eachSolveR \) and \( \eqit \). If concurrent model evaluation is not possible due to computing limitations, bisection is the most efficient as shown in Table \ref{tab:equiiter} in terms of the number of model evaluations\footnote{Model evaluation refers to the number of times the model has to be solved for up to and including integrating over productivity types.} required to reach a fixed level of accuracy. Bisection achieves higher precision with 9 model evaluations than \( \eachSolveR = 4 \) (4 concurrent model evaluations) achieves with 12 model evaluations. When computing resources are available, as shown in Table \ref{tab:equiiter}, to achieve about 0.5 percentage point equilibrium \ITR accuracy, \( \eachSolveR = 8 \) requires only 2 iterations, using one-third of the time as bi-section.

\subsection{Estimation Algorithm\label{sec:estialgo}}
\subsubsection{Algorithm}
\label{subsec:nestedestimation}
\newcommand{\FOOTnestedestiOne}{Given \(\totAWSparam = 15\) and \(\totAWSpolyDegree = 3\), then \(\totAWSpoly=816\). Full third order polynomial means all polynomial combinations with degree less than or equal to \(3\). The number of polynomial terms equal to: \begin{equation}
  \totAWSpoly =
  \underbrace{\left(0\right)}_{
    \substack{
    0^{\text{th}} \text{ Deg.}\\
    \text{i.e.: } 1\\
    }
    } +
  \underbrace{\left(\totAWSparam\right)}_{
    \substack{
    1^{\text{st}} \text{ Deg.}\\
    \text{i.e.: } x \text{, } y\\
    }
  } +
  \underbrace{\left(\totAWSparam + \binom{\totAWSparam}{2}\right)}_{
    \substack{
    2^{\text{nd}} \text{ Degree}\\
    \text{i.e.: } x^2 \text{, } x \cdot y \text{, } y^2\\
    }
  } +
  \underbrace{\left(\totAWSparam + \totAWSparam\cdot\left(\totAWSparam-1\right)+ \binom{\totAWSparam}{3}\right)}_{
    \substack{
    3^{\text{rd}} \text{ Degree}\\
    \text{i.e.: } x^3 \text{, } x^2\cdot y \text{, ... } x \cdot y^2 \text{, } y^3 \\
    }
  }
\end{equation} Standard machine learning packages have routines that generate full polynomials, I use the \textit{PolynomialFeatures} module from \textit{scikits.learn}.}

I describe in this section the four steps of the \textit{more parallel} estimation algorithm summarized in Section \ref{subsec:summalgo}. First, I draw \(\totAWSdraws\) sets of parameters:
\[
\label{eq:nestestione}
\left\{ \ESTIALL_{\eachAWSdraws} \right\}_{\eachAWSdraws=1}^{\totAWSdraws}
=
\left\{
 \left\{
   \ESTIeach_{\eachAWSdraws\eachAWSparam}
 \right\}_{\eachAWSparam = 1}^{\totAWSparam}
\right\}_{\eachAWSdraws=1}^{\totAWSdraws}
\]
Draws are made uniformly within upper and lower bounds for each parameter, \(\ESTIeach_{\eachAWSdraws\eachAWSparam} \in \left[\ESTIeach_{\eachAWSparam}^{\text{min}}, \ESTIeach_{\eachAWSparam}^{\text{max}} \right] \). For each draw, I evaluate the model at \(\totAWSobjs\) relevant components of the estimation objective function.\footnote{These could be different moments for moment estimation, or they could be different individual components of the likelihood function. For my estimation here, \(\totAWSobjs\) equals the number of households in each region plus the gaps between observed and model generated period specific equilibrium informal interest rates. This means that there is a separate polynomial approximation for each household's contribution to the overall likelihood.} This creates a \(\totAWSdraws {\times} \totAWSobjs\) matrix \(\modelObjMatrix\), where each element \(\ONEOBJ\) is determined by the \(\eachAWSobjs^{\text{th}}\) component of the estimation objective function, the \(\eachAWSdraws^{\text{th}}\) set of randomly drawn parameters, and  \(\eachAWSobjs\) specific data matrix \(\dataMatrix_\eachAWSobjs\).

Second, I approximate the estimation objective function using polynomials.\footnote{Solving the dynamic programming problem often involves interpolation, sometimes using polynomials \autocite{KeaneWolpinRestat}. In my solution algorithm, I used multi-linear spline rather than polynomials to preserve the relative expected returns of the safe and risky assets in Section \ref{subsec:vfi}. The idea here is to use polynomials to approximate potentially not only the value function but also components of the estimation objective function.} For each set of \(\totAWSparam\) parameters from the \(\totAWSdraws\) draws, I create a \(\totAWSpolyDegree^{\text{th}}\) order full polynomial with \(\totAWSpoly\) polynomial terms.\footnote{\FOOTnestedestiOne} This creates a \(\totAWSdraws {\times} \totAWSpoly\) matrix of polynomials. I regress each \(\eachAWSobjs\) column of matrix \(\modelObjMatrix\) on the polynomial matrix. \(\regCoefMatrix\) denotes the \(\totAWSobjs {\times} \totAWSpoly\) coefficient matrix. For \(\totAWSpolyDegree=3\), for each  \(\eachAWSobjs\), \(\regCoefMatrix_0^{\eachAWSobjs,0}\) is the intercept, \(\regCoefMatrix^{\eachAWSobjs,1}\) is the coefficients vector for the linear terms, \(\regCoefMatrix^{\eachAWSobjs,2}\) is the coefficients vector for the quadratic terms, and \(\regCoefMatrix^{\eachAWSobjs,3}\) is the coefficients vector for the cubic terms. Specifically:
\begin{equation}
  \label{eq:nestestitwo}
  \begin{split}
      \ONEOBJ
      = &
      \regCoefMatrix_0^{\eachAWSobjs,0}
      +
      \SUMA_{\eachAWSdrawsONE=1}^{\totAWSparam}
      \left(
      \regCoefMatrix_{\eachAWSdrawsONE}^{\eachAWSobjs,1}
      \ESTIeach_{\eachAWSdraws\eachAWSdrawsONE}
      \right)\\
      & +
      \SUMA_{\eachAWSdrawsONE=1}^{\totAWSparam}
      \left(
      \SUMA_{\eachAWSdrawsTWO=1}^{\eachAWSdrawsTWO \LESEQ \eachAWSdrawsONE}
      \left(
      \regCoefMatrix_{\eachAWSdrawsONE\eachAWSdrawsTWO}^{\eachAWSobjs,2}
      \ESTIeach_{\eachAWSdraws\eachAWSdrawsONE}
      \ESTIeach_{\eachAWSdraws\eachAWSdrawsTWO}
      \right)
      \right)\\
      & +
      \SUMA_{\eachAWSdrawsONE=1}^{\totAWSparam}
      \left(
      \SUMA_{\eachAWSdrawsTWO=1}^{\eachAWSdrawsTWO \LESEQ \eachAWSdrawsONE}
      \left(
      \SUMA_{\eachAWSdrawsTHREE=1}^{\eachAWSdrawsTHREE \LESEQ \eachAWSdrawsTWO}
      \left(
      \regCoefMatrix_{\eachAWSdrawsONE\eachAWSdrawsTWO\eachAWSdrawsTHREE}^{\eachAWSobjs,3}
      \cdot
      \ESTIeach_{\eachAWSdraws\eachAWSdrawsONE}
      \ESTIeach_{\eachAWSdraws\eachAWSdrawsTWO}
      \ESTIeach_{\eachAWSdraws\eachAWSdrawsTHREE}
      \right)
      \right)
      \right)
      \\
      & + \residualEle_{\eachAWSdraws}^{\eachAWSobjs} \\
  \end{split}
\end{equation}
After  \(\totAWSobjs\) regressions, I obtain a \(\totAWSobjs {\times} \totAWSpoly\) matrix of regression coefficients:
\begin{equation}
  \regCoefMatrixHat(
  \left\{
    \ESTIeach_{\eachAWSdraws\eachAWSparam}
  \right\}_{\eachAWSparam = 1}^{\totAWSparam}
  ,
  \dataMatrix
  ) =
  \left[\begin{array}{@{}c|ccc|ccc|ccc@{}}
      \regCoefMatrixHat^{1,0}
      &
      \regCoefMatrixHat^{1,1}_{1}
      &
      \hdots
      &
      \regCoefMatrixHat^{1,1}_{\totAWSparam}
      &
      \regCoefMatrixHat^{1,2}_{1,1}
      &
      \hdots
      &
      \regCoefMatrixHat^{1,2}_{\totAWSparam,\totAWSparam}
      &
      \regCoefMatrixHat^{1,3}_{1,1,1}
      &
      \hdots
      &
      \regCoefMatrixHat^{1,3}_{\totAWSparam,\totAWSparam,\totAWSparam}
      \\
      \vdots & \vdots & \vdots & \vdots & \vdots & \vdots & \vdots & \vdots & \vdots& \vdots
      \\
      \regCoefMatrixHat^{\totAWSobjs,0}
      &
      \regCoefMatrixHat^{\totAWSobjs,1}_{1}
      &
      \hdots
      &
      \regCoefMatrixHat^{\totAWSobjs,1}_{\totAWSparam}
      &
      \regCoefMatrixHat^{\totAWSobjs,2}_{1,1}
      &
      \hdots
      &
      \regCoefMatrixHat^{\totAWSobjs,2}_{\totAWSparam,\totAWSparam}
      &
      \regCoefMatrixHat^{\totAWSobjs,3}_{1,1,1}
      &
      \hdots
      &
      \regCoefMatrixHat^{\totAWSobjs,3}_{\totAWSparam,\totAWSparam,\totAWSparam}
    \end{array}\right]
\end{equation}

Third, I estimate the model now by evaluating the objective function using the polynomial approximation matrix \(\regCoefMatrixHat\):
\begin{equation}
  \label{eq:nestestithree}
  \min_{
  \substack{
    \ESTIALL\\
    \text{initialize at: } \ESTIALL_{\eachAWSdraws}\\
  }
  }
  \modelObjMatrixHat
  \left(
    \left\{
      \modelObjMatrixHat^{\eachAWSobjs}
      \left(
        \left\{
          \ESTIeach_{\eachAWSparam}
        \right\}_{\eachAWSparam = 1}^{\totAWSparam}
        ,
        \regCoefMatrixHat^{\eachAWSobjs}
          \left(
            \dataMatrix_\eachAWSobjs
          \right)
      \right)
    \right\}_{\eachAWSobjs=1}^{\totAWSobjs}
  \right)
\end{equation}
I estimate this approximated objective function at the \(\totAWSdraws\) sets of parameter draws. The parameter space should have the same bounds as before to avoid extrapolation. Evaluating \(\modelObjMatrixHat\) is fast, and derivative-based numerical algorithms can efficiently and stably obtain local minima given polynomials. Initializing estimation at each \(\ESTIALL_{\eachAWSdraws}\) draws, I obtain a vector of parameters,
\(
\left\{
  \ESTIAPPROX{\eachAWSdraws}
\right\}_{\eachAWSdraws=1}^{\totAWSdraws}
\), which locally minimize \(\modelObjMatrixHat\) given different starting points.

Fourth, I estimate the model with the non-polynomial approximated objective functions using results from earlier steps as starting points. Specifically:
\begin{equation}
  \label{eq:nestestifour}
  \begin{split}
    \min_{
    \substack{
      \ESTIALL\\
      \text{initialize at: }
      \ESTIALL^{\text{init}}\\
    }
    }
  \modelObjMatrix
  \left(
    \left\{
      \modelObjMatrix^{\eachAWSobjs}
      \left(
        \left\{
          \ESTIeach_{\eachAWSparam}
        \right\}_{\eachAWSparam = 1}^{\totAWSparam}
        \condi
        \dataMatrix_\eachAWSobjs
      \right)
    \right\}_{\eachAWSobjs=1}^{\totAWSobjs}
  \right)\\
  \text{where: }
  \ESTIALL^{\text{init}} =
  \argmin_{
    \left\{
      \ESTIAPPROX{\eachAWSdraws}
    \right\}_{\eachAWSdraws=1}^{\totAWSdraws}
  }
  \left\{
   \modelObjMatrixHat
     \left(
     \ESTIAPPROX{1}
     \right)
   \text{, ..., }
   \modelObjMatrixHat
     \left(
     \ESTIAPPROX{\totAWSdraws}
     \right)
  \right\}\\
  \end{split}
\end{equation}
The first three steps provide Equation \eqref{eq:nestestifour} with starting values that are potentially close to the global minimizers. For this step, one could implement standard routines and use parallel resources for multi-start. One could also use parallel iterative optimizing algorithms \autocite{guvenen_macroeconomics_2011}. If Equation \eqref{eq:nestestitwo} has very high explanatory power, then \(\ESTIALL^{\text{init}}\) would be close to global minimizers, and convergence should be fast. If Equation \eqref{eq:nestestitwo} has low explanatory power, the first three steps of this algorithm might not improve convergence speed for the fourth step. When that is the case, it indicates potentially that convergence to a global minimum using any estimation routine could be difficult, and the econometrician might need to adjust the estimation objective function or transform the estimation parameter domain.

The first three steps described above are parallelizable. To implement, the model would be solved concurrently across \(\totAWSdraws\) parameter draws. In the first step, the time it takes to solve the model \(\totAWSdraws\) times is the same as solving the model once. The second step involves \(\totAWSobjs\) linear projections. The third step is also fully parallelizable, and each task has a minimal computational footprint. The fourth step is computationally intensive, but with the first three steps providing the final step with potentially good initial guesses, the number of computationally-intensive full estimation tasks is potentially dramatically lowered.
 \subsubsection{Container Orchestration}
\label{subsec:containerorchestration}

Container orchestration allows for large-scale multi-start simulation or estimation. The idea is to first "containerize" the model codes and dependencies, then specify the CPU and memory requirements of the container and run separate processes across containers. Before proceeding to the polynomial approximation step, all containerized tasks need to be completed. Under managed container services, depending on the CPU-to-memory ratio as well as the price and availability of EC2 units, AWS determines which sets of machines to concurrently start to complete the containerized tasks.
 \subsubsection{AWS Cluster and Containers}
\label{subsec:awsbatchqueue}

I create computing clusters with the maximum CPU count set at 2560. For each container, memory requirement is set at the maximum level required for solving the model. Computing resources can be used more efficiently when the model solution code has consistent and flat memory usages. I set the CPU requirement for each containerized task at 8, and the memory requirement at 45 GB. An EC2 instance with 64 CPUs and 360 GB of memory can run eight containers concurrently.
 \subsubsection{Costs}
\label{subsec:awsprice}

AWS launches a combination of these types of instances during my estimation runs: \textit{m4.16xlarge} instances (each with 64 virtual CPUs and 256 GB of memory), \textit{m5.24xlarge} instances (each with 96 virtual CPUs and 384 GB of memory), and \textit{r4.16xlarge} instances (each with 64 virtual CPUs and 488 GB of memory). Using Spot pricings, these three main instance types in fall of 2018 generally cost \$1 per dollar (in the fall of 2018).

On average, the model algorithm, described previously, takes 2 minutes and 30 seconds to complete using 8 CPUs. For estimation, given each set of parameters, I need to run the model four times for the two periods and two regions. This means that given each set of parameters, finding the overall estimation objective takes 10 minutes for each container.

For the initial step of the estimation process described in Section \ref{subsec:nestedestimation}, the estimation objective is solved at 12800 random initial points (6400 for each region). Given this, the first three steps of the estimation algorithm take about 3 hours with total cluster maximum CPU count set at 2560, and costs between \$100 to \$150. Holding the model fixed, if the econometrician decides to change which components to include in the estimation objective or how they are weighted, this first costly step does not need to be repeated.

\pagebreak


\end{document}